\documentclass{article}

\usepackage{PRIMEarxiv}

\usepackage[utf8]{inputenc} 
\usepackage[T1]{fontenc}    
\usepackage{hyperref}       
\usepackage{url}            
\usepackage{booktabs}       
\usepackage{amsfonts}       
\usepackage{nicefrac}       
\usepackage{microtype}      
\usepackage{lipsum}
\usepackage{fancyhdr}       
\usepackage{graphicx}       
\usepackage{amsmath}

\usepackage[%
    binary-units=true,
    prefixes-as-symbols=false,
]{siunitx}
\newcommand*{\kT}{k_{\textup{B}}\textup{T}}

\DeclareSIUnit{\microsecond}{\SIUnitSymbolMicro s}

\title{Fitting Coarse-Grained Models to Macroscopic Experimental Data via Automatic Differentiation
}

\author{
  Ryan K. Krueger \\
  School of Engineering and Applied Sciences \\
  Harvard University \\
  Cambridge, MA\\
  \texttt{ryan\_krueger@g.harvard.edu} \\
   \And
  Megan C. Engel \\
  Department of Biological Sciences \\
  University of Calgary \\
  Calgary, AB, Canada \\
  \texttt{megan.engel@ucalgary.ca} \\
  \And
  Ryan Hausen \\
  Institute for Data Intensive Engineering and Science \\
  Johns Hopkins University \\
  Baltimore, MD \\
  \texttt{rhausen@jhu.edu} \\
  \And
  Michael P. Brenner \\
  School of Engineering and Applied Sciences \\
  Harvard University \\
  Cambridge, MA \\[3pt] 
  Department of Physics\\ 
  Harvard University\\
  Cambridge, MA\\
  \texttt{brenner@seas.harvard.edu} \\
}

\begin{document}
\maketitle

\begin{abstract}
Developing physics-based models for molecular simulation requires fitting many unknown parameters to diverse experimental datasets. 
Traditionally, this process is piecemeal and difficult to reproduce, leading to a fragmented landscape of models. Here, we establish a systematic, extensible framework for fitting coarse-grained molecular models to macroscopic experimental data by leveraging recently developed methods for computing low-variance gradient estimates with automatic differentiation. 
Using a widely validated DNA force field as an exemplar, we develop methods for optimizing structural, mechanical, and thermodynamic properties across a range of simulation techniques, including enhanced sampling and external forcing, spanning micro- and millisecond timescales.
We highlight how gradients enable efficient sensitivity analyses that yield physical insight.
We then demonstrate the broad applicability of these techniques by optimizing diverse biomolecular systems, including RNA and DNA-protein hybrid models.
We show how conflict-free gradient methods from multi-task learning can be adapted to impose multiple constraints simultaneously without compromising accuracy. 
This approach provides a foundation for transparent, reproducible, community-driven force field development, accelerating progress in molecular modeling.
\end{abstract}


\section{Introduction}

Physics-based molecular simulations using coarse-grained models are a cornerstone of \emph{in silica} biophysical modeling.
When sufficiently accurate, they provide a unique way to study a biophysical system beyond available experimental data~\cite{drorBiomolecularSimulationComputational2012}.
Such models have provided deep insights ranging from viral assembly~\cite{perlmutter2015mechanisms, huber2017multiscale, ruiz2015simulations, ode2012molecular} to protein folding~\cite{scheraga2007protein, swope2004describing, lindorff2011fast} to chromosomal organization~\cite{rosa2008structure, di2016transferable}.

Recent years have seen extraordinary progress in the accuracy of large-scale machine learning (ML) models, from large language models \cite{achiam2023gpt,team2024gemini} to protein structure prediction \cite{jumper2021highly}.
Donoho hypothesized \cite{donoho2024data} that {\sl frictionless reproducibility}, the ease of reproducing and iteratively improving models, is the major factor leading to dramatic ML model improvement.
Within ML, model parameters are almost universally fit via gradient-based optimization with respect to well-defined objective functions, enabling seamless reproducibility, extensibility, and composability with models and fitting protocols made available via open-source platforms (e.g. GitHub, Hugging Face).
These standardized and transparent fitting procedures enable rapid technological innovation~\cite{krizhevsky2012imagenet, brown2020language, devlin2019bert, lecun2015deep}.

Comparatively, the accuracy of coarse-grained potentials has stagnated.
Despite resounding successes in specific molecular models, force fields are often the product of ``black magic'' fitting procedures that are difficult to reproduce, typically involving significant hand design.
Even relatively principled fitting procedures such as Iterative Boltzmann Inversion~\cite{maffeoCoarseGrainedModelUnstructured2014,assenzaAccurateSequenceDependentCoarseGrained2022}, maximum likelihood methods~\cite{heOptimizationNucleicAcids2015}, or Newton Inversion~\cite{naomeSolventMediatedCoarseGrainedModel2014} are often performed piecemeal, with each interaction parameterized individually rather than performing a global optimization of the entire parameter set.
As such, existing models cannot be easily adapted to account for new effects or incorporate new experimental data.
A case in point is nucleic acid force field development, with (i) expanded parameterizations of DNA/RNA models appearing years after the original model release~\cite{vsulc2012sequence, vsulc2014nucleotide, procyk2021coarse, snodin2015introducing, ratajczyk2024coarse} and (ii) entirely new models developed to resolve existing shortcomings~\cite{assenzaAccurateSequenceDependentCoarseGrained2022},
attesting to the difficulty and laboriousness of refining models where parameterization is nontransparent and unsystematic.
The ultimate result is a fragmented ``zoo'' of models that each require months or years of trial-and-error parameter fitting despite attempting to represent similar datasets.

\begin{figure}[t!]
\begin{center}
\centerline{\includegraphics[width=0.85\textwidth]{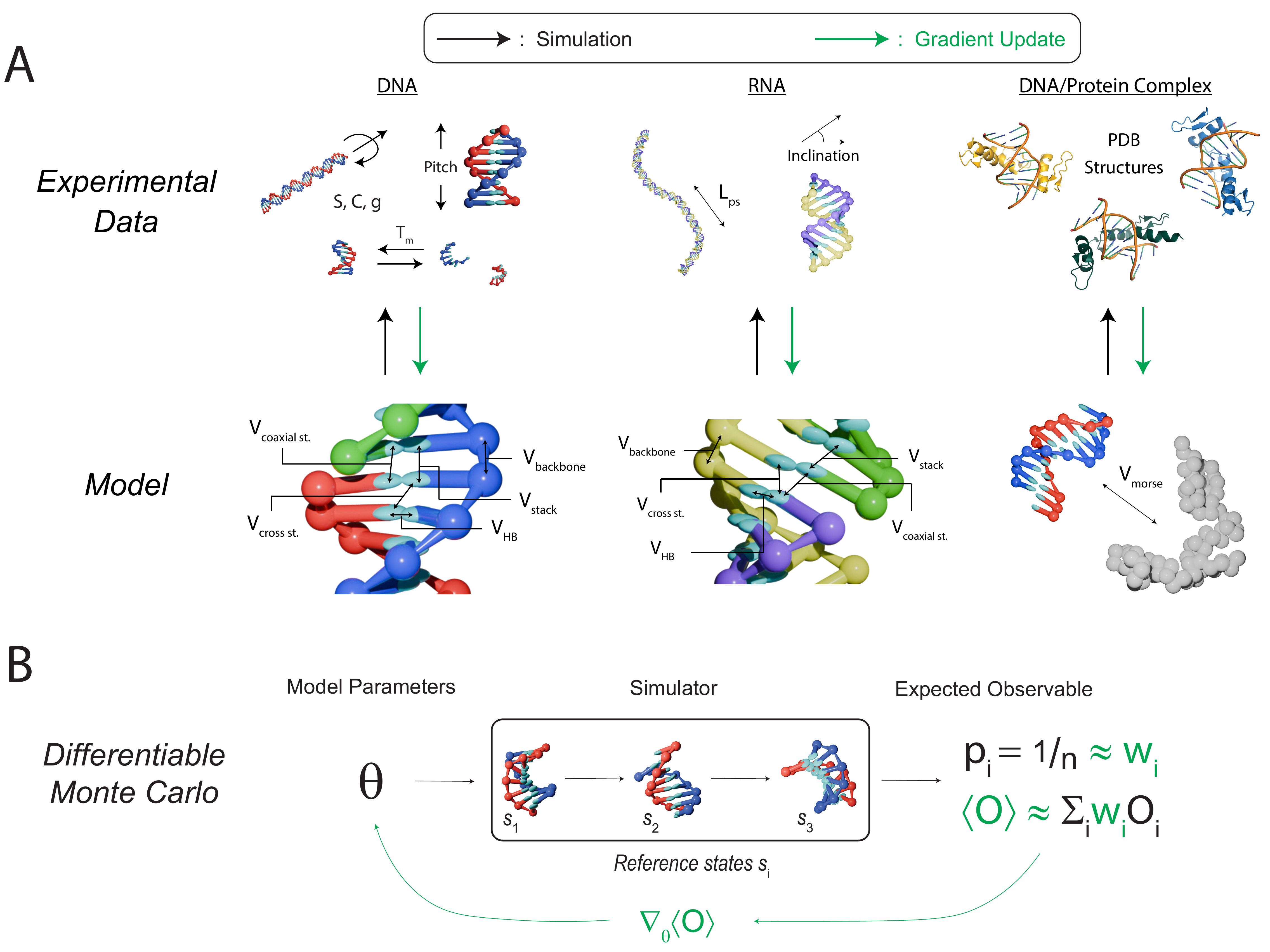}}
\caption{
Automatically fitting coarse-grained models to experimental data.
\textbf{A.} For a given molecular system (e.g. from left to right, DNA, RNA, DNA-protein complexes), a loss function is defined that characterizes agreement between simulated macroscopic properties and experimental data.
The gradient of this loss function is computed via automatic differentiation to update the parameters of the energy function.
The depiction of the DNA and RNA models are adapted from Refs. \cite{doye2023oxdna} and \cite{vsulc2014nucleotide}.
\textbf{B.} An overview of differentiable Monte Carlo (or ``DiffTRE''), a recently developed means of stochastic gradient estimation that operates at the level of unnormalized probability distributions~\cite{thaler2021learning, zhang2023automatic}. 
Reference states are sampled from the steady state distribution and the expected value of the target (time-independent) observable, $\langle O \rangle_{\theta}$, is reweighted via a set of differentiable weights.
This provides a direct functional relationship between $\theta$ and $\langle O \rangle_{\theta}$ without differentiating through the simulation~\cite{thaler2021learning}; see \emph{Materials and Methods} for details.
}
\label{fig:overview}
\end{center}
\vspace{-4mm}
\end{figure}

Here, we establish how coarse-grained, physics-based potentials can be fit to a broad class of experimental data with ``frictionless reproducibility'' (Figure \ref{fig:overview}A).
We leverage \emph{differentiable programming}, the backbone of machine learning algorithms whereby gradients are computed via automatic differentiation (AD).
Traditional differentiable molecular dynamics (MD) is plagued by three core problems.
First, computing the gradient through long unrolled simulations imposes significant memory overhead and numerical instabilities~\cite{metz2021gradients, greener2024differentiable}.
Second, many fundamental algorithms in molecular simulation are not end-to-end differentiable~\cite{whitelam2007avoiding, ruuvzivcka2014collective, vitalis2009methods}.
Third, decades of research have been dedicated to optimizing MD algorithms to increase accessible timescales, but many of these optimizations are lost upon reimplementation in an automatic differentiation framework~\cite{schoenholz2020jax}.

We surmount these limitations using a novel method for stochastic gradient estimation \cite{zhang2023automatic,thaler2021learning}, making it possible to use gradient-descent to fit force field parameters to three broad classes of experimental data: structural, mechanical, and thermodynamic.
Each class of data poses a unique set of challenges that are not present in toy models previously studied with differentiable MD.
For example, calculation of mechanical properties, such as persistence length or stretch modulus, typically requires micro- and millisecond-scale simulations to sample many uncorrelated states, line-fitting procedures to infer summary statistics, and the application of external forces and torques~\cite{ouldridge2010dna, vsulc2014nucleotide, assenzaAccurateSequenceDependentCoarseGrained2022, hinckley2013experimentally, ouldridgeStructuralMechanicalThermodynamic2011a, uusitalo2015martini, naskar2021mechanical, braun2014determining, franz2020advances, yao2001mechanical, tsai2010characterizing, root2016predicting, chen2005mechanical, gautieri2010coarse}.
Similarly, thermodynamic (e.g. melting temperature) calculations require particularly complex simulations, involving enhanced sampling, non-differentiable Monte Carlo algorithms, and temperature extrapolation~\cite{ouldridge2010dna, vsulc2014nucleotide, zerze2021thermodynamics, wong2008pathway, ouldridgeStructuralMechanicalThermodynamic2011a, hinckley2013experimentally, aimoli2014force, kumar1992weighted, shirts2008statistically, deublein2011ms2}.
We devise frameworks for differentiating each such class of complex simulations.
We also demonstrate how the calculated gradients can be adapted for sensitivity analyses and therefore the rapid derivation of physical insights.

Finally, we combine the methods developed for each individual class of data to fit parameters to {\sl single} objective functions that jointly target structural, mechanical, and thermodynamic properties, allowing for the first time a systematic, simultaneous parameter optimization for different loss functions. 
We demonstrate how \emph{conflict-free gradients} can be adapted from multi-task learning~\cite{yu2020gradient, liu2021towards, long2017learning, kirkpatrick2017overcoming, liu2024config} to impose individual hard constraints despite expressing a global objective function.

In what follows, we demonstrate our approach by focusing first on an illustrative case study: nucleic acid force field development within oxDNA~\cite{sengar2021primer}, a popular model of DNA that has enjoyed extensive experimental validation~\cite{ouldridge2010dna, rovigatti2014gels, romano2015switching}. 
To reinforce the applicability of our approach to any coarse-grained model, we then detail sample extensions of our method to RNA and DNA-protein systems.

\begin{figure}[t!]
\begin{center}
\centerline{\includegraphics[width=0.9\textwidth]{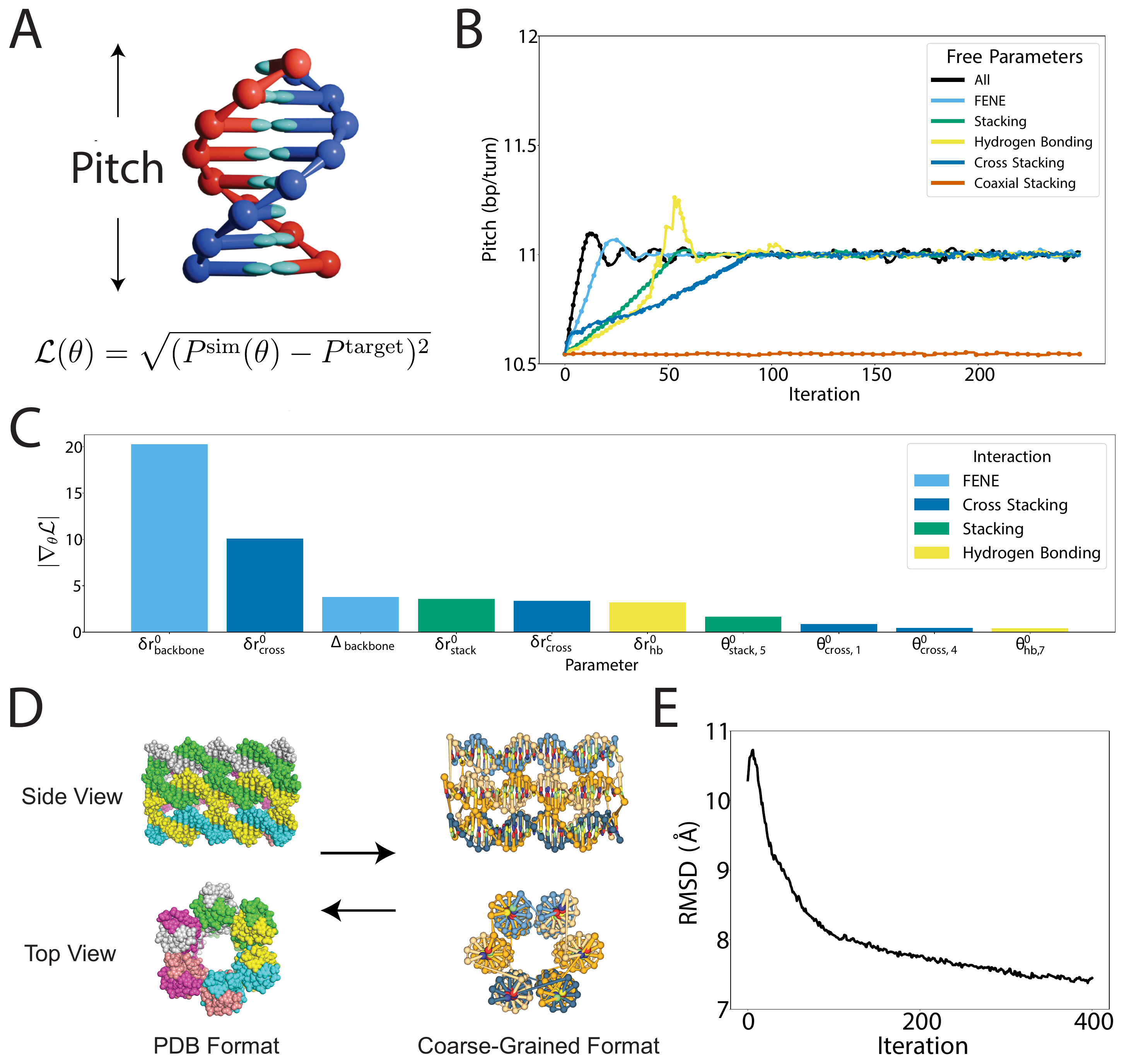}}
\caption{Parameter optimization for target structural properties via trajectory reweighting. 
\textbf{A.} 
A depiction of pitch, defined as the axial distance over which the helix makes one complete turn, and our loss function for fitting energy function parameters $\theta$ to achieve a target pitch.
We define the loss as the root mean squared error between the simulated pitch $P^{\text{sim}}(\theta)$ and the target pitch $P^{\text{target}}$.
\textbf{B.} The pitch as a function of iteration when using the gradients calculated via DiffTRE for gradient-based optimization. 
Iterations at which reference states were resampled are represented by scatter points.
Varying only coaxial stacking parameters (orange line) fails because coaxial stacking -- the stacking interaction between contiguous but distinct strands on a single side of a DNA duplex -- was absent in these simulations. 
\textbf{C.} The ten most significant parameters ranked by the absolute value of their gradients.
\textbf{D.} The gated DNA nanopore introduced in Ref. \cite{burns2016biomimetic}. 
The structure is depicted both in PDB format and in the coarse-grained representation employed by oxDNA.
\textbf{E.} The average RMSD of the simulated structure compared to the target structure when applying gradient-based optimization to maximize agreement between simulation and the ground-truth  sequence/structure pair.
This optimization uses the sequence-dependent variant of oxDNA2.
}
\label{fig:good-structural}
\end{center}
\end{figure}

\section{Results}

\subsection{Differentiable Force Field Implementation}

The first step towards gradient-based optimization of a molecular force field is implementing it in an AD library.
We implement the oxDNA energy function in JAX, a state-of-the-art AD library for scientific computing~\cite{jax2018github}, enabling the automatic calculation of arbitrary higher-order derivatives with respect to the control parameters.
Unless otherwise stated, we focus our results on the latest sequence-averaged version of oxDNA, oxDNA2~\cite{snodin2015introducing}. 
The oxDNA2 model comprises 103 total parameters governing potential energy functions that capture excluded volume, base pairing, base stacking, and electrostatic interactions. 
These parameters were originally fit to structural, mechanical, and thermodynamic experimental data; we treat each in turn below. 
For further details of the model, see \textit{Materials and Methods}.
Using our implementation of the energy function, we can simulate systems of DNA using JAX-MD, an end-to-end differentiable molecular dynamics engine written in JAX~\cite{schoenholz2020jax}.
We benchmark these simulations against the standalone oxDNA code~\cite{poppleton2023oxdna, rovigatti2015comparison}, involving both highly-optimized C++ and CUDA implementations, and find that simulation via JAX-MD is 2-3x slower (Figure \ref{fig:benchmark}) -- a good compromise for the increased flexibility offered by the differentiable framework.

The JAX implementation offers two key advantages.
First, since we write our energy function in an AD framework, forces are computed automatically from energy functions.
This significantly reduces the barrier to modifying the functional form of the energy function and also permits the application of arbitrary (continuous and differentiable) bias potentials.
Second, since the code can be compiled to a range of hardware accelerators, we can immediately run simulations on GPU and TPU without reimplementing the force field in a separate framework.
Therefore, molecular simulation in an AD framework can be seen as complementary to highly optimized MD frameworks: beyond permitting gradient-based optimization, it enables flexibility and extensibility for force field development.

\subsection{Structural Optimization}
With a differentiable force field implementation in hand, we turn to optimizing parameters. 
Structural properties are generally the most amenable to optimization via differentiable MD since these properties converge even in relatively short simulations of small systems.
As a first step, we fit a subset of oxDNA parameters to target pitch (Figure \ref{fig:good-structural}A) and propeller twist (Figure \ref{fig:ptwist}) values. 
We found that the two traditional means of stochastic gradient estimation for differentiable MD (Figure \ref{fig:traditional-grad-est}) were not sufficient for optimization in this case.
Firstly, the pathwise estimator~\cite{mohamed2020monte}, whereby gradients are computed directly through an unrolled numerical simulation, imposes significant time and memory overhead and leads to exploding gradients~\cite{metz2021gradients}.
The alternative, the score function estimator~\cite{mohamed2020monte} -- by which batches of simulations are performed and the probability of such simulations are increased or decreased according to the calculated observable -- did not yield sufficient signal for optimization, likely due to unwieldy variance.
See Figure \ref{fig:benchmark} for benchmarking and Figure \ref{fig:reinforce_and_reparam} for these optimization results with varying hyperparameters.

We instead turn to a recently developed method of stochastic gradient estimation that directly operates at the level of unnormalized probability distributions (Figure \ref{fig:overview}B).
This method was first introduced for differentiable Monte Carlo in the context of variational quantum Monte Carlo by Zhang et al.~\cite{zhang2023automatic} and for  machine-learned molecular simulation potentials~\cite{thaler2021learning}, where it was termed ``differentiable trajectory reweighting'' (DiffTRE). 
The crux of the method is establishing a direct functional relationship between $\theta$, the control parameters of a distribution $p(x; \theta)$, and the expected value of a state-level observable $O$, $\mathbb{E}_{p(x; \theta)}[O(x; \theta)]$, that is independent of the sampling procedure.
This is achieved via a perturbative calculation that expresses the expected value via a differentiable set of weights; see \textit{Materials and Methods} for details.
This estimator circumvents the memory and numerical issues associated with traditional differentiable MD by only differentiating the \textit{energies} of reference states sampled from a set of simulations, rather than full trajectories.

We find that for fitting structural properties to oxDNA, DiffTRE provides a numerically stable, scalable, and efficient method for optimization via gradient descent.
Unlike the pathwise and score function estimators, this method yielded oxDNA parameters for specified pitch values within 50 iterations (Figure \ref{fig:good-structural}B), and the procedure was stable across various target pitch values. 
It is important to note that DiffTRE does not require that the sampling procedure itself is differentiable: reference states can be collected with simulations of any standalone MD code -- in our case, the standalone oxDNA implementation~\cite{poppleton2023oxdna}.

\begin{figure}[t!]
\begin{center}
\centerline{\includegraphics[width=0.95\textwidth]{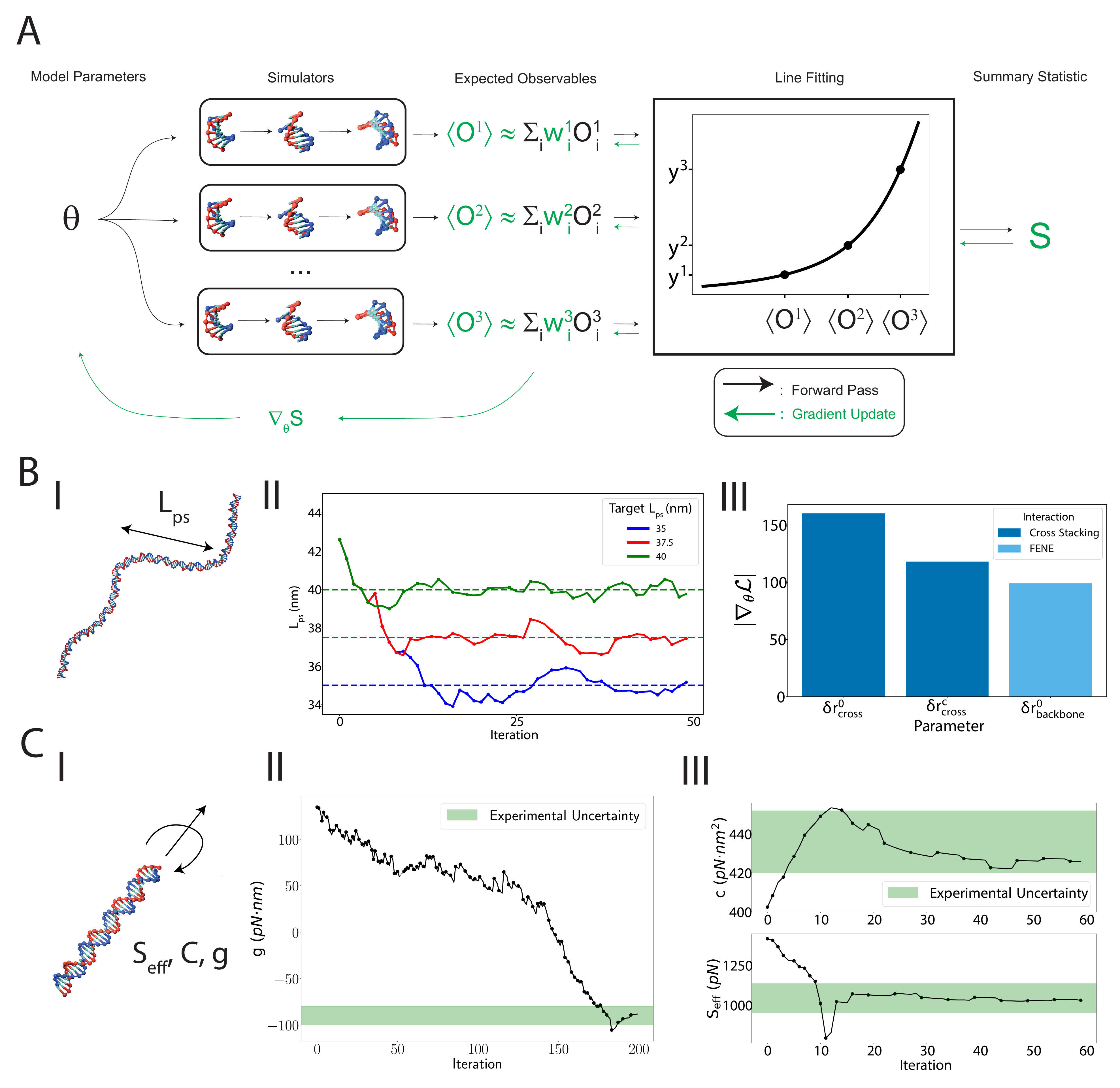}}
\caption{
Parameter optimization for target mechanical properties via trajectory reweighting. 
\textbf{A.} Our method for parameter optimization for complex target mechanical properties via DiffTRE and implicit derivatives.
Individual ensembles are generated corresponding to the system's response to a varying mechanical stimulus.
In practice, multiple observables may be computed from a single ensemble (e.g. correlations at different distances in the passive calculation of persistence length).
These values are then used to infer a summary statistic via an arbitrary line-fitting procedure.
Individual observables are differentiated via DiffTRE~\cite{thaler2021learning} and line-fitting procedures are differentiated via implicit differentiation~\cite{blondel2022efficient}.
\textbf{B.} Parameter optimization for target values of persistence length ($L_{ps}$). \textbf{I.} A depiction of $L_{ps}$. \textbf{II.} Optimization of all oxDNA parameters to match various target $L_{ps}$ values. \textbf{III.} The top three most important parameters ranked via the absolute value of the gradient in the initial optimization step.
\textbf{C.} Parameter optimization for target values of moduli characterizing response to external force and torque. 
\textbf{I.} A depiction of a duplex under external force and torque. \textbf{II-III.} Optimization of all oxDNA parameters to match the twist-stretch coupling, torsional modulus, and effective stretching modulus, all of which can be brought to experimental agreement (green bars) within error.
Outliers in the twist-stretch coupling optimization at iterations 52 and 53 are omitted and replaced with a dashed line.
See Figure \ref{fig:moduli_sensitivity} for corresponding sensitivity analyses.
For \textbf{B}-\textbf{C}, iterations at which reference states were resampled are represented by scatter points.
}
\label{fig:mech}
\end{center}
\end{figure}

For a physics-based model, the gradient contains information about the most important physical interactions for a target property, accessible by ranking parameters based on the magnitude of their gradients (Figure \ref{fig:good-structural}C). 
In this work, we consider the gradient magnitudes in the first step of the optimization procedure, but more sophisticated analysis schemes are possible.
For pitch, we find that the most significant parameter is the equilibrium distance of the backbone sites, as previously found~\cite{snodin2015introducing}. More surprisingly, we find that cross-stacking interactions are quite important: this interaction is most often implicated for controlling duplex melting temperatures given its stabilizing role~\cite{ouldridge2012coarse, swart2007pi, vsponer2006nature}.
Sensitivity analysis for propeller twist (Figure \ref{fig:ptwist}) also reveals that the two most important parameters are involved in the cross-stacking interaction.

Finally, we can fit to macroscopic structural properties rather than those defined locally.
As an example, we consider the gated DNA nanopore introduced in Ref. \cite{burns2016biomimetic}.
We minimize the root mean square deviation (RMSD) between the simulated structure and the ground-truth structure deposited in Nanobase~\cite{poppleton2022nanobase}, directly improving agreement between simulation and experiment (Figure \ref{fig:good-structural}D-E and Figure \ref{fig:structure_opt}).

\subsection{Mechanical Optimization}

A second broad class of data for fitting coarse-grained potentials is mechanical data, which characterize the response of a molecular system to force and torque. 
Computing mechanical properties introduces a host of challenges: while structural properties are typically defined locally and thus only require short ($\sim$ 10 \si{\pico\second}) simulations of small (10s of particles) systems for convergence, mechanical properties converge over significantly longer time and length scales.
For example, oxDNA simulations to compute the persistence length ($L_{ps}$) of dsDNA typically require $\sim$10 \textmu s simulations of 100s of nucleotides. 
Such simulations are well beyond the reach of traditional differentiable MD.
Second, unlike structural properties, mechanical properties cannot be expressed as an arithmetic mean of state-level observables.
Instead, we typically calculate mechanical properties fitting the ensemble response to force or torque, such as fitting to a worm-like chain model. 
The parameters from this fit then give summary statistics to match experimental data.

To solve these challenges, we extend DiffTRE to compute derivatives with respect to arbitrary functions of expected observables (Figure \ref{fig:mech}A) as follows.
Consider a set of observables $\overrightarrow{O} = \{O^1\text{, }O^2\text{, } \ldots\text{, } {O^m}\}$ where the $i^{th}$ observable corresponds to the response of the system to the $i^{th}$ value used to infer an ensemble response, e.g. $O^i$ corresponds to the extension of a duplex subject to the $i^{th}$ value of external force.
Given all $m$ values, we compute the summary statistic of interest $S$ via an arbitrary line-fitting procedure $\ell$, i.e. $\ell(\overrightarrow{O}) = S$.
Figure \ref{fig:mech}A gives an overview of our procedure for computing $\nabla_{\theta}S$ for control parameters $\theta$. We use implicit differentiation~\cite{blondel2022efficient} on the fitting parameters and combine this with trajectory reweighting, rendering the entire calculation of mechanical properties end-to-end differentiable. Details of our implementation can be found in the Supporting Information. 
Trajectory reweighting naturally accommodates ensembles involving constant external forces and torques since their corresponding effect on the probability of a microstate does not depend explicitly on $\theta$ (see Equation S\ref{eq:Uext_cancels}).

\begin{figure}[t!]
\begin{center}
\centerline{\includegraphics[width=0.95\textwidth]{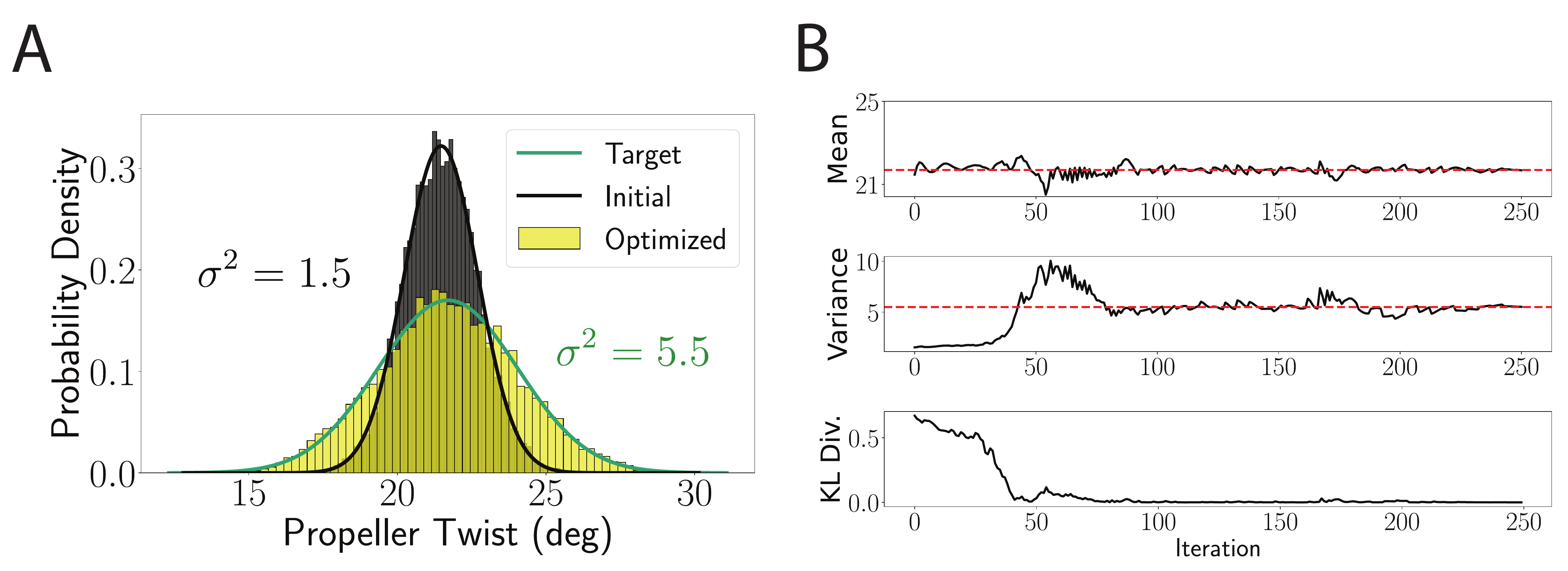}}
\caption{
Fitting to a target propeller twist distribution.
\textbf{A.} We optimize the parameters to achieve a target distribution of propeller twists.
Here, we consider the case of increasing the variance ($\sigma^2 = 5.5$) while maintaining the mean ($\mu = 21.7^{\circ}$).
The loss is measured as the KL-divergence between the simulated distribution and the target distribution.
The initial and target distributions are represented as Gaussians.
We also plot histograms of the sampled propeller twists with the initial parameters and the optimized parameters.
The initial parameters yield a distribution with $\sigma^2 = 1.5$.
\textbf{B.} The progression of the mean, variance, and KL-divergence (i.e. the loss) throughout the optimization.
Target values are depicted as dashed red lines.
}
\label{fig:ptwist-dist}
\end{center}
\end{figure}

As a specific example, $L_{ps}$ can be calculated passively by treating DNA as an infinitely long, semi-flexible polymer (Figure \ref{fig:mech}B-I).
Under this assumption, correlations in alignment decay exponentially with separation and $L_{ps}$ obeys the following equality:
\begin{align}
    \langle \mathbf{l_m} \cdot \mathbf{l_0} \rangle = \exp (-m \langle l_0 \rangle / L_{ps})
\end{align}
where $\mathbf{l_0}$ denotes the vector between the first two monomers and $\langle \mathbf{l_m} \cdot \mathbf{l_0} \rangle$ the correlation between the tangent vectors corresponding to the first and $n^{th}$ monomers~\cite{ouldridgeStructuralMechanicalThermodynamic2011a,cantor1980biophysical}.
As shown in Figure \ref{fig:mech}B-II, our framework provides a robust means of optimizing $L_{ps}$;
see the Supporting Information for simulation and calculation details.
As in the structural case, sensitivity analysis also reveals a strong dependence of $L_{ps}$ on the cross-stacking interaction (Figure \ref{fig:mech}B-III); this is surprising given that previous work focused on hydrogen-bonding and stacking interactions when fitting new persistence length data~\cite{snodin2015introducing}.

We demonstrate the generality of our approach by fitting to additional mechanical properties: the torsional modulus $C$, the effective stretch modulus $S_{\text{eff}}$, and the twist-stretch coupling $g$ (Figure \ref{fig:mech}C-I). 
Computing these properties requires fitting the linear dependence of changes in twist and extension as a function of applied force and torque with analytical formulas derived from the equipartition theorem~\cite{assenzaAccurateSequenceDependentCoarseGrained2022, sassi2017shape}.
For these optimizations, we sample reference states using the LAMMPS implementation of oxDNA~\cite{henrich2018coarse} given existing simulation protocols~\cite{assenzaAccurateSequenceDependentCoarseGrained2022}.
Strikingly, our method yields a parameter set that successfully models the overwinding of DNA under low force, represented by a negative value of $g$, within experimental error (Figure \ref{fig:mech}C-II).
The original oxDNA model incorrectly predicts \textit{under}winding.
In addition, we are able to achieve $C$ and $S_{\text{eff}}$ values that agree with experiments (within experimental uncertainty); see Figure \ref{fig:mech}C-III.
Again, sensitivity analysis reveals a strong dependence of all three quantities on the cross-stacking interaction (Figure \ref{fig:moduli_sensitivity}).

\begin{figure}[t!]
\begin{center}
\centerline{\includegraphics[width=0.95\textwidth]{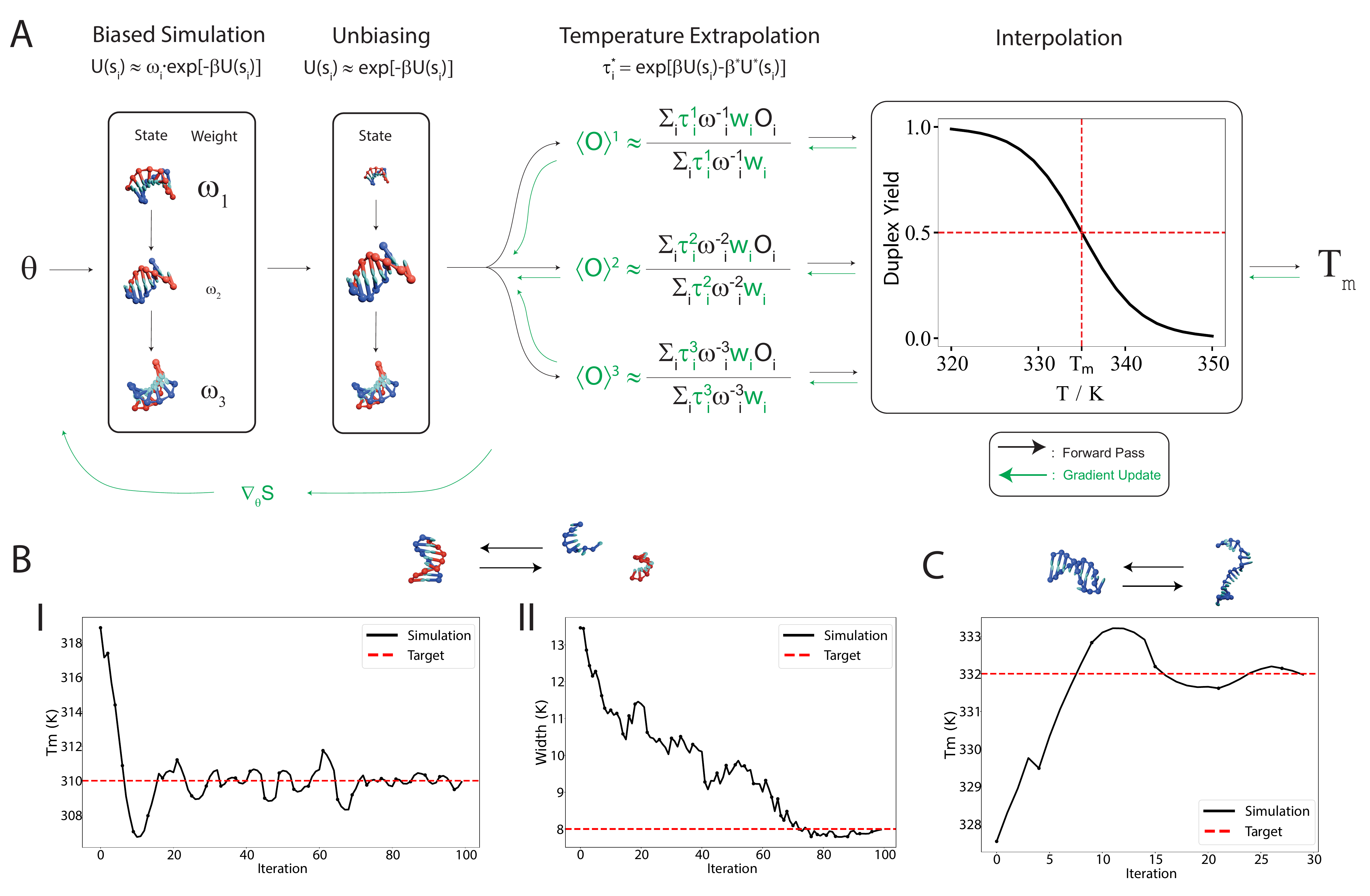}}
\caption{Parameter optimization for target thermodynamic properties via trajectory reweighting. 
\textbf{A.} 
Our method for fitting energy functions to complex thermodynamic properties via enhanced sampling, DiffTRE~\cite{thaler2021learning}, and implicit derivatives~\cite{blondel2022efficient}.
A biased ensemble is generated via equilibrium enhanced sampling methods (e.g. umbrella sampling) at temperature $T$ with biasing weights $\omega_i$ and subsequently reweighted to obtain unbiased statistics.
For a different temperature $T^*$, these biasing weights combined with temperature-dependent extrapolation weights $\tau^*_i$ define unnormalized probabilities of the sampled states at $T^*$.
We can therefore extrapolate the expected observable to a range of temperature values, all of which are differentiable via DiffTRE.
As with mechanical properties, summary statistics of the temperature response can be computed via arbitrary line fitting procedures, which are differentiable via implicit differentiation.
We apply this scheme to optimize the hydrogen bonding parameters in oxDNA to achieve \textbf{B.I.} a melting temperature of $T_m = 310\text{ K}$ and \textbf{B.II.} a melting transition width of $8\text{ K}$ for an 8 base pair duplex at a salt concentration of $0.5 \text{ M}$ with both strands having a total concentration of $3.3 \times 10^{-4} \text{ M}$, and \textbf{C.} a melting temperature of $T_m = 332\text{ K}$ for a hairpin with a stem length of 6 base pairs and a loop consisting of 6 unpaired nucleotides at a salt concentration of $0.25 \text{ M}$.
The duplex and hairpin optimizations are performed using oxDNA1 and oxDNA2, respectively.
For \textbf{B}-\textbf{C}, iterations at which reference states were resampled are represented by scatter points.
}
\label{fig:thermo}
\end{center}
\end{figure}

The framework outlined above can also be applied to fit \emph{distributions} of equilibrium structural properties, as this requires fitting the functional form of a probability distribution to a histogram of sampled quantities. 
To illustrate this, we fit to a target equilibrium distribution of propeller twists (Figure \ref{fig:ptwist-dist}).
Likewise, calculating \emph{worm-like chain} (WLC) properties requires fitting a non-linear force-extension response~\cite{odijk1995stiff}.
Though the WLC fit is highly sensitive to small changes in measured extensions and therefore challenging to optimize directly, we can successfully optimize the stretch modulus $S$ by constraining the fit to the value of $L_{ps}$ computed via passive simulations (Figure \ref{fig:wlc}).
Lastly, \emph{sequence-specificity} can substantially affect response to force and torque. 
Sequence-specificity is modelled in oxDNA by fitting multiplicative parameters for the stacking and hydrogen-bonding interactions.
We fit these parameters to an arbitrarily chosen $L_{ps}$ (Figure \ref{fig:lp-ss}), highlighting the flexibility of our method across sequence-averaged and sequence-specific considerations.

\begin{figure}[t!]
\begin{center}
\centerline{\includegraphics[width=0.95\textwidth]{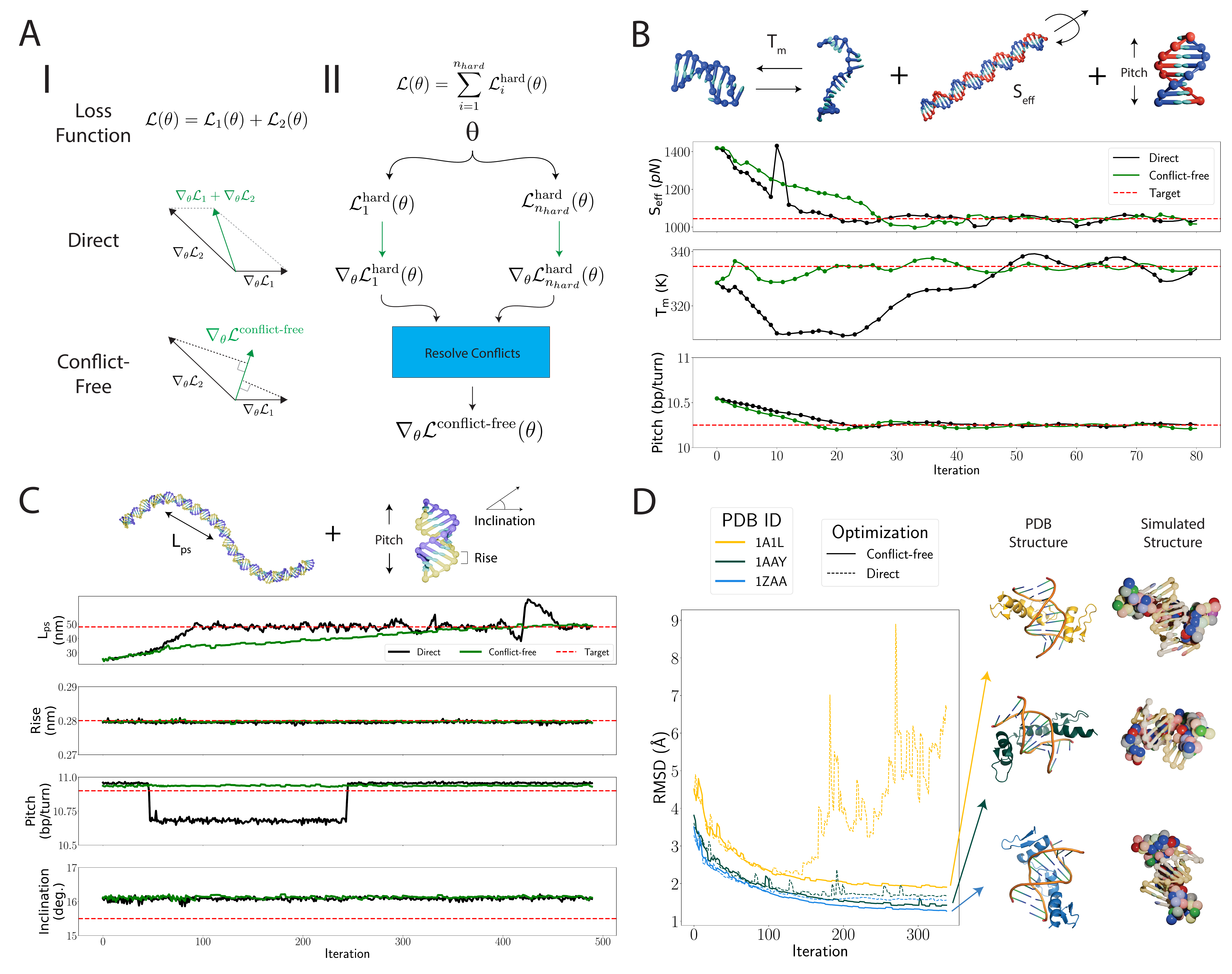}}
\caption{
Jointly fitting to multiple target properties simultaneously via conflict-free gradients.
\textbf{A.} Conflict-free gradients provide a means of imposing \emph{hard constraints}.
\textbf{I.} Given a loss function expressed as the sum of individual loss terms, directly differentiating the total loss may yield a gradient that conflicts with the gradients of individual terms.
Instead, the conflicting components of individual gradients can be projected out to yield a \emph{conflict-free} gradient that is analytically guaranteed to minimize all subterms.
This figure is adapted from Ref. \cite{liu2024config}.
\textbf{II.} A detailed schematic of a conflict-free update for an arbitrary number of hard constraints, in which one first computes the gradients corresponding to individual loss terms and then maps these individual gradients to a single conflict-free gradient via a conflict-resolution scheme.
There are many choices of conflict-resolution schemes developed in the context of multi-task learning.
We use the ConFIG method of Zhang et al.~\cite{liu2024config}.
In \textbf{B}-\textbf{D}, all optimizations are performed with and without conflict-free updates, using identical hyperparameters.
\textbf{B.} Fitting oxDNA parameters to $S_{\text{eff}}$, pitch, and the $T_m$ of a hairpin with a 6 nucleotide loop and 6 base-pair stem.
The target pitch is $10.25$ bp/turn, the target $S_{\text{eff}}$ is $1045$ pN, and the target $T_m$ is $334.5$ K.
Iterations in which reference states are resampled are represented as scatter points.
Reference state resampling information is omitted from \textbf{C}-\textbf{D}.
\textbf{C.} Fitting oxRNA parameters to $L_{ps}$, pitch, inclination, and rise.
The target $L_{ps}$ is $48$ nm, the target rise is $0.28$ nm, the target pitch is $10.9$ bp/turn, and the target inclination is $15.5^{\circ}$.
\textbf{D.} Fitting the parameters of a custom energy function for DNA-protein complexes to minimize the root mean square deviation (RMSD) between simulated structures and three experimentally-derived zinc finger structures from the Protein Data Bank (PDB).
Under this energy function, DNA energetics are described by oxDNA, protein energetics are described by an anisotropic network model (ANM), and DNA-protein interactions are described by a sequence-specific Morse potential (see \emph{Materials and Methods}).
}
\label{fig:joint-opt}
\end{center}
\end{figure}

\subsection{Thermodynamic Optimization}

The third class of data to which coarse-grained models are fit is thermodynamic data, e.g. melting temperatures.
This presents a host of challenges for traditional differentiable MD beyond those discussed so far.
For example, melting temperatures usually require fitting melting curves to expected values of \emph{discrete} order parameters (e.g. number of base pairs), manifestly not differentiable via the pathwise estimator.
Additionally, melting temperature calculations typically require enhanced sampling methods such as umbrella sampling given the rare nature of the binding/unbinding events and the necessity of sampling multiple such transitions in order to amass sufficient statistics. 
Methods for performing such biased simulations historically involve non-differentiable Monte Carlo schemes, e.g. oxDNA applies a bespoke cluster-move Monte Carlo algorithm~\cite{ruuvzivcka2014collective}.
Lastly, statistics sampled at a given temperature are typically extrapolated to derive statistics at a range of temperatures, necessitating similar unbiasing principles.

To overcome these challenges, we extend the DiffTRE formalism to accommodate such complex calculations.
In umbrella sampling, states $s_i$ are sampled from a biased Markov chain with unnormalized probability given by
\begin{align}
    p(s_i) = \omega_i \exp(-\beta U(s_i)) \label{eqn:umbrella-unnorm}
\end{align}
where the weights $\omega_i$ are chosen to promote configurations along some order parameter(s) that would otherwise be virtually inaccessible to an unweighted Monte Carlo sampling.
Similarly, a simulation at temperature $T$ can be treated as a biased sample of the ensemble at a different temperature $T'$ and the expected value of a target observable $O$ at temperature $T'$ can be computed as
\begin{align}
    \langle O(s) \rangle_{T'} = \langle O(s) \exp \left(U(s, T) / kT - U(s, T') / kT'\right) \rangle_T \label{eqn:temp-unnorm}
\end{align}
Note that this is a special case of Equation \ref{eqn:umbrella-unnorm} where $\omega_i = \exp (U(s_i, T) / kT)$.
Summary statistics of the temperature response may then be computed by line-fitting procedures and differentiated via implicit differentiation, as with mechanical properties.
See Figure \ref{fig:thermo}A for an overview of this framework.

This framework enables the differentiable calculation of the melting temperature $T_m$ of duplexes and hairpins in oxDNA.
This is achieved by calculating the expected ratio of bound to unbound states at a range of temperatures and computing a melting curve to extract $T_m$ (see Supporting Information).
We successfully apply our method to optimize melting temperatures of both DNA duplexes and hairpins.
Since oxDNA matches experimental duplex melting temperatures with high accuracy~\cite{snodin2015introducing, ouldridge2012coarse}, we optimize for an arbitrarily different value of $T_m$ for an 8 base-pair duplex (Figure \ref{fig:thermo}B-I).
In principle, we can fit \emph{any} (continuous and differentiable) function of the melting curve.
To demonstrate this flexibility, we also fit to an arbitrarily chosen width of the melting curve for the 8 base-pair duplex (Figure \ref{fig:thermo}B-II).
Unlike for duplexes, oxDNA underestimates the melting temperatures of hairpins.
We therefore fine-tune the parameters to match the melting temperature of a hairpin with a 6 base-pair stem with a loop comprised of 6 unpaired nucleotides as predicted by the Santa-Lucia model (Figure \ref{fig:thermo}C)~\cite{santalucia2004thermodynamics}.

Sensitivity analysis once more reveals a pronounced dependence on the cross-stacking interaction (Figure \ref{fig:thermo_grads}).
While cross-stacking is generally understood to stabilize duplexes and therefore affect melting, the relative contribution of cross-stacking compared to hydrogen-bonding and stacking in the model is surprising.
Taken together with the sensitivity analyses for structural and mechanical properties, these results suggest that the cross-stacking interaction is far more influential than previously understood in the parameterization of oxDNA.

\subsection{Joint Optimization}

Until now, we have considered fitting parameters to individual target properties -- either structural, mechanical, or thermodynamic.  
Ideally, parameters should be \emph{simultaneously} optimized using a joint loss function that describes multiple targets.
To carry this out, we first define the objective as the simple linear sum of root mean squared errors (RMSEs) for individual properties, though our framework naturally accommodates more sophisticated objective functions (see Supporting Information).
However, this vanilla formulation has the property that gradients of one target property may conflict with another, resulting in the improvement of one term at the cost of another.

To address this, we turn to conflict-free gradient methods from multi-task learning problems in machine learning.
These methods compute an update direction by projecting out conflicting components of individual gradients, ensuring that multiple loss terms can be minimized simultaneously without interference (Figure \ref{fig:joint-opt}A). 
We use the conflict-resolution method of Liu et al.~\cite{liu2024config}, though there are many alternatives~\cite{yu2020gradient, liu2021towards, dong2022gdod, javaloy2021rotograd}. 
In the context of fitting coarse-grained potentials, this allows us to define a notion of \emph{hard constraints} by which we ensure that each term included in the conflict-free update decreases monotonically (up to momentum artifacts in the numerical optimizer).
See \textit{Materials and Methods} for details on constructing a conflict-free update.

To illustrate this approach, we jointly fit $S_{\text{eff}}$, pitch, and the $T_m$ for the hairpin described above. 
This is motivated by oxDNA's current overestimate of $S_{\text{eff}}$ and the difficulty of maintaining structural and thermodynamic properties given strong parameter interdependence.
We show the progression of each property throughout the optimization in Figure \ref{fig:joint-opt}B.
Without conflict-free updates, the optimization begins by decreasing $S_{\text{eff}}$ and pitch at the cost of the hairpin $T_m$, subsequently recovering the hairpin $T_m$ while maintaining $S_{\text{eff}}$.
Conflict-free updates make no such sacrifices, converging to the target values in roughly half as many gradient updates.
Second, we jointly fit $S_{\text{eff}}$, $C$, $L_{ps}$, and pitch (Figure \ref{fig:seff-lp-c-pitch}).
This is motivated by difficulty in the original parameterization of capturing $S$ without sacrificing accuracy in other mechanical properties.
Again, we find that conflict-free updates yield a more stable and efficient means of optimization.

\subsection{Extension to other models}

We have focused so far on reparameterizing the oxDNA model as a representative case study of our framework, but anticipate our approach can be extended to a broad family of coarse-grained biomolecular models. 
To illustrate this, we conclude by treating two other example systems: RNA (Figure \ref{fig:joint-opt}C), a notoriously challenging subject of molecular modeling~\cite{smith2018benchmarking, stamatis2023benchmarking, yildirim2011benchmarking, condon2015stacking, bergonzo2015highly, schrodt2015large, mlynsky2024can}, and DNA-protein complexes (Figure \ref{fig:joint-opt}D), a subject of broad therapeutic interest~\cite{barrangou2016applications, praetorius2017self, singh2010recent}.
These systems highlight the utility of conflict-free gradients beyond merely accelerating optimization.

For RNA, we consider the oxRNA model, a sister model to oxDNA that captures its structure and duplex thermodynamics but poorly models mechanical properties (see \textit{Materials and Methods}).
We optimize oxRNA to match the experimentally-derived $L_{ps}$ while maintaining key structural properties already captured in the model (i.e. rise, pitch, and inclination).
All quantities are computed as in Ref. \cite{vsulc2014nucleotide}.
We find that conflict-free gradient updates yield significantly more stable optimization than vanilla gradients.
Using direct gradients, agreement with the target pitch value is degraded while optimizing towards the target $L_{ps}$ value and, after recovering the target pitch, $L_{ps}$ oscillates around the target value, likely due to frustration with other loss terms.
Conflict-free gradient updates, however, yield a stable optimization and a parameter set that matches the target $L_{ps}$ with low variance and without compromising existing agreement with structural properties, as shown in Figure \ref{fig:joint-opt}.
To the best of our knowledge, these optimized updated parameters represent the most coarse-grained RNA model to capture these four properties simultaneously~\cite{li2021rna, vsponer2018rna, cruz2018coarse, uusitalo2017martini}.

For DNA-protein complexes, we introduce a new force field that parameterizes DNA-protein interactions via a custom sequence-specific Morse potential (see \emph{Materials and Methods}).
This is an extension of the model introduced in Ref. \cite{procyk2021coarse}, in which DNA and protein molecules interact solely via hard-sphere repulsion.
We circumvent the need to manually derive forces by simulating in JAX-MD, underscoring the flexibility offered by our framework for experimenting with custom functional forms.
We fit model parameters directly to structures deposited in the Protein Data Bank, motivated by recently developed machine-learning models that fit directly to such structures~\cite{jumper2021highly, lin2023evolutionary}.
We target three zinc fingers with the loss for each zinc finger defined as the average RMSD between the simulated structure and the coarse-grained representation generated by oxView~\cite{bohlin2022design}.
Despite the optimization problem being highly overparameterized, vanilla gradients become trapped in a local minimum which improves agreement with two structures while severely degrading performance with the third.
Alternatively, conflict-free gradients serve as stable updates that yield a substantially lower total loss without compromising the accuracy of a single structure (Figure \ref{fig:joint-opt}).

\section{Discussion}

Differentiable programming is an emerging paradigm in scientific computing that holds great promise for bringing the successes of machine learning to physics-based modeling.
However, current methods are limited to relatively simple systems and practitioners have not yet converged on best practices for fitting complex, experimentally-relevant models.

In this work, we demonstrate how state-of-the-art methods in differentiable programming can be combined with highly complex simulation routines.
We successfully optimize structural, mechanical, and thermodynamic properties of DNA, exceeding the accuracy of the original parameterization for a range of targets and introducing and successfully fitting to qualitatively new objectives, such as the width of melting curves. 
Accomplishing this requires differentiating complex simulation methods such as enhanced sampling, use of external forces and torques, and bespoke Monte Carlo algorithms, as well as downstream calculations such as line-fitting procedures.
Our approach scales to long simulations of large systems, is both memory-efficient and numerically stable, and directly makes use of existing molecular simulation software.
By adapting conflict-free gradient methods from reinforcement learning, we provide a framework for systematically fitting to multiple data sources simultaneously, a principle which has heretofore been lacking in the molecular modeling community. 
We demonstrate the generality and flexibility of our method by performing joint optimizations for models of RNA and DNA-protein complexes.

There is a host of existing numerical methods for fitting coarse-grained force field parameters.
For example, methods like force matching, relative entropy minimization, and Boltzmann inversion are often cast as variational problems for deriving coarse-grained potentials.
However, each of these methods is limited in its generality.
For example, force matching is sensitive to nonlinear dependencies on the force field parameters and therefore the functional form is often restricted to spline interpolation~\cite{izvekov2005multiscale}.
Similarly, methods like Boltzmann inversion and relative entropy (RE) minimization are effective for matching a ground truth distribution (often derived from atomistic simulations), but the reliance on Lagrange multipliers to constrain ensemble averages (specifically in the case of RE methods) does not accommodate frustrated optimization problems (i.e. those with mutually incompatible data)~\cite{cesari2018using}.
In addition, the existing suite of methods cannot be naturally composed, necessitating independent parameter optimizations depending on the applicable method for a given target property.

Alternatively, gradient-based methods for individual target properties can be naturally composed, either (i) by combining individual objective functions into a global objective function or (ii) via more sophisticated methods like the conflict resolution scheme employed in this work. 
Moreover, gradient-based methods can immediately be integrated with the rich suite of tools developed by the machine learning community (e.g. momentum-based optimizers, overparameterization via neural networks) for navigating complex optimization landscapes (e.g. frustrated systems).
Recent work by Fuchs \textit{et al.} demonstrates the effectiveness of these tools for optimizing molecular force fields by developing an ML-inspired ecosystem for fitting neural-network (NN) potentials to biomolecular experimental and simulation data~\cite{fuchsChemtrainLearningDeep2024}.

There is also existing work on scaling differentiable MD to complex, experimentally-relevant force fields.
Greener uses differentiable simulation with 5 ns trajectories to develop an improved force field for disordered proteins that better reproduces experimentally-measured secondary structure and radius of gyration~\cite{greener2024differentiable},  employing gradient clipping to mitigate numerical instabilities like those observed in Figure \ref{fig:benchmark}D.
Yet, in this formulation, each epoch required one day of compute on 12 GPUs, likely due to the large time cost of differentiating unrolled trajectories (Figure \ref{fig:benchmark}B), restricting the optimized parameter set to only five epochs of training.  
The methods outlined in the present work avoid these difficulties: all optimizations presented here were performed on standard high-performance computing CPU nodes, and gradient updates take, at most, the duration of the longest forward simulation. 
This enhanced efficiency permits tens to hundreds of training epochs within several hours or days at low cost.

There are several limitations of our approach.
Foremost, DiffTRE is restricted to ensembles for which unnormalized probabilities of states are known and therefore we only fit to target equilibrium properties.
In future work, we plan to exploit fundamental results in non-equilibrium physics (e.g. the Jarzynski equality~\cite{jarzynski_nonequilibrium_1997}) to differentiate dynamical properties such as reaction rates and response to external work protocols. 
In the spirit of actor-critic algorithms in reinforcement learning, we also envision a more general probabilistic treatment of out-of-equilibrium properties in which the attractor of the dynamical system is parameterized and jointly inferred.
There is also great promise in approximate methods of gradient estimation that may enable tractable gradient calculation through unrolled trajectories~\cite{oktay2020randomized}.

There are many additional opportunities for future work.
For example, we recently introduced a method for directly designing biopolymers via simulation~\cite{krueger2024generalized} and our JAX implementations of oxDNA and oxRNA are amenable to this method.
Second, existing coarse-grained models are often developed to reproduce models that are themselves approximations, such as the nearest-neighbor model and atomistic simulations.
In our framework, one could fit a model of DNA or RNA directly to experimental data (e.g. the optical melting experiments underlying the nearest-neighbor model) rather than the model itself.
Finally, we are excited to work with experts in biomolecular modelling to fit new coarse-grained models incorporating full datasets of structural, mechanical, and thermodynamic data.
Such efforts will benefit from a standardized library of experimental target behaviors and corresponding simulation protocols for direct integration with our optimization framework.


\section{Materials and Methods}

Here we provide details on the DNA, RNA, and DNA-protein models optimized in this work as well as the method for stochastic gradient estimation given knowledge of unnormalized probabilities developed independently by Zhang et al.~\cite{zhang2023automatic} and Thaler and Zavadlav~\cite{thaler2021learning}.

\subsection{Force Fields}

\subsubsection{oxDNA and oxRNA}
The oxDNA force field is a coarse-grained model of DNA first introduced in 2010 by Thomas Ouldridge~\cite{ouldridge2012coarse}. Originally intended to capture DNA properties salient to DNA nanotechnological devices -- hybridization, mechanical properties of both single- and double-stranded DNA -- oxDNA has since been profitably used in a host of biophysical contexts~\cite{mosayebiRoleLoopStacking2014c,Schreck2015,NomidisTwistBendCoupling2019} in addition to becoming a workhorse in the DNA nanotechnology community~\cite{sengar2021primer, poppleton2023oxdna, poppleton2022nanobase, ouldridge2015dna}. 
The model, widely validated by experiments, has been used to study phenomena ranging from small-scale strand displacement reactions~\cite{ouldridge2010dna, machinek2014programmable, haley2020design} to the gelation and crystallization of DNA origami~\cite{rovigatti2014gels, romano2015switching, liu2024inverse, centola2024rhythmically}.
In its original form, the only sequence specificity in oxDNA was the restriction of hydrogen bonds to valid base pairs (i.e. A-T and G-C). Since its inception, oxDNA has been extended to include base-pair specific hydrogen-bonding and stacking parameters~\cite{vsulc2012sequence} and major and minor grooving~\cite{snodin2015introducing}.

In the oxDNA model, DNA is modelled as a string of rigid nucleotides.
Each nucleotide is represented as a rigid body consisting of a backbone repulsion site, a stacking site, and a hydrogen-bonding site.
All interactions are pairwise and the model considers six types of interactions: backbone connectivity, excluded volume, hydrogen bonding, stacking, cross stacking, and coaxial stacking.
More specifically, the potential energy is given by
\begin{align}
    V_{\text{oxdna}} &= \sum_{nn} \left( V_{\text{backbone}} + V_{\text{stack}} + V'_{exc} \right) \\ 
    &+ \sum_{\text{other pairs}} \left( V_{\text{HB}} + V_{\text{cross\_stack}} + V_{\text{coaxial\_stack}} + V_{\text{exc}} \right)
\end{align}
where $nn$ denotes the set of consecutive bases within strands; see \cite{ouldridge2012coarse} for details.
The latest version of oxDNA, oxDNA2, contains an additional Debye-Hückel term to model electrostatics.
All optimizations presented in this work use the oxDNA2 model unless otherwise stated.

The parameters of this potential were fit to reproduce three types of properties of single- and double-stranded B-DNA: (i) structural, (ii) mechanical, and (iii) thermodynamic.
Examples of structural properties include helical radius, pitch, and propellor twist.
Mechanical properties include the persistence length and force-extension properties of both single- and double-stranded DNA, as well as the torsional modulus for the double-stranded case.
Regarding thermodynamics, the model is largely fit to reproduce sequence-averaged duplex and hairpin melting temperatures as well as the single-stranded stacking transition.
The sequence-averaged parameters of the original model were fit by hand, and the exact properties to which the model was fit versus the properties on which it was evaluated is not recorded.

oxDNA simulations, both in our JAX-MD implementation and the original C++/CUDA implementations, use simulation units as follows~\cite{ouldridge2010dna,snodin2015introducing}: 1 unit length = 0.8518 \si{\nano\metre}; 1 unit force = 48.63 \si{\pico\newton}; 1 unit time = 3.03 \si{\pico\second}; 1 unit temperature = 3000 \si{\kelvin}; 1 unit energy = 41.42 \si{\pico\newton \nano\metre}. 
We have used an MD time step of dt = 0.005 simulation units = 15.15 \si{\pico\second} for all simulations in this work with the exception of the sequence-structure optimization in Figure \ref{fig:structure_opt}, which used dt = 0.003 simulation units = 9.09 \si{\pico\second}. 
We use a linear diffusion coefficient of D = 2.5 and a rotational diffusion coefficient $D_r = 3D$, which enables significantly faster diffusion than that seen in experiments and accelerates convergence of average properties accordingly.

oxRNA is a companion to oxDNA for the coarse-grained simulation of RNA~\cite{vsulc2014nucleotide}.
The model was similarly parameterized to reproduce structural, mechanical, and thermodynamic properties of RNA, though is generally less accurate than oxDNA.
The functional form is nearly identical to that of oxDNA but differs slightly in its stacking and cross-stacking interactions.
Applications of the model range from toehold-mediated strand displacement~\cite{vsulc2015modelling} to supercoiling~\cite{matek2015coarse} to viral self-assembly~\cite{mattiotti2024molecular}.
Simulation units differ slightly from those in oxDNA: 1 unit length = 0.84 \si{\nano\metre}; 1 unit force = 49.3 \si{\pico\newton}; 1 unit time = 3.06 \si{\pico\second}. 
See Ref. \cite{vsulc2014nucleotide} for a complete introduction to oxRNA.

\subsubsection{DNA-Protein Model}

In Ref. \cite{procyk2021coarse}, Procyk et al. introduce a coarse grained model of DNA-protein hybrids.
The combined model includes energetic descriptions of (i) DNA-DNA, (ii) protein-protein, and (iii) DNA-protein interactions. 
DNA-DNA interactions are described by the oxDNA model.
Protein-protein interactions are given by an Anisotropic Network Model (ANM), which represents a protein with a known structure as beads connected by spring.
Each residue is modelled as a single bead.
To capture fluctuations of the protein backbone, spring constants are chosen to best fit B-factors of the $\alpha-\text{carbons}$ given in the original PDB structure.
This is performed automatically via oxView~\cite{bohlin2022design}.
DNA-protein interactions are simply modelled as excluded volume effects, neglecting any attractive or non-steric repulsive interactions.
Procyk additionally develop more complex protein-protein interactions that include torsional and bending potentials, though we focus on the ANM model in this work. 

We extend this model by parameterizing DNA-protein interactions via a sequence-specific Morse potential.
Specifically, given an interacting pair consisting of a DNA nucleotide of type $i$ (corresponding to one of four possible nucleotides) and an amino acid of type $j$ (corresponding to one of 20 possible amino acids), we introduce an attractive interaction:
\begin{align}
    V_{\text{morse}}^{ij} = \epsilon^{ij} \left( 1 - \exp \left(-\alpha^{ij} \left(r - r^{ij}_0\right)\right)\right)^2
\end{align}
where $r$ is the inter-particle distance, $r^{ij}_0$ is the equilibrium bond length, $\epsilon^{ij}$ is the sequence-specific strength, and $\alpha^{ij}$ is the sequence-specific width.
Thus, this new interaction is parameterized via $\boldsymbol\epsilon, \boldsymbol\alpha, \mathbf{r_0} \in \mathbb{R}^{4 \times 20}$.
In practice, we parameterize individual Morse potentials for both the backbone site and hydrogen bonding site and therefore fit six $4 \times 20$ matrices (three for each site).

\subsection{Differentiable Monte Carlo (DiffTRE)}\label{sec:difftre}

Zhang et al.~\cite{zhang2023automatic} and Thaler and Zavadlav~\cite{thaler2021learning} independently developed a method for computing a low variance gradient estimate of an expected value, given knowledge of the unnormalized probabilities of individual samples.
For an equilibrium system in the canonical ensemble, the probability of an individual microstate $\overrightarrow{x}_i$ follows the Boltzmann distribution:
\begin{align}
    p(\overrightarrow{x}_i) = \frac{e^{-\beta U(\overrightarrow{x}_i)}}{Z}
\end{align}
where $\beta$ is the inverse temperature, $U(\overrightarrow{x}_i)$ is the potential energy of $\overrightarrow{x}_i$, and $Z=\sum_j e^{-\beta U(\overrightarrow{x}_j)}$ is the partition function.

Consider a set of states sampled from this distribution given some control parameters $\boldsymbol{\theta}$, $X_{\theta} = \{ \overrightarrow{x}_1, \overrightarrow{x}_2, \cdots \overrightarrow{x}_N \}$, e.g. via standard MD or MC algorithms.
The expectation of an arbitrary state-level observable $O(\overrightarrow{x}, \boldsymbol{\theta})$ is
\begin{align}
    \langle O(\overrightarrow{x}, \boldsymbol{\theta}) \rangle_{\overrightarrow{x}_i \in X} = \frac{1}{N}\sum_i O(\overrightarrow{x}_i, \boldsymbol{\theta}) \label{eqn:state-expectation}
\end{align}
Following thermodynamic perturbation theory, the key insight is to rewrite Eq. \ref{eqn:state-expectation} as a weighted sum over sampled states, then adjust this weighting as the governing parameters are changed from reference set $\hat{\boldsymbol{\theta}}$ to $\boldsymbol{\theta}$, assuming sufficient overlap between the respective probability distributions: 
\begin{align}
    \langle O(\overrightarrow{x}, \theta) \rangle_{\overrightarrow{x}_i \in X} = \sum_i w_iO(\overrightarrow{x}_i, \theta)
    \label{eqn:traj_reweigt}
\end{align}
where
\begin{align}
    w_i = \frac{p_{\theta}(\overrightarrow{x}_i) / p_{\hat{\theta}}(\overrightarrow{x}_i)}{\sum_j p_{\theta}(\overrightarrow{x}_j) / p_{\hat{\theta}}(\overrightarrow{x}_j)} \label{eqn:weight}
\end{align}
and $\hat{\theta}$ is the reference potential under which $X_{\theta}$ was sampled~\cite{thaler2021learning}.
Crucially, Equation \ref{eqn:weight} only requires unnormalized probabilities as normalizing factors cancel, e.g. in the case of the canonical ensemble
\begin{align}
    w_i = \frac{e^{-\beta (U_{\theta}(\overrightarrow{x}_i) - U_{\hat{\theta}}(\overrightarrow{x}_i))}}{\sum_j e^{-\beta (U_{\theta}(\overrightarrow{x}_j) - U_{\hat{\theta}}(\overrightarrow{x}_j))}}, \label{eqn:weight-boltz}
\end{align}
When $\boldsymbol{\theta} = \hat{\boldsymbol{\theta}}$, Equation \ref{eqn:traj_reweigt} exactly reduces to Equation \ref{eqn:state-expectation} as $w_i = \frac{1}{N}$ but $\nabla_{\boldsymbol{\theta}} \log(p(\overrightarrow{x}_i)) \neq 0$.

Zhang \textit{et al.}~\cite{zhang2023automatic} were the first to calculate gradients of expectation values to circumvent full trajectory unrolling in the context of Monte Carlo simulations; Thaler and Zavadlav adapted the approach, which they call `differentiable trajectory reweighting (DiffTRE)', for molecular dynamics simulations~\cite{thaler2021learning}.
Thaler and Zavadlav also introduced the notion that reference states collected via $\hat{\boldsymbol{\theta}}$ can be reused for small differences between $\boldsymbol{\theta}$ and $\hat{\boldsymbol{\theta}}$.
The effective sample size is
\begin{align}
    N_{\text{eff}} = e^{-\sum_{i=1}^Nw_i\ln(w_i)} \label{eqn:neff}
\end{align}
and, given a total number of reference states $N_{\text{ref}}$, reference states are sampled when $N_{\text{eff}} < \lambda N_{\text{ref}}$ where $\lambda$ is a hyperparameter.

\subsection{Conflict-Free Gradients}

Here we generically describe conflict-free gradient methods, and subsequently the ConFIG method of Liu et al. employed in this work~\cite{liu2024config}.
We largely follow the notation and presentation of Ref. \cite{liu2024config}.

Consider an optimization problem in which the goal is to find $\theta$ that simultaneously minimizes $m$ individual loss functions, $\{\mathcal{L}_1, \mathcal{L}_2, \ldots, \mathcal{L}_m\}$ where the total loss function is expressed as
\begin{align}
    \mathcal{L}(\theta) = \sum_{i=1}^m \mathcal{L}_i(\theta)
\end{align}
Naively, direct differentiation of this objective yields
\begin{align}
    \nabla_{\theta}\mathcal{L}(\theta) = \sum_{i=1}^m \nabla_{\theta}\mathcal{L}_i(\theta) \label{eqn:direct-grad-main}
\end{align}
However, in such problems, it is typical that optimization becomes stuck in a local minimum of a specific loss term due to the conflict between losses.
Consider a gradient update $\mathbf{g}^{\text{total}}$, e.g. $\mathbf{g}^{\text{total}} = \nabla_{\theta}\mathcal{L}(\theta)$ as in Equation \ref{eqn:direct-grad-main}.
A gradient update will conflict with the decrease of $\mathcal{L}_i$ if $\mathbf{g}_i^{\top}\mathbf{g}^{\text{total}} < 0$ where $\mathbf{g}_i = \nabla_{\theta}\mathcal{L}_i(\theta)$~\cite{liu2024config, riemer2018learning, du2018adapting}.
One solution is to construct $\mathbf{g}^{\text{total}}$ such that
\begin{align}
    \mathbf{g}^{\text{total}} = [\mathbf{g}_i, \mathbf{g}_2, \ldots, \mathbf{g}_m]^{-\top}\mathbf{w}
\end{align}
where $w_i > 0$ for all $i$ and $M^{-\top}$ is the pseudoinverse of the transposed matrix $M^{\top}$.

In this work, we apply the recently-developed Conflict-Free Inverse Gradients (ConFIG) method of Liu et al.~\cite{liu2021towards}.
In this scheme, $\mathbf{g}^{\text{total}}$ is constructed as follows:
\begin{align}
    \mathbf{g}^{\text{total}} &= \left( \sum_{i=1}^m \mathbf{g}_i^{\top}\mathbf{g}_u \right) \mathbf{g}_u \label{eqn:g-config} \\
    \mathbf{g}_u &= \mathcal{U}\left[ \left[ \mathcal{U}(\mathbf{g}_1), \mathcal{U}(\mathbf{g}_2), \ldots, \mathcal{U}(\mathbf{g}_m) \right]^{-\top} \mathbf{1}_m \right]
\end{align}
where $\mathcal{U}(\mathbf{g}_i)$ is a normalization operator and $\mathbf{1}_m$ is a unit vector with $m$ components that ensures a uniform rate of decrease across all loss terms.
This unit vector can be replaced with custom direction weights $\hat{\mathbf{w}}$ to control the relative importance of individual loss terms, but we set $\hat{\mathbf{w}} = \mathbf{1}_m$ in this work.
See Ref. \cite{liu2021towards} for a complete derivation.

\section*{Acknowledgments}
We thank Petr Sulc, Tom Ouldridge, Jonathan Doye, Lorenzo Rovigatti, Erik Poppleton, and Ard Louis for support and helpful discussions surrounding the oxDNA force field and the oxDNA ecosystem.
We also thank Petr Sulc and Erik Poppleton for inspiring sequence-structure optimization.
We thank Salvatore Assenza and Rubén Pérez for their helpful feedback regarding the calculation of mechanical properties, including providing simulation protocols for stretch-torsion simulations in LAMMPS.
We thank Jamie Smith for helpful conversations relating to stochastic gradient estimation.
R.K.K. thanks Stefania Ketzetzi for helpful feedback on the manuscript.
This material is based upon work supported by the NSF AI Institute of Dynamic Systems (\#2112085) and the Office of Naval Research (N00014-17-1-3029).

\bibliographystyle{unsrt}  
\bibliography{main}

\begin{thebibliography}{100}

\bibitem{drorBiomolecularSimulationComputational2012}
Ron~O. Dror, Robert~M. Dirks, J.P. Grossman, Huafeng Xu, and David~E. Shaw.
\newblock Biomolecular {{Simulation}}: {{A Computational Microscope}} for {{Molecular Biology}}.
\newblock {\em Annu. Rev. Biophys.}, 41(1):429--452, June 2012.

\bibitem{perlmutter2015mechanisms}
Jason~D Perlmutter and Michael~F Hagan.
\newblock Mechanisms of virus assembly.
\newblock {\em Annual review of physical chemistry}, 66(1):217--239, 2015.

\bibitem{huber2017multiscale}
Roland~G Huber, Jan~K Marzinek, Daniel~A Holdbrook, and Peter~J Bond.
\newblock Multiscale molecular dynamics simulation approaches to the structure and dynamics of viruses.
\newblock {\em Progress in Biophysics and Molecular Biology}, 128:121--132, 2017.

\bibitem{ruiz2015simulations}
Teresa Ruiz-Herrero and Michael~F Hagan.
\newblock Simulations show that virus assembly and budding are facilitated by membrane microdomains.
\newblock {\em Biophysical journal}, 108(3):585--595, 2015.

\bibitem{ode2012molecular}
Hirotaka Ode, Masaaki Nakashima, Shingo Kitamura, Wataru Sugiura, and Hironori Sato.
\newblock Molecular dynamics simulation in virus research.
\newblock {\em Frontiers in microbiology}, 3:258, 2012.

\bibitem{scheraga2007protein}
Harold~A Scheraga, Mey Khalili, and Adam Liwo.
\newblock Protein-folding dynamics: overview of molecular simulation techniques.
\newblock {\em Annu. Rev. Phys. Chem.}, 58(1):57--83, 2007.

\bibitem{swope2004describing}
William~C Swope, Jed~W Pitera, and Frank Suits.
\newblock Describing protein folding kinetics by molecular dynamics simulations. 1. theory.
\newblock {\em The Journal of Physical Chemistry B}, 108(21):6571--6581, 2004.

\bibitem{lindorff2011fast}
Kresten Lindorff-Larsen, Stefano Piana, Ron~O Dror, and David~E Shaw.
\newblock How fast-folding proteins fold.
\newblock {\em Science}, 334(6055):517--520, 2011.

\bibitem{rosa2008structure}
Angelo Rosa and Ralf Everaers.
\newblock Structure and dynamics of interphase chromosomes.
\newblock {\em PLoS computational biology}, 4(8):e1000153, 2008.

\bibitem{di2016transferable}
Michele Di~Pierro, Bin Zhang, Erez~Lieberman Aiden, Peter~G Wolynes, and Jos{\'e}~N Onuchic.
\newblock Transferable model for chromosome architecture.
\newblock {\em Proceedings of the National Academy of Sciences}, 113(43):12168--12173, 2016.

\bibitem{achiam2023gpt}
Josh Achiam, Steven Adler, Sandhini Agarwal, Lama Ahmad, Ilge Akkaya, Florencia~Leoni Aleman, Diogo Almeida, Janko Altenschmidt, Sam Altman, Shyamal Anadkat, et~al.
\newblock Gpt-4 technical report.
\newblock {\em arXiv preprint arXiv:2303.08774}, 2023.

\bibitem{team2024gemini}
Gemini Team, Petko Georgiev, Ving~Ian Lei, Ryan Burnell, Libin Bai, Anmol Gulati, Garrett Tanzer, Damien Vincent, Zhufeng Pan, Shibo Wang, et~al.
\newblock Gemini 1.5: Unlocking multimodal understanding across millions of tokens of context.
\newblock {\em arXiv preprint arXiv:2403.05530}, 2024.

\bibitem{jumper2021highly}
John Jumper, Richard Evans, Alexander Pritzel, Tim Green, Michael Figurnov, Olaf Ronneberger, Kathryn Tunyasuvunakool, Russ Bates, Augustin {\v{Z}}{\'\i}dek, Anna Potapenko, et~al.
\newblock Highly accurate protein structure prediction with alphafold.
\newblock {\em nature}, 596(7873):583--589, 2021.

\bibitem{donoho2024data}
David Donoho.
\newblock Data science at the singularity.
\newblock {\em Harvard Data Science Review}, 6(1), 2024.

\bibitem{krizhevsky2012imagenet}
Alex Krizhevsky, Ilya Sutskever, and Geoffrey~E Hinton.
\newblock Imagenet classification with deep convolutional neural networks.
\newblock {\em Advances in neural information processing systems}, 25, 2012.

\bibitem{brown2020language}
Tom Brown, Benjamin Mann, Nick Ryder, Melanie Subbiah, Jared~D Kaplan, Prafulla Dhariwal, Arvind Neelakantan, Pranav Shyam, Girish Sastry, Amanda Askell, et~al.
\newblock Language models are few-shot learners.
\newblock {\em Advances in neural information processing systems}, 33:1877--1901, 2020.

\bibitem{devlin2019bert}
Jacob Devlin, Ming-Wei Chang, Kenton Lee, and Kristina Toutanova.
\newblock Bert: Pre-training of deep bidirectional transformers for language understanding.
\newblock In {\em Proceedings of the 2019 conference of the North American chapter of the association for computational linguistics: human language technologies, volume 1 (long and short papers)}, pages 4171--4186, 2019.

\bibitem{lecun2015deep}
Yann LeCun, Yoshua Bengio, and Geoffrey Hinton.
\newblock Deep learning.
\newblock {\em nature}, 521(7553):436--444, 2015.

\bibitem{maffeoCoarseGrainedModelUnstructured2014}
Christopher Maffeo, Thuy T.~M. Ngo, Taekjip Ha, and Aleksei Aksimentiev.
\newblock A {{Coarse-Grained Model}} of {{Unstructured Single-Stranded DNA Derived}} from {{Atomistic Simulation}} and {{Single-Molecule Experiment}}.
\newblock {\em J. Chem. Theory Comput.}, 10(8):2891--2896, August 2014.

\bibitem{assenzaAccurateSequenceDependentCoarseGrained2022}
Salvatore Assenza and Rub{\'e}n P{\'e}rez.
\newblock Accurate {{Sequence-Dependent Coarse-Grained Model}} for {{Conformational}} and {{Elastic Properties}} of {{Double-Stranded DNA}}.
\newblock {\em J. Chem. Theory Comput.}, 18(5):3239--3256, May 2022.

\bibitem{heOptimizationNucleicAcids2015}
Yi~He, Adam Liwo, and Harold~A. Scheraga.
\newblock Optimization of a {{Nucleic Acids}} united-{{RESidue}} 2-{{Point}} model ({{NARES-2P}}) with a maximum-likelihood approach.
\newblock {\em The Journal of Chemical Physics}, 143(24):243111, December 2015.

\bibitem{naomeSolventMediatedCoarseGrainedModel2014}
Aymeric Na{\^o}m{\'e}, Aatto Laaksonen, and Daniel~P. Vercauteren.
\newblock A {{Solvent-Mediated Coarse-Grained Model}} of {{DNA Derived}} with the {{Systematic Newton Inversion Method}}.
\newblock {\em J. Chem. Theory Comput.}, 10(8):3541--3549, August 2014.

\bibitem{vsulc2012sequence}
Petr {\v{S}}ulc, Flavio Romano, Thomas~E Ouldridge, Lorenzo Rovigatti, Jonathan~PK Doye, and Ard~A Louis.
\newblock Sequence-dependent thermodynamics of a coarse-grained dna model.
\newblock {\em The Journal of chemical physics}, 137(13), 2012.

\bibitem{vsulc2014nucleotide}
Petr {\v{S}}ulc, Flavio Romano, Thomas~E Ouldridge, Jonathan~PK Doye, and Ard~A Louis.
\newblock A nucleotide-level coarse-grained model of rna.
\newblock {\em The Journal of chemical physics}, 140(23), 2014.

\bibitem{procyk2021coarse}
Jonah Procyk, Erik Poppleton, and Petr {\v{S}}ulc.
\newblock Coarse-grained nucleic acid--protein model for hybrid nanotechnology.
\newblock {\em Soft Matter}, 17(13):3586--3593, 2021.

\bibitem{snodin2015introducing}
Benedict~EK Snodin, Ferdinando Randisi, Majid Mosayebi, Petr {\v{S}}ulc, John~S Schreck, Flavio Romano, Thomas~E Ouldridge, Roman Tsukanov, Eyal Nir, Ard~A Louis, et~al.
\newblock Introducing improved structural properties and salt dependence into a coarse-grained model of dna.
\newblock {\em The Journal of chemical physics}, 142(23), 2015.

\bibitem{ratajczyk2024coarse}
Eryk~J Ratajczyk, Petr {\v{S}}ulc, Andrew~J Turberfield, Jonathan~PK Doye, and Ard~A Louis.
\newblock Coarse-grained modeling of dna--rna hybrids.
\newblock {\em The Journal of Chemical Physics}, 160(11), 2024.

\bibitem{doye2023oxdna}
Jonathan~PK Doye, Hannah Fowler, Domen Pre{\v{s}}ern, Joakim Bohlin, Lorenzo Rovigatti, Flavio Romano, Petr {\v{S}}ulc, Chak~Kui Wong, Ard~A Louis, John~S Schreck, et~al.
\newblock The oxdna coarse-grained model as a tool to simulate dna origami.
\newblock In {\em DNA and RNA Origami: Methods and Protocols}, pages 93--112. Springer, 2023.

\bibitem{thaler2021learning}
Stephan Thaler and Julija Zavadlav.
\newblock Learning neural network potentials from experimental data via differentiable trajectory reweighting.
\newblock {\em Nature communications}, 12(1):6884, 2021.

\bibitem{zhang2023automatic}
Shi-Xin Zhang, Zhou-Quan Wan, and Hong Yao.
\newblock Automatic differentiable monte carlo: Theory and application.
\newblock {\em Physical Review Research}, 5(3):033041, 2023.

\bibitem{metz2021gradients}
Luke Metz, C~Daniel Freeman, Samuel~S Schoenholz, and Tal Kachman.
\newblock Gradients are not all you need.
\newblock {\em arXiv preprint arXiv:2111.05803}, 2021.

\bibitem{greener2024differentiable}
Joe~G Greener.
\newblock Differentiable simulation to develop molecular dynamics force fields for disordered proteins.
\newblock {\em Chemical Science}, 15(13):4897--4909, 2024.

\bibitem{whitelam2007avoiding}
Stephen Whitelam and Phillip~L Geissler.
\newblock Avoiding unphysical kinetic traps in monte carlo simulations of strongly attractive particles.
\newblock {\em The Journal of chemical physics}, 127(15), 2007.

\bibitem{ruuvzivcka2014collective}
{\v{S}}t{\v{e}}p{\'a}n R{\r{u}}{\v{z}}i{\v{c}}ka and Michael~P. Allen.
\newblock Collective translational and rotational monte carlo cluster move for general pairwise interaction.
\newblock {\em Physical Review E}, 90(3):033302, 2014.

\bibitem{vitalis2009methods}
Andreas Vitalis and Rohit~V Pappu.
\newblock Methods for monte carlo simulations of biomacromolecules.
\newblock {\em Annual reports in computational chemistry}, 5:49--76, 2009.

\bibitem{schoenholz2020jax}
Samuel Schoenholz and Ekin~Dogus Cubuk.
\newblock Jax md: a framework for differentiable physics.
\newblock {\em Advances in Neural Information Processing Systems}, 33:11428--11441, 2020.

\bibitem{ouldridge2010dna}
Thomas~E Ouldridge, Ard~A Louis, and Jonathan~PK Doye.
\newblock Dna nanotweezers studied with a coarse-grained model of dna.
\newblock {\em Physical Review Letters}, 104(17):178101, 2010.

\bibitem{hinckley2013experimentally}
Daniel~M Hinckley, Gordon~S Freeman, Jonathan~K Whitmer, and Juan~J De~Pablo.
\newblock An experimentally-informed coarse-grained 3-site-per-nucleotide model of dna: Structure, thermodynamics, and dynamics of hybridization.
\newblock {\em The Journal of chemical physics}, 139(14), 2013.

\bibitem{ouldridgeStructuralMechanicalThermodynamic2011a}
Thomas~E. Ouldridge, Ard~A. Louis, and Jonathan P.~K. Doye.
\newblock Structural, mechanical, and thermodynamic properties of a coarse-grained {{DNA}} model.
\newblock {\em The Journal of Chemical Physics}, 134(8):085101, February 2011.

\bibitem{uusitalo2015martini}
Jaakko~J Uusitalo, Helgi~I Ingo{\'o}lfsson, Parisa Akhshi, D~Peter Tieleman, and Siewert~J Marrink.
\newblock Martini coarse-grained force field: extension to dna.
\newblock {\em Journal of chemical theory and computation}, 11(8):3932--3945, 2015.

\bibitem{naskar2021mechanical}
Supriyo Naskar and Prabal~K Maiti.
\newblock Mechanical properties of dna and dna nanostructures: Comparison of atomistic, martini and oxdna models.
\newblock {\em Journal of Materials Chemistry B}, 9(25):5102--5113, 2021.

\bibitem{braun2014determining}
Anthony~R Braun and Jonathan~N Sachs.
\newblock Determining structural and mechanical properties from molecular dynamics simulations of lipid vesicles.
\newblock {\em Journal of chemical theory and computation}, 10(9):4160--4168, 2014.

\bibitem{franz2020advances}
Florian Franz, Csaba Daday, and Frauke Gr{\"a}ter.
\newblock Advances in molecular simulations of protein mechanical properties and function.
\newblock {\em Current Opinion in Structural Biology}, 61:132--138, 2020.

\bibitem{yao2001mechanical}
Zhenhua Yao, Chang-Chun Zhu, Min Cheng, and Junhua Liu.
\newblock Mechanical properties of carbon nanotube by molecular dynamics simulation.
\newblock {\em Computational Materials Science}, 22(3-4):180--184, 2001.

\bibitem{tsai2010characterizing}
Jia-Lin Tsai and Jie-Feng Tu.
\newblock Characterizing mechanical properties of graphite using molecular dynamics simulation.
\newblock {\em Materials \& Design}, 31(1):194--199, 2010.

\bibitem{root2016predicting}
Samuel~E Root, Suchol Savagatrup, Christopher~J Pais, Gaurav Arya, and Darren~J Lipomi.
\newblock Predicting the mechanical properties of organic semiconductors using coarse-grained molecular dynamics simulations.
\newblock {\em Macromolecules}, 49(7):2886--2894, 2016.

\bibitem{chen2005mechanical}
Nan Chen, Mark~T Lusk, Adri~CT Van~Duin, and William~A Goddard~III.
\newblock Mechanical properties of connected carbon nanorings via molecular dynamics simulation.
\newblock {\em Physical Review B—Condensed Matter and Materials Physics}, 72(8):085416, 2005.

\bibitem{gautieri2010coarse}
Alfonso Gautieri, Antonio Russo, Simone Vesentini, Alberto Redaelli, and Markus~J Buehler.
\newblock Coarse-grained model of collagen molecules using an extended martini force field.
\newblock {\em Journal of Chemical Theory and Computation}, 6(4):1210--1218, 2010.

\bibitem{zerze2021thermodynamics}
G{\"u}l~H Zerze, Frank~H Stillinger, and Pablo~G Debenedetti.
\newblock Thermodynamics of dna hybridization from atomistic simulations.
\newblock {\em The Journal of Physical Chemistry B}, 125(3):771--779, 2021.

\bibitem{wong2008pathway}
Ka-Yiu Wong and B~Montgomery Pettitt.
\newblock The pathway of oligomeric dna melting investigated by molecular dynamics simulations.
\newblock {\em Biophysical journal}, 95(12):5618--5626, 2008.

\bibitem{aimoli2014force}
Cassiano~G Aimoli, Edward~J Maginn, and Charlles~RA Abreu.
\newblock Force field comparison and thermodynamic property calculation of supercritical co2 and ch4 using molecular dynamics simulations.
\newblock {\em Fluid Phase Equilibria}, 368:80--90, 2014.

\bibitem{kumar1992weighted}
Shankar Kumar, John~M Rosenberg, Djamal Bouzida, Robert~H Swendsen, and Peter~A Kollman.
\newblock The weighted histogram analysis method for free-energy calculations on biomolecules. i. the method.
\newblock {\em Journal of computational chemistry}, 13(8):1011--1021, 1992.

\bibitem{shirts2008statistically}
Michael~R Shirts and John~D Chodera.
\newblock Statistically optimal analysis of samples from multiple equilibrium states.
\newblock {\em The Journal of chemical physics}, 129(12), 2008.

\bibitem{deublein2011ms2}
Stephan Deublein, Bernhard Eckl, J{\"u}rgen Stoll, Sergey~V Lishchuk, Gabriela Guevara-Carrion, Colin~W Glass, Thorsten Merker, Martin Bernreuther, Hans Hasse, and Jadran Vrabec.
\newblock ms2: A molecular simulation tool for thermodynamic properties.
\newblock {\em Computer Physics Communications}, 182(11):2350--2367, 2011.

\bibitem{yu2020gradient}
Tianhe Yu, Saurabh Kumar, Abhishek Gupta, Sergey Levine, Karol Hausman, and Chelsea Finn.
\newblock Gradient surgery for multi-task learning.
\newblock {\em Advances in neural information processing systems}, 33:5824--5836, 2020.

\bibitem{liu2021towards}
Liyang Liu, Yi~Li, Zhanghui Kuang, J~Xue, Yimin Chen, Wenming Yang, Qingmin Liao, and Wayne Zhang.
\newblock Towards impartial multi-task learning.
\newblock iclr, 2021.

\bibitem{long2017learning}
Mingsheng Long, Zhangjie Cao, Jianmin Wang, and Philip~S Yu.
\newblock Learning multiple tasks with multilinear relationship networks.
\newblock {\em Advances in neural information processing systems}, 30, 2017.

\bibitem{kirkpatrick2017overcoming}
James Kirkpatrick, Razvan Pascanu, Neil Rabinowitz, Joel Veness, Guillaume Desjardins, Andrei~A Rusu, Kieran Milan, John Quan, Tiago Ramalho, Agnieszka Grabska-Barwinska, et~al.
\newblock Overcoming catastrophic forgetting in neural networks.
\newblock {\em Proceedings of the national academy of sciences}, 114(13):3521--3526, 2017.

\bibitem{liu2024config}
Qiang Liu, Mengyu Chu, and Nils Thuerey.
\newblock Config: Towards conflict-free training of physics informed neural networks.
\newblock {\em arXiv preprint arXiv:2408.11104}, 2024.

\bibitem{sengar2021primer}
Aditya Sengar, Thomas~E Ouldridge, Oliver Henrich, Lorenzo Rovigatti, and P~{\v{S}}ulc.
\newblock A primer on the oxdna model of dna: when to use it, how to simulate it and how to interpret the results.
\newblock {\em Frontiers in Molecular Biosciences}, 8:693710, 2021.

\bibitem{rovigatti2014gels}
Lorenzo Rovigatti, Frank Smallenburg, Flavio Romano, and Francesco Sciortino.
\newblock Gels of dna nanostars never crystallize.
\newblock {\em ACS nano}, 8(4):3567--3574, 2014.

\bibitem{romano2015switching}
Flavio Romano and Francesco Sciortino.
\newblock Switching bonds in a dna gel: An all-dna vitrimer.
\newblock {\em Physical Review Letters}, 114(7):078104, 2015.

\bibitem{burns2016biomimetic}
Jonathan~R Burns, Astrid Seifert, Niels Fertig, and Stefan Howorka.
\newblock A biomimetic dna-based channel for the ligand-controlled transport of charged molecular cargo across a biological membrane.
\newblock {\em Nature nanotechnology}, 11(2):152--156, 2016.

\bibitem{jax2018github}
James Bradbury, Roy Frostig, Peter Hawkins, Matthew~James Johnson, Chris Leary, Dougal Maclaurin, George Necula, Adam Paszke, Jake Vander{P}las, Skye Wanderman-{M}ilne, and Qiao Zhang.
\newblock {JAX}: composable transformations of {P}ython+{N}um{P}y programs, 2018.

\bibitem{poppleton2023oxdna}
Erik Poppleton, Michael Matthies, Debesh Mandal, Flavio Romano, Petr {\v{S}}ulc, and Lorenzo Rovigatti.
\newblock oxdna: coarse-grained simulations of nucleic acids made simple.
\newblock {\em Journal of Open Source Software}, 8(81):4693, 2023.

\bibitem{rovigatti2015comparison}
Lorenzo Rovigatti, Petr {\v{S}}ulc, Istv{\'a}n~Z Reguly, and Flavio Romano.
\newblock A comparison between parallelization approaches in molecular dynamics simulations on gpus.
\newblock {\em Journal of computational chemistry}, 36(1):1--8, 2015.

\bibitem{mohamed2020monte}
Shakir Mohamed, Mihaela Rosca, Michael Figurnov, and Andriy Mnih.
\newblock Monte carlo gradient estimation in machine learning.
\newblock {\em The Journal of Machine Learning Research}, 21(1):5183--5244, 2020.

\bibitem{blondel2022efficient}
Mathieu Blondel, Quentin Berthet, Marco Cuturi, Roy Frostig, Stephan Hoyer, Felipe Llinares-L{\'o}pez, Fabian Pedregosa, and Jean-Philippe Vert.
\newblock Efficient and modular implicit differentiation.
\newblock {\em Advances in neural information processing systems}, 35:5230--5242, 2022.

\bibitem{ouldridge2012coarse}
Thomas~E Ouldridge.
\newblock {\em Coarse-grained modelling of DNA and DNA self-assembly}.
\newblock Springer Science \& Business Media, 2012.

\bibitem{swart2007pi}
Marcel Swart, Tushar van~der Wijst, C{\'e}lia Fonseca~Guerra, and F~Matthias Bickelhaupt.
\newblock $\pi$-$\pi$ stacking tackled with density functional theory.
\newblock {\em Journal of molecular modeling}, 13:1245--1257, 2007.

\bibitem{vsponer2006nature}
Ji{\v{r}}{\'\i} {\v{S}}poner, Petr Jure{\v{c}}ka, Ivan Marchan, F~Javier Luque, Modesto Orozco, and Pavel Hobza.
\newblock Nature of base stacking: reference quantum-chemical stacking energies in ten unique b-dna base-pair steps.
\newblock {\em Chemistry--A European Journal}, 12(10):2854--2865, 2006.

\bibitem{poppleton2022nanobase}
Erik Poppleton, Aatmik Mallya, Swarup Dey, Joel Joseph, and Petr {\v{S}}ulc.
\newblock Nanobase. org: a repository for dna and rna nanostructures.
\newblock {\em Nucleic acids research}, 50(D1):D246--D252, 2022.

\bibitem{cantor1980biophysical}
Charles~R Cantor and Paul~Reinhard Schimmel.
\newblock {\em Biophysical chemistry: Part III: the behavior of biological macromolecules}.
\newblock Macmillan, 1980.

\bibitem{sassi2017shape}
Alberto~S Sassi, Salvatore Assenza, and Paolo De~Los~Rios.
\newblock Shape of a stretched polymer.
\newblock {\em Physical review letters}, 119(3):037801, 2017.

\bibitem{henrich2018coarse}
Oliver Henrich, Yair~Augusto Guti{\'e}rrez~Fosado, Tine Curk, and Thomas~E Ouldridge.
\newblock Coarse-grained simulation of dna using lammps: An implementation of the oxdna model and its applications.
\newblock {\em The European Physical Journal E}, 41:1--16, 2018.

\bibitem{odijk1995stiff}
Theo Odijk.
\newblock Stiff chains and filaments under tension.
\newblock {\em Macromolecules}, 28(20):7016--7018, 1995.

\bibitem{santalucia2004thermodynamics}
John SantaLucia~Jr and Donald Hicks.
\newblock The thermodynamics of dna structural motifs.
\newblock {\em Annu. Rev. Biophys. Biomol. Struct.}, 33(1):415--440, 2004.

\bibitem{dong2022gdod}
Xin Dong, Ruize Wu, Chao Xiong, Hai Li, Lei Cheng, Yong He, Shiyou Qian, Jian Cao, and Linjian Mo.
\newblock Gdod: Effective gradient descent using orthogonal decomposition for multi-task learning.
\newblock In {\em Proceedings of the 31st ACM International Conference on Information \& Knowledge Management}, pages 386--395, 2022.

\bibitem{javaloy2021rotograd}
Adri{\'a}n Javaloy and Isabel Valera.
\newblock Rotograd: Gradient homogenization in multitask learning.
\newblock {\em arXiv preprint arXiv:2103.02631}, 2021.

\bibitem{smith2018benchmarking}
Louis~G Smith, Zhen Tan, Aleksandar Spasic, Alan Grossfield, and David~H Mathews.
\newblock Benchmarking rna force fields using hairpin loop folding free energy change.
\newblock {\em Biophysical Journal}, 114(3):435a, 2018.

\bibitem{stamatis2023benchmarking}
Dimitris Stamatis, Chandra Verma, and Jonathan Essex.
\newblock Benchmarking rna all-atom force fields using hairpin motifs.
\newblock {\em ChemRxiv}, 2023.

\bibitem{yildirim2011benchmarking}
Ilyas Yildirim, Harry~A Stern, Jason~D Tubbs, Scott~D Kennedy, and Douglas~H Turner.
\newblock Benchmarking amber force fields for rna: Comparisons to nmr spectra for single-stranded r (gacc) are improved by revised $\chi$ torsions.
\newblock {\em The journal of physical chemistry B}, 115(29):9261--9270, 2011.

\bibitem{condon2015stacking}
David~E Condon, Scott~D Kennedy, Brendan~C Mort, Ryszard Kierzek, Ilyas Yildirim, and Douglas~H Turner.
\newblock Stacking in rna: Nmr of four tetramers benchmark molecular dynamics.
\newblock {\em Journal of chemical theory and computation}, 11(6):2729--2742, 2015.

\bibitem{bergonzo2015highly}
Christina Bergonzo, Niel~M Henriksen, Daniel~R Roe, and Thomas~E Cheatham.
\newblock Highly sampled tetranucleotide and tetraloop motifs enable evaluation of common rna force fields.
\newblock {\em Rna}, 21(9):1578--1590, 2015.

\bibitem{schrodt2015large}
Michael~V Schrodt, Casey~T Andrews, and Adrian~H Elcock.
\newblock Large-scale analysis of 48 dna and 48 rna tetranucleotides studied by 1 $\mu$s explicit-solvent molecular dynamics simulations.
\newblock {\em Journal of chemical theory and computation}, 11(12):5906--5917, 2015.

\bibitem{mlynsky2024can}
Vojtech Mlynsky, Petra K{\"u}hrov{\'a}, Martin Pykal, Miroslav Krepl, Petr Stadlbauer, Michal Otyepka, Pavel Banas, and Jiri Sponer.
\newblock Can we ever develop an ideal rna force field? lessons learned from simulations of the uucg rna tetraloop and other systems.
\newblock {\em Journal of Chemical Theory and Computation}, 2024.

\bibitem{barrangou2016applications}
Rodolphe Barrangou and Jennifer~A Doudna.
\newblock Applications of crispr technologies in research and beyond.
\newblock {\em Nature biotechnology}, 34(9):933--941, 2016.

\bibitem{praetorius2017self}
Florian Praetorius and Hendrik Dietz.
\newblock Self-assembly of genetically encoded dna-protein hybrid nanoscale shapes.
\newblock {\em Science}, 355(6331):eaam5488, 2017.

\bibitem{singh2010recent}
Yashveer Singh, Pierre Murat, and Eric Defrancq.
\newblock Recent developments in oligonucleotide conjugation.
\newblock {\em Chemical Society Reviews}, 39(6):2054--2070, 2010.

\bibitem{li2021rna}
Jun Li and Shi-Jie Chen.
\newblock Rna 3d structure prediction using coarse-grained models.
\newblock {\em Frontiers in Molecular Biosciences}, 8:720937, 2021.

\bibitem{vsponer2018rna}
Ji{\v{r}}{\'\i} {\v{S}}poner, Giovanni Bussi, Miroslav Krepl, Pavel Ban{\'a}{\v{s}}, Sandro Bottaro, Richard~A Cunha, Alejandro Gil-Ley, Giovanni Pinamonti, Sim{\'o}n Poblete, Petr Jure{\v{c}}ka, et~al.
\newblock Rna structural dynamics as captured by molecular simulations: a comprehensive overview.
\newblock {\em Chemical Reviews}, 118(8):4177, 2018.

\bibitem{cruz2018coarse}
Sergio Cruz-Le{\'o}n, Alvaro V{\'a}zquez-Mayagoitia, Simone Melchionna, Nadine Schwierz, and Maria Fyta.
\newblock Coarse-grained double-stranded rna model from quantum-mechanical calculations.
\newblock {\em The Journal of Physical Chemistry B}, 122(32):7915--7928, 2018.

\bibitem{uusitalo2017martini}
Jaakko~J Uusitalo, Helgi~I Ing{\'o}lfsson, Siewert~J Marrink, and Ignacio Faustino.
\newblock Martini coarse-grained force field: extension to rna.
\newblock {\em Biophysical journal}, 113(2):246--256, 2017.

\bibitem{lin2023evolutionary}
Zeming Lin, Halil Akin, Roshan Rao, Brian Hie, Zhongkai Zhu, Wenting Lu, Nikita Smetanin, Robert Verkuil, Ori Kabeli, Yaniv Shmueli, et~al.
\newblock Evolutionary-scale prediction of atomic-level protein structure with a language model.
\newblock {\em Science}, 379(6637):1123--1130, 2023.

\bibitem{bohlin2022design}
Joakim Bohlin, Michael Matthies, Erik Poppleton, Jonah Procyk, Aatmik Mallya, Hao Yan, and Petr {\v{S}}ulc.
\newblock Design and simulation of dna, rna and hybrid protein--nucleic acid nanostructures with oxview.
\newblock {\em Nature protocols}, 17(8):1762--1788, 2022.

\bibitem{izvekov2005multiscale}
Sergei Izvekov and Gregory~A Voth.
\newblock A multiscale coarse-graining method for biomolecular systems.
\newblock {\em The Journal of Physical Chemistry B}, 109(7):2469--2473, 2005.

\bibitem{cesari2018using}
Andrea Cesari, Sabine Rei{\ss}er, and Giovanni Bussi.
\newblock Using the maximum entropy principle to combine simulations and solution experiments.
\newblock {\em Computation}, 6(1):15, 2018.

\bibitem{fuchsChemtrainLearningDeep2024}
Paul Fuchs, Stephan Thaler, Sebastien R{\"o}cken, and Julija Zavadlav.
\newblock Chemtrain: {{Learning Deep Potential Models}} via {{Automatic Differentiation}} and {{Statistical Physics}}, August 2024.

\bibitem{jarzynski_nonequilibrium_1997}
C.~Jarzynski.
\newblock Nonequilibrium {Equality} for {Free} {Energy} {Differences}.
\newblock {\em Physical Review Letters}, 78(14):2690--2693, April 1997.
\newblock Publisher: American Physical Society.

\bibitem{oktay2020randomized}
Deniz Oktay, Nick McGreivy, Joshua Aduol, Alex Beatson, and Ryan~P Adams.
\newblock Randomized automatic differentiation.
\newblock {\em arXiv preprint arXiv:2007.10412}, 2020.

\bibitem{krueger2024generalized}
Ryan Krueger, Michael~P Brenner, and Krishna Shrinivas.
\newblock Generalized design of sequence-ensemble-function relationships for intrinsically disordered proteins.
\newblock {\em bioRxiv}, pages 2024--10, 2024.

\bibitem{mosayebiRoleLoopStacking2014c}
Majid Mosayebi, Flavio Romano, Thomas~E. Ouldridge, Ard~A. Louis, and Jonathan P.~K. Doye.
\newblock The {{Role}} of {{Loop Stacking}} in the {{Dynamics}} of {{DNA Hairpin Formation}}.
\newblock {\em J. Phys. Chem. B}, 118(49):14326--14335, December 2014.

\bibitem{Schreck2015}
John~S. Schreck, Thomas~E. Ouldridge, Flavio Romano, Petr {\v S}ulc, Liam~P. Shaw, Ard~A. Louis, and Jonathan~P.K. Doye.
\newblock {{DNA}} hairpins destabilize duplexes primarily by promoting melting rather than by inhibiting hybridization.
\newblock {\em Nucleic Acids Res.}, 43(13):6181--6190, July 2015.

\bibitem{NomidisTwistBendCoupling2019}
Stefanos~K. Nomidis, Enrico Skoruppa, Enrico Carlon, and John~F. Marko.
\newblock Twist-bend coupling and the statistical mechanics of the twistable wormlike-chain model of dna: Perturbation theory and beyond.
\newblock {\em Phys. Rev. E}, 99:032414, Mar 2019.

\bibitem{ouldridge2015dna}
Thomas~E Ouldridge.
\newblock Dna nanotechnology: understanding and optimisation through simulation.
\newblock {\em Molecular Physics}, 113(1):1--15, 2015.

\bibitem{machinek2014programmable}
Robert~RF Machinek, Thomas~E Ouldridge, Natalie~EC Haley, Jonathan Bath, and Andrew~J Turberfield.
\newblock Programmable energy landscapes for kinetic control of dna strand displacement.
\newblock {\em Nature communications}, 5(1):5324, 2014.

\bibitem{haley2020design}
Natalie~EC Haley, Thomas~E Ouldridge, Ismael Mullor~Ruiz, Alessandro Geraldini, Ard~A Louis, Jonathan Bath, and Andrew~J Turberfield.
\newblock Design of hidden thermodynamic driving for non-equilibrium systems via mismatch elimination during dna strand displacement.
\newblock {\em Nature communications}, 11(1):2562, 2020.

\bibitem{liu2024inverse}
Hao Liu, Michael Matthies, John Russo, Lorenzo Rovigatti, Raghu~Pradeep Narayanan, Thong Diep, Daniel McKeen, Oleg Gang, Nicholas Stephanopoulos, Francesco Sciortino, et~al.
\newblock Inverse design of a pyrochlore lattice of dna origami through model-driven experiments.
\newblock {\em Science}, 384(6697):776--781, 2024.

\bibitem{centola2024rhythmically}
Mathias Centola, Erik Poppleton, Sujay Ray, Martin Centola, Robb Welty, Juli{\'a}n Valero, Nils~G Walter, Petr {\v{S}}ulc, and Michael Famulok.
\newblock A rhythmically pulsing leaf-spring dna-origami nanoengine that drives a passive follower.
\newblock {\em Nature nanotechnology}, 19(2):226--236, 2024.

\bibitem{vsulc2015modelling}
Petr {\v{S}}ulc, Thomas~E Ouldridge, Flavio Romano, Jonathan~PK Doye, and Ard~A Louis.
\newblock Modelling toehold-mediated rna strand displacement.
\newblock {\em Biophysical journal}, 108(5):1238--1247, 2015.

\bibitem{matek2015coarse}
Christian Matek, Petr {\v{S}}ulc, Ferdinando Randisi, Jonathan~PK Doye, and Ard~A Louis.
\newblock Coarse-grained modelling of supercoiled rna.
\newblock {\em The Journal of chemical physics}, 143(24), 2015.

\bibitem{mattiotti2024molecular}
Giovanni Mattiotti, Manuel Micheloni, Lorenzo Petrolli, Lorenzo Rovigatti, Luca Tubiana, Samuela Pasquali, and Raffaello Potestio.
\newblock Molecular dynamics characterization of the free and encapsidated rna2 of ccmv with the oxrna model.
\newblock {\em Macromolecular Rapid Communications}, 45(24):2400639, 2024.

\bibitem{riemer2018learning}
Matthew Riemer, Ignacio Cases, Robert Ajemian, Miao Liu, Irina Rish, Yuhai Tu, and Gerald Tesauro.
\newblock Learning to learn without forgetting by maximizing transfer and minimizing interference.
\newblock {\em arXiv preprint arXiv:1810.11910}, 2018.

\bibitem{du2018adapting}
Yunshu Du, Wojciech~M Czarnecki, Siddhant~M Jayakumar, Mehrdad Farajtabar, Razvan Pascanu, and Balaji Lakshminarayanan.
\newblock Adapting auxiliary losses using gradient similarity.
\newblock {\em arXiv preprint arXiv:1812.02224}, 2018.

\bibitem{engel2023optimal}
Megan~C Engel, Jamie~A Smith, and Michael~P Brenner.
\newblock Optimal control of nonequilibrium systems through automatic differentiation.
\newblock {\em Physical Review X}, 13(4):041032, 2023.

\bibitem{davidchack2015new}
Ruslan~L Davidchack, TE~Ouldridge, and MV~Tretyakov.
\newblock New langevin and gradient thermostats for rigid body dynamics.
\newblock {\em The Journal of chemical physics}, 142(14), 2015.

\bibitem{leimkuhler2013rational}
Benedict Leimkuhler and Charles Matthews.
\newblock Rational construction of stochastic numerical methods for molecular sampling.
\newblock {\em Applied Mathematics Research eXpress}, 2013(1):34--56, 2013.

\bibitem{poppleton2020design}
Erik Poppleton, Joakim Bohlin, Michael Matthies, Shuchi Sharma, Fei Zhang, and Petr {\v{S}}ulc.
\newblock Design, optimization and analysis of large dna and rna nanostructures through interactive visualization, editing and molecular simulation.
\newblock {\em Nucleic acids research}, 48(12):e72--e72, 2020.

\bibitem{marko1995stretching}
John~F Marko and Eric~D Siggia.
\newblock Stretching dna.
\newblock {\em Macromolecules}, 28(26):8759--8770, 1995.

\bibitem{moritz2018ray}
Philipp Moritz, Robert Nishihara, Stephanie Wang, Alexey Tumanov, Richard Liaw, Eric Liang, Melih Elibol, Zongheng Yang, William Paul, Michael~I Jordan, et~al.
\newblock Ray: A distributed framework for emerging $\{$AI$\}$ applications.
\newblock In {\em 13th USENIX symposium on operating systems design and implementation (OSDI 18)}, pages 561--577, 2018.

\bibitem{bryantStructuralTransitionsElasticity2003}
Zev Bryant, Michael~D. Stone, Jeff Gore, Steven~B. Smith, Nicholas~R. Cozzarelli, and Carlos Bustamante.
\newblock Structural transitions and elasticity from torque measurements on {{DNA}}.
\newblock {\em Nature}, 424(6946):338--341, July 2003.

\bibitem{goreDNAOverwindsWhen2006}
Jeff Gore, Zev Bryant, Marcelo N{\"o}llmann, Mai~U. Le, Nicholas~R. Cozzarelli, and Carlos Bustamante.
\newblock {{DNA}} overwinds when stretched.
\newblock {\em Nature}, 442(7104):836--839, August 2006.

\end{thebibliography}

\newpage

\renewcommand{\thefigure}{S\arabic{figure}}
\setcounter{figure}{0}  

\renewcommand{\thetable}{S\arabic{table}}
\setcounter{table}{0}  

\appendix

\section{Preliminaries}

\subsection{Stochastic Gradient Estimation}

The challenge of computing the gradient of an expectation with respect to the parameters defining the distribution over which it is integrated is a well-explored topic in machine learning literature (for an in-depth review, refer to \cite{mohamed2020monte}). 
Consider a loss function \( \mathcal{L}(\theta) \) defined as:
\begin{align}
\mathcal{L}(\theta) &:= \int p(x; \theta) O(x; \theta) \, dx = \mathbb{E}_{p(x; \theta)}[f(x; \theta)]
\end{align}
Here, \( p(x; \theta) \) denotes a probability distribution over a random variable \( x \) that depends on the parameter \( \theta \), and \( O(x; \theta) \) represents an observable of interest. 
For instance, in minimizing the RMSD between a measured helical diameter with a target value, \( O(x; \theta) \) could represent the helical diameter for a particular configuration, and \( \mathcal{L}(\theta) \) captures the difference between the target value and the expected diameter over an ensemble of such configurations. 

Optimizing \( \mathcal{L}(\theta) \) often requires computing its gradient:
\begin{align}
\nabla_{\theta} \mathcal{L}(\theta) &= \nabla_{\theta} \left( \int p(x; \theta) O(x; \theta) \, dx \right) \\
&= \int p(x; \theta) \nabla_{\theta} O(x; \theta) \, dx + \int \nabla_{\theta} p(x; \theta) O(x; \theta) \, dx \\ 
&=  \mathbb{E}_{p(x; \theta)}[\nabla_{\theta} O(x; \theta)] + \mathbb{E}_{p(x; \theta)}[\nabla_{\theta} \log p(x; \theta) O(x; \theta)]
\end{align}
via the identity \( \nabla_{\theta} p(x; \theta) = p(x; \theta) \nabla_{\theta} \log p(x; \theta) \).
In standard supervised learning, random sampling of training examples ensures that \( \nabla_{\theta} \log p(x; \theta) = 0 \) and simplifies the gradient to \( \mathbb{E}_{p(x; \theta)}[\nabla_{\theta} O(x; \theta)] \).
Conversely, in reinforcement learning, the distribution \( p(x; \theta) \) -- determined by the policy -- is parameter-dependent, but the observable \( O(x; \theta) \) is usually not. 
To address this, the gradient is often reparameterized to isolate the source of randomness. In the \emph{pathwise gradient estimator}, samples \( x \) are generated from a distribution \( q_\epsilon(\theta) \), independent of \( \theta \), and are mapped deterministically to \( x(\epsilon; \theta) \). This results in:
\begin{align}
\nabla_{\theta} \mathcal{L}(\theta) &= \mathbb{E}_{p(\epsilon)}[\nabla_{\theta} O(x(\epsilon; \theta))]
\end{align}
This approach, known as the \emph{reparameterization trick}, allows the gradient to be computed as if it were a supervised learning problem. 
Alternatively, reinforcement learning employs the score-function gradient estimator:
\begin{align}
\nabla_{\theta} \mathcal{L}(\theta) &= \mathbb{E}_{p(x; \theta)}[\nabla_{\theta} \log p(x; \theta) O(x; \theta)]
\end{align}
Commonly known as the REINFORCE algorithm~\cite{mohamed2020monte}, this formulation provides a signal that increases the probability of high-value states and decreases the probability of low-value ones.

Each gradient estimation technique comes with its own advantages and limitations. 
The reparameterization trick is beneficial when the function or system being optimized can be differentiated through its entire process, allowing for gradient estimation that resembles standard backpropagation in supervised learning. 
This approach often results in lower variance estimates and more efficient optimization. 
However, its main drawback is the requirement for a differentiable path through the process, which may not always be feasible for complex or discrete simulations.
On the other hand, the score-function estimator (REINFORCE algorithm) provides more flexibility, as it only requires differentiating the probability distribution itself, making it applicable in cases where direct differentiation through the process is impossible. 
The trade-off, however, is that this method often has higher variance, which can lead to less stable and slower convergence during optimization.

\subsection{Differentiable Molecular Dynamics}

In standard molecular dynamics simulations, a system comprising $n$ interacting particles, represented by a vector $\mathbf{x} \in \mathcal{R}^{6n}$ that encodes their positions and momenta, is propagated iteratively over time using an integration step function $\mathcal{S}$:
\begin{align*}
    \mathbf{x}_{t+1} = \mathcal{S}(\mathbf{x}_t, \theta)
\end{align*}
Here, $\mathcal{S}$ is defined by the energy function and the numerical integration method, with $\theta$ representing control parameters.
For a given number of steps $N$, the final state $\mathbf{x}_N$ can be represented via a single function evaluation:
\begin{align}
    \mathcal{T}(\mathbf{x}_0) \equiv \mathcal{S}(\cdots \mathcal{S}(\mathcal{S}(\mathbf{x}_0))\cdots ) = \mathbf{x}_N
\end{align}
where $\mathcal{S}$ is applied $N$ times and $\mathbf{x}_0$ is the initial state.
This implies that an MD trajectory can be represented as the result of one numerical computation.

\begin{figure}[th!]
\begin{center}
\centerline{\includegraphics[width=0.85\textwidth]{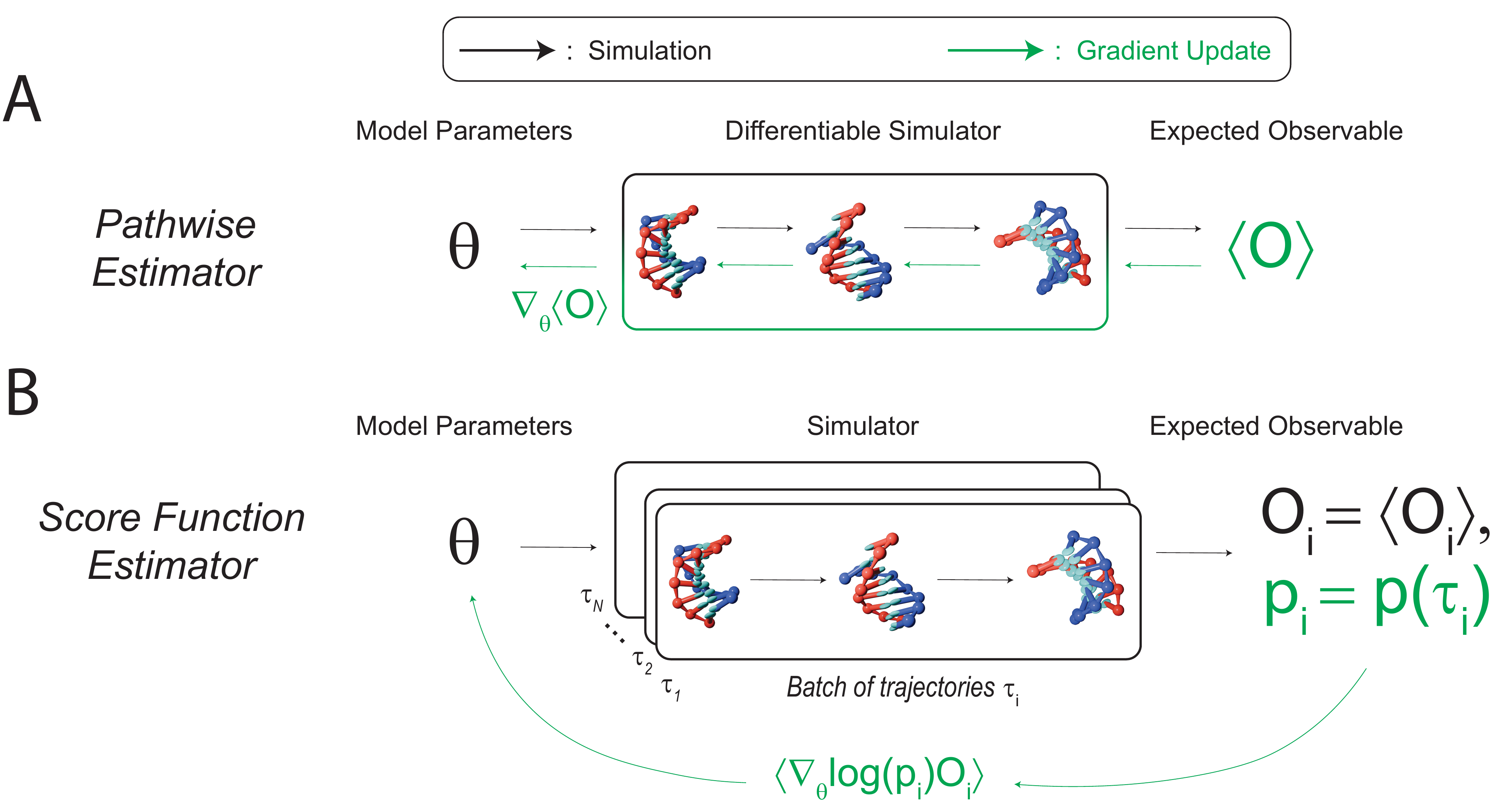}}
\caption{
Traditional means of stochastic gradient estimation for differentiable molecular dynamics (MD).
\textbf{A.} Under the \emph{pathwise estimator}, gradients are directly calculated via a differentiable simulator.
\textbf{B.} Alternatively, the \emph{score function estimator} scales the gradients of (log) probabilities of individual trajectories by a reward defined for each trajectory.
Differentiable (MD) typically relies on the pathwise estimator, though Engel et al. also used the score function estimator~\cite{engel2023optimal}.
}
\label{fig:traditional-grad-est}
\end{center}
\end{figure}

When embedded within an automatic differentiation framework, gradients of this computation can be efficiently computed with respect to $\theta$. 
Assume an observable dependent on the final state $\mathbf{x}_N$ and control parameters $\theta$, denoted $O(\mathbf{x}_N, \theta)$.
Given the stochastic nature of MD trajectories, the common objective is the expectation of such functions:
\begin{align}
    \mathcal{L}(\theta) \equiv \langle O(\mathbf{x}_{N}, \theta)\rangle_{\rho \in \mathrm{R}} &= \langle O(\mathcal{T}(\mathbf{x}_{0}), \theta)\rangle_{\rho \in \mathrm{R}} \\ 
    &= \frac{1}{|\mathrm{R}|}\sum_{\rho \in \mathrm{R}} O(\mathcal{T}_{\rho}(\mathbf{x}_{0}), \theta) \label{eqn:trajectory-expectation}
\end{align}
where $\mathrm{R}$ represents a set of random seeds initializing the trajectories and $\mathbf{x}_{N, \rho}$ is the final state for seed $\rho \in \mathrm{R}$.
Gradients are typically computed via the reparameterization trick (c.f. \cite{engel2023optimal}):
\begin{align}
    \nabla_{\theta}\mathcal{L}(\theta) &\approx \langle \nabla_{\theta} O(\mathcal{T}_{\rho}(\mathbf{x}_{0}), \theta)\rangle_{\rho \in \mathrm{R}}
    \label{eqn:basic-grad}
\end{align}
It is important to note that (i) objective functions can also be defined to depend on the entire trajectory rather than solely on the final state, and (ii) in equilibrium systems with long simulations, states separated by sufficiently large time intervals can be interpreted as distinct trajectories under standard assumptions of ergodicity. 
However, for clarity, this discussion will focus on the above formulation.
Figure \ref{fig:traditional-grad-est} depicts the methods for stochastic gradient estimation described above in the context of differentiable MD.

Since gradients must be computed through the simulation process, which requires repeated application of $\mathcal{S}$, computational and memory limitations may arise. 
To illustrate this, consider the gradient in Equation \ref{eqn:basic-grad} for one-dimensional space (see Ref. \cite{metz2021gradients} for details):
\begin{align}
    \frac{d O_N}{d \theta} = \frac{\partial{O_N}}{\partial{\theta}} + \sum_{k=1}^N \frac{\partial{O_N}}{\partial{\mathbf{x_N}}} \left(  
    \prod_{i=k}^N\frac{\partial{\mathbf{x_i}}}{\partial{\mathbf{x}_{i-1}}} \right) \frac{\partial{\mathbf{x_k}}}{\partial{\theta}} \label{eqn:expensive-grad}
\end{align}
where $O_N$ is the objective function evaluated at the final state.
A critical aspect here is the Jacobian of the dynamical system, represented by the matrix of partial derivatives $\frac{\partial{\mathbf{x_i}}}{\partial{\mathbf{x}_{i-1}}}$.
The magnitude of the Jacobian's eigenvalues must generally be less than one for the gradient calculation to be stable.

\subsection{Implicit Differentiation}

The typical paradigm in differentiable programming is to treat a program as an analytical calculation that is subject to the chain rule.
However, complications arise for iterative programs such as optimization algorithms (e.g. line fitting procedures).
Optimization algorithms can not typically be represented as an explicit formula in terms of their inputs; instead, one must differentiate through an unrolled iterative procedure, which imposes significant numerical and memory issues, or apply heuristics such as using the final iteration as a proxy for the optimization
problem solution.

An alternative approach to differentiating through optimization procedures is to implicitly relate the solution of an optimization problem to its inputs using optimality conditions. 
Consider an optimization procedure $y : \mathbb{R}^n \to \mathbb{R}^m$ that maps an input $x \in \mathbb{R}^n$ to a solution $y(x) \in \mathbb{R}^m$ that satisfies an optimality condition described by the relation $R(y(x), x) = 0$.
When the relation $R$ is continuously differentiable and $\partial R_y$ evaluated at $(y(x), x)$ is square invertible, the implicit function theorem provides expression for the implicit relationship between $x$ and $y(x)$:
\begin{align}
    \partial_x y(x) = \frac{\partial R_x(y(x), x)}{\partial R_y(y(x), x)} \label{eqn:implicit-diff}
\end{align}
Crucially, Equation \ref{eqn:implicit-diff} provides an alternative means of computing $\partial_x y(x)$ to unrolling $y$ that is exact while numerically and computationally stable.

Historically, this approach required complicated, case-specific derivations that made implicit differentiation difficult to use in practice.
However, Blondel et al.~\cite{blondel2022efficient} developed a method for automatic implicit differentiation that uses automatic differentiation to compute $\partial_x y(x)$ without deriving $\partial R_x(y(x), x)$ and $\partial R_y(y(x), x)$ by hand.
We use this method in our work to compute derivatives of line fitting procedures involved in the calculation of thermodynamic and mechanical properties of DNA.

\begin{figure}[t!]
\begin{center}
\centerline{\includegraphics[width=1.0\columnwidth]{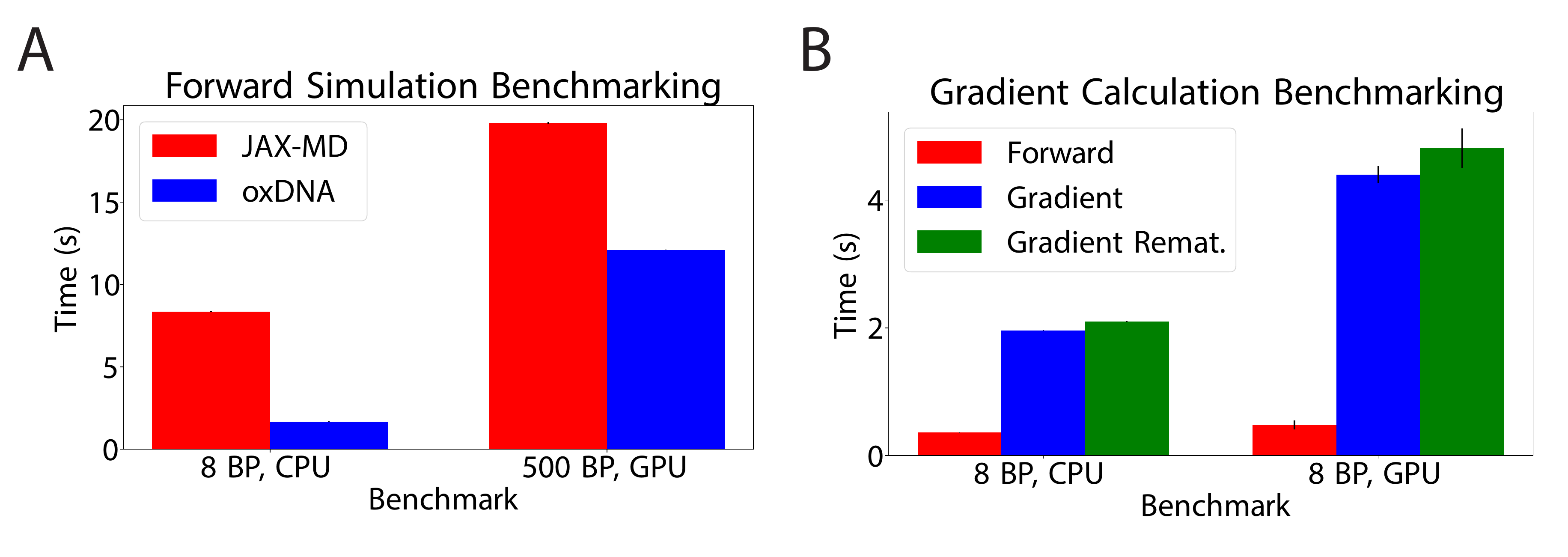}}
\caption{
Benchmarking the overhead cost for \textbf{A.} forward simulations and \textbf{B.} gradient calculations in JAX-MD. 
In \textbf{A.}, a small system (8 base pair duplex) and large system (500 base pair duplex) were simulated for 50,000 timesteps using both the oxDNA standalone code and JAX-MD. 
The 8 base pair system was simulated on CPU without neighbor lists while the 500 base pair system was simulated on NVIDIA A100 GPU with Verlet lists.
JAX-MD achieves comparable performance for the large system, taking $19.81$ seconds vs. the $12.11$ seconds with the standalone code.
For short simulations of small systems, the overhead of JAX's JIT-compilation exacerbates this difference.
In \textbf{B.}, we measure the time to simulate the 8 base pair system described above with and without gradient calculation. 
Gradient calculation through the unrolled trajectory imposes a significant time cost ($0.36 \text{ s}$ vs $1.96 \text{ s}$ on CPU, $0.47 \text{ s}$ vs $4.40 \text{ s}$ on GPU).
Gradient rematerialization enables gradient calculation through longer trajectories but at an additional time cost ($1.96 \text{ s}$ vs $2.10 \text{ s}$ on CPU, $4.40 \text{ s}$ vs $4.81 \text{ s}$ on GPU).
Observe that the small system runs faster on CPU.
\textbf{C.} The memory cost per pathwise estimator gradient calculation as a function of simulation length for a 60 bp duplex, the prototypical system for evaluating structural properties, revealing the maximum simulation length ($\sim$ 2500 steps) feasible on currently available hardware without gradient rematerialization..
\textbf{D.} The (log) average absolute value of the gradient of the RMSE of the measured pitch vs. the target pitch as a function of simulation length for simulations of a 60 bp duplex. 
Gradient rematerialization was used to permit gradient calculations at longer simulation lengths at the cost of runtime.
}
\label{fig:benchmark}
\end{center}
\end{figure}

\section{Model Details}\label{sec:notation}

For a given topology (i.e. a set of bonded pairs), an oxDNA configuration of $n$ nucleotides is represented in a space-fixed frame by the center of masses of each nucleotide, $\mathbf{R} \in \mathbb{R}^{n \times 3}$, and the orientations of each particle represented by a quaternion, $\mathbf{Q} \in \mathbb{R}^{n \times 4}$.
For a given nucleotide, a local reference frame can also be defined with basis vectors $\mathbf{a_1}$ (a vector directed from the backbone site to the base site), $\mathbf{a_2}$ (a vector normal to the plane of the base), and $\mathbf{a_3}$ ($\mathbf{a_3} = \mathbf{a_1} \times \mathbf{a_2}$).
For the $i^{th}$ nucleotide with orientation described by $\mathbf{q} = \mathbf{Q_i} \in \mathbb{R}^{4}$, the local nucleotide reference frame can be computed as:
\begin{align}
    a_1 = 
    \begin{bmatrix}
        q_0^2 + q_1^2 - q_2^2 - q_3^2, & 2(q_1q_2 + q_0q_3), & 2(q_1q_3 - q_0q_2)
    \end{bmatrix} \\
    a_2 = 
    \begin{bmatrix}
        2(q_1q_2 - q_0q_3), & q_0^2 - q_1^2 + q_2^2 - q_3^2, & 2(q_2q_3 + q_0q_1)
    \end{bmatrix} \\
    a_3 = 
    \begin{bmatrix}
        2(q_1q_3 + q_0q_2), & 2(q_2q_3 - q_0q_1), & q_0^2 - q_1^2 - q_2^2 + q_3^2
    \end{bmatrix}
\end{align}

In oxDNA, a nucleotide has three interaction sites used in the energy calculation: a backbone repulsion site, a stacking site, and a hydrogen bonding site. For a particular nucleotide with centre of mass $\mathbf{r} = \mathbf{R_i} \in \mathbb{R}^{3}$, the locations of these interaction sites in the \emph{space-fixed frame} are given by:
\begin{align}
    \mathbf{r_{back}} &= \mathbf{r} + d_{1,\text{back}} \mathbf{a_1} + d_{2,\text{back}} \mathbf{a_2} + d_{3,\text{back}} \mathbf{a_3}  \\ \mathbf{r_{stack}} &= \mathbf{r} + d_{1,\text{stack}} \mathbf{a_1} + d_{2,\text{stack}} \mathbf{a_2} + d_{3,\text{stack}} \mathbf{a_3}  \\
    \mathbf{r_{hb}} &= \mathbf{r} + d_{1,\text{hb}} \mathbf{a_1} + d_{2,\text{hb}} \mathbf{a_2} + d_{3,\text{hb}} \mathbf{a_3}  \\
\end{align}
where vectors $\mathbf{d_X} = (\text{d}_{1,X}, \text{d}_{2,X}, \text{d}_{3,X})$ are displacements from the nucleotide centre-of-mass to interaction sites $X$ in the local nucleotide frame; these are model-specific parameters controlling relative placement of interaction sites.
The values of $\mathbf{d_X}$ are as follows:
\begin{align}
    \mathbf{d_\text{back}^1} &= 
    \begin{bmatrix}
        -0.4, & 0.0, & 0.0
    \end{bmatrix} \\
    \mathbf{d_\text{back}^2} &= 
    \begin{bmatrix}
        -0.34, & 0.3408, & 0.0
    \end{bmatrix} \\
    \mathbf{d_\text{stack}} &= 
    \begin{bmatrix}
        0.34, & 0.0, & 0.0
    \end{bmatrix} \\
    \mathbf{d_\text{hb}} &= 
    \begin{bmatrix}
        0.4, & 0.0, & 0.0
    \end{bmatrix} \\
\end{align}
where $\mathbf{d_\text{back}^1}$ and $\mathbf{d_\text{back}^2}$ are the displacement vectors for the backbone site in oxDNA1 and oxDNA2, respectively.
In oxDNA2 (and oxRNA), several additional ``ghost'' sites are used in energy subterms (e.g. stacking) that do not have excluded volume; see Refs. \cite{snodin2015introducing} and \cite{vsulc2012sequence} for details. 

An oxDNA (or oxRNA) energy function is a function that maps a collection of center of masses $\mathbf{R} \in \mathbb{R}^{n \times 3}$, and orientations $\mathbf{Q} \in \mathbb{R}^{n \times 4}$ to a scalar -- it is the role of the energy function to determine the position of interaction sites.
Thus, the parameters describing the relative position of the interaction sites can be thought of as parameters of the energy function that can be optimized via our method just as the parameters describing angular and radial terms.
Our method could naturally be used to optimize these parameters though we present no such experiments in this work.

\begin{figure}[t!]
\begin{center}
\centerline{\includegraphics[width=1.0\columnwidth]{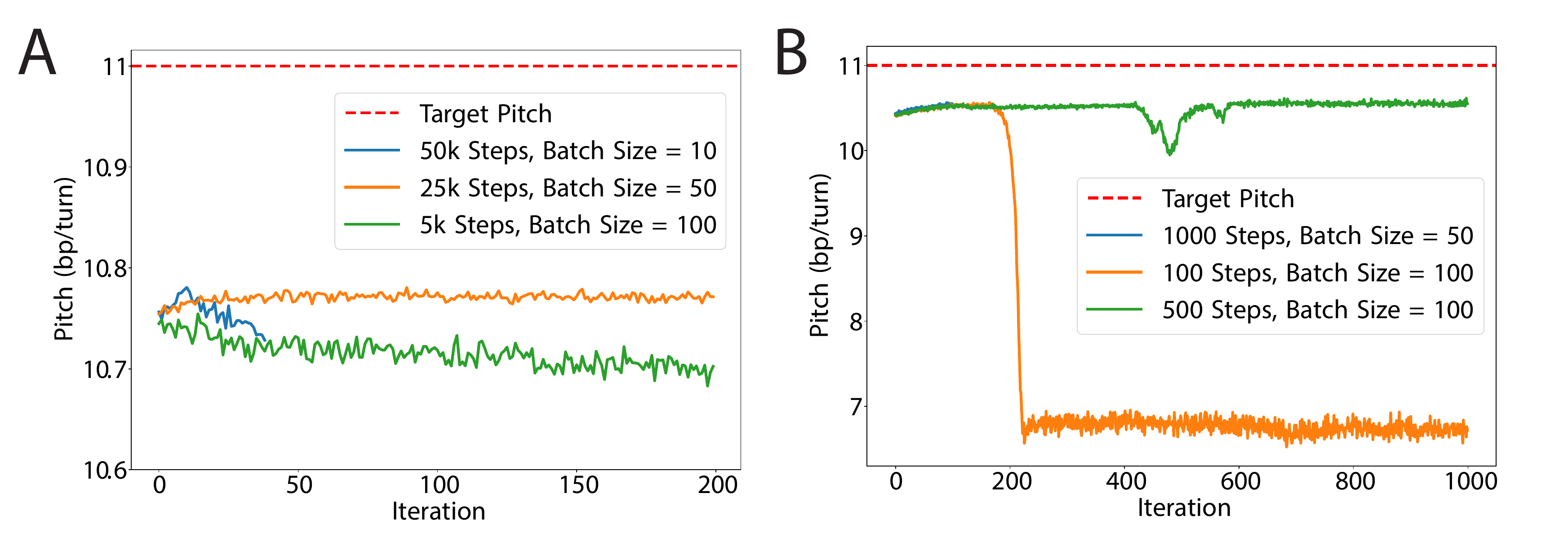}}
\caption{
Benchmarking the performance of traditional stochastic gradient estimators for pitch optimization in oxDNA.
Plots depict the pitch over time for optimizations using either the \textbf{A.} pathwise estimator or \textbf{B.} score function estimator, using a range of batch sizes and simulation lengths.
Optimizations using the pathwise estimator and simulations of length $50,000$ timesteps yielded \texttt{nan} values before the maximum number of iterations.
}
\label{fig:reinforce_and_reparam}
\end{center}
\end{figure}

\section{Benchmarking}\label{sec:benchmarking}

We first sought to benchmark performance for forward simulations via a pure JAX implementation of oxDNA vs. the standalone implementation.
Our pure JAX version involves an implementation of the oxDNA energy function in JAX and relies on JAX-MD for integration and force calculation via automatic differentiation.
We consider a small system (8 base pairs) and a large system (500 base pairs) on CPU and GPU, respectively; in Ref. \cite{sengar2021primer}, Sengar et al. highlight that small systems do not benefit from GPU acceleration.
We also employ neighbor lists to accelerate the force calculation for the 500 base pair system.
We observe a difference in runtime commensurate with the statistics reported by Schoenholz and Cubuk in their original presentation of JAX-MD~\cite{schoenholz2020jax}.
For the large system, simulated with GPU, we observe a 64\% increase in runtime relative to the standalone oxDNA CUDA code (see Fig. \ref{fig:benchmark}A).
We observe a significantly larger runtime cost to simulate the small system in JAX-MD due to the fixed overhead of JIT-compilation.

We next sought to benchmark the overhead of gradient calculation via our JAX implementation.
We considered only the 8 base pair system on both CPU and GPU.
We also compared vanilla backpropagation through the unrolled trajectory with checkpointed gradients (every 100 timesteps), a method which can enable gradient calculation through longer simulations by reducing memory at the cost of runtime.
For this system, vanilla backpropagation imposes a significant runtime overhead compared to forward simulations (approximately 450\% and 850\% on CPU and GPU, respectively); see Fig.~\ref{fig:benchmark}B.
Gradient checkpointing imposes an additional 7-9\% runtime increase.
Note that this cost may be larger for longer time scales or more complex systems, but this is impossible to evaluate empirically as gradients cannot be computed for such simulations without checkpointing.

\section{Optimizing Structural Properties}\label{sec:struct-appendix}

In this section, we describe the details of our structural optimization including simulation protocols and definitions of observables.

\subsection{Pitch}

\subsubsection{Definition}

The \emph{pitch} of a double-stranded DNA (or RNA) duplex describes the height of one complete helix turn, measured parallel to the helical axis; this is typically reported as the number of base pairs per helical turn.
Consider a duplex comprised of two single strands, each with $n$ nucleotides. 
This duplex contains $n$ base pairs where the $i^{th}$ base pair involves the pairing of the $i^{th}$ nucleotide strand 1 and the $(n-i)^{th}$ nucleotide of strand 2.
For two adjacent base pairs $i$ and $j$, we define the local helix axis as
\begin{align}
    \hat{\mathbf{h}} = \frac{\overline{\mathbf{hb}}_i - \overline{\mathbf{hb}}_j}{||\overline{\mathbf{hb}}_i - \overline{\mathbf{hb}}_j||}
\end{align}
where $\overline{\mathbf{hb}}_i$ denotes the mean hydrogen bonding site of the $i^{th}$ base pair.
We can then define $\theta_{i,j}$, the angle between the projections of two base-base vectors in a plane perpendicular to the helical axis, as
\begin{align}
    \theta_{i,j} = \arccos (\hat{\mathbf{back}}_i^{proj} \cdot \hat{\mathbf{back}}_{j}^{proj}),
\end{align}
where $\mathbf{back}_i$ denotes the vector between the backbone sites in the $i^{th}$ base pair and 
\begin{align}
    \mathbf{back}_i^{proj} = \mathbf{back}_i - \left(\mathbf{back}_i \cdot \hat{\mathbf{h}}\right)\hat{\mathbf{h}}
\end{align}
Let $O : \mathbb{R}^{2n \times 3} \to \mathbb{R}$ denote a map that computes the average angle $\theta_{ij}$ for all adjacent base pairs,
\begin{align}
    O(\mathbf{R}) = \langle \theta_{i, i+1} \rangle_{i \in [n-1]}
\end{align}
where $\mathbf{R}$ denotes a single configuration. 
In practice, the two base pairs nearest each end of the duplex are not included in the calculation to remove the effects of fraying.
We can then define the (differentiable) average $\theta$ across many equilibrium states
\begin{align}
    \overline{\theta_{i, i+1}} = \mathbb{E}[O(\mathbf{R})]_{\mathbf{R} \sim \exp(-\beta U(\mathbf{R}))}
\end{align}
Finally, the expected pitch $P$ is calculated as
\begin{align}
    P = \frac{2\pi}{\overline{\theta_{i,i+1}}} \text{ base pairs / turn} \label{eq:pitch}
\end{align}
Note that $P$ is differentiable via the differentiability of $\overline{\theta_{i,i+1}}$.
For pitch optimizations, we minimize the RMSE between Eq.~\ref{eq:pitch} and a chosen target value.

\begin{figure}[t!]
\begin{center}
\centerline{\includegraphics[width=1.0\columnwidth]{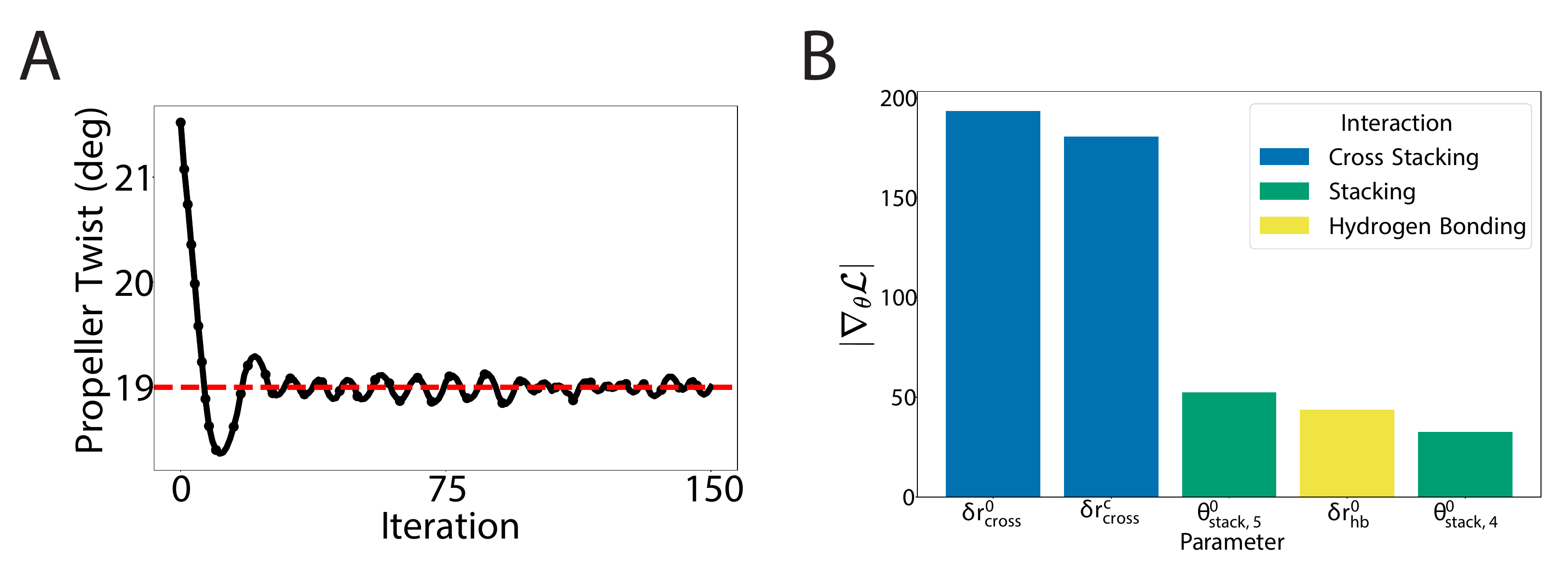}}
\caption{
Fitting oxDNA parameters to a target propeller twist via trajectory reweighting. 
\textbf{A.} The loss values over time, defined as the RMSE between the calculated propeller twist and a target propeller twist of 19\textdegree.
Iterations in which reference state are resampled are represented as scatter points.
\textbf{B.} The five most significant parameters ranked by the absolute value of their gradients.
}
\label{fig:ptwist}
\end{center}
\end{figure}

\subsubsection{Optimization via Reparameterization Gradient}

Despite the memory cost and numerical instabilities arising from differentiating through a simulation, we sought to evaluate the traditional pathwise estimator for pitch optimization.
We defined the loss as the RMSE between the measured pitch and the target pitch.
For a given simulation length and batch size (i.e. number of replicate simulations), we simulated a 60 base pair duplex using the Langevin integrator in JAX-MD and used the gradients computed via the pathwise estimator for gradient descent.
Averages were computed from states sampled every 100 timesteps and all simulations were initialized with states obtained from $10,000$ equilibration steps.
All optimizations used a learning rate of $0.001$.
We found that the resultant gradient estimates did not yield optimized parameters for a range of simulation lengths and batch sizes (Figure \ref{fig:reinforce_and_reparam}A).

As structural properties are defined locally and therefore have particularly fast relaxation timescales, in practice we could mitigate these numerical instabilities by drastically increasing the friction coefficient, tantamount to decreasing the simulation time-step.
However, this is not a tenable solution for many observables of interest as such an approach requires significantly longer simulations.
We also note that numerical instabilities can be mitigated by gradient clipping but this method imposes its own numerical approximations and also requires significant time and memory overhead.

\subsubsection{Optimization via Score-Function Estimator}

We then sought to evaluate the performance of the score function estimator.
This requires obtaining a differentiable expression for the probability of each trajectory.
The Langevin integrator in JAX-MD implements the integration scheme of Davidchak et al.~\cite{davidchack2015new} that follows a ``BAOAB'' splitting scheme~\cite{leimkuhler2013rational}, one choice for splitting the Langevin equation into deterministic and stochastic components.
We represent the probability of a given timestep as the probability of the stochastic step in the integration scheme, and therefore the probability of a trajectory as the product of the probabilities over all timesteps.
We extract this probability from JAX-MD and it is differentiable via JAX.
All simulations were initialized with states obtained from $1,000$ equilibration steps and all optimizations used a learning rate of $0.001$.

Given a simulation length and batch size, we use this probability to obtain gradient estimates for pitch optimization for optimization.
We evaluate a range of simulation lengths and batch sizes with a bias towards shorter simulations and higher batch sizes given the high variance introduced by longer simulations.
In practice, we do not find any hyperparameter set that yields optimized parameter values.
See Figure \ref{fig:reinforce_and_reparam}B for our results.

\subsubsection{Optimization via DiffTRE}

The gradient estimator for the optimizations depicted in Figure 2 requires a set of reference states at each iteration.
To sample reference states for a given set of parameters $\hat{\theta}$, we perform 10 parallel simulations of length $250,000$ timesteps and sample states every $1,000$ steps.
The total number of reference states is therefore $N_{\text{ref}} = 2,500$.
We resample reference states when (i) $N_{\text{eff}} < \lambda N_{\text{ref}}$ with $\lambda = 0.95$ or (ii) after five gradient updates.
We sampled reference states using the standalone oxDNA code rather than JAX-MD, and automatically recompiled the executable for each updated parameter set.
All optimizations used a learning rate of $0.001$.

\begin{figure}[t!]
\begin{center}
\centerline{\includegraphics[width=1.0\columnwidth]{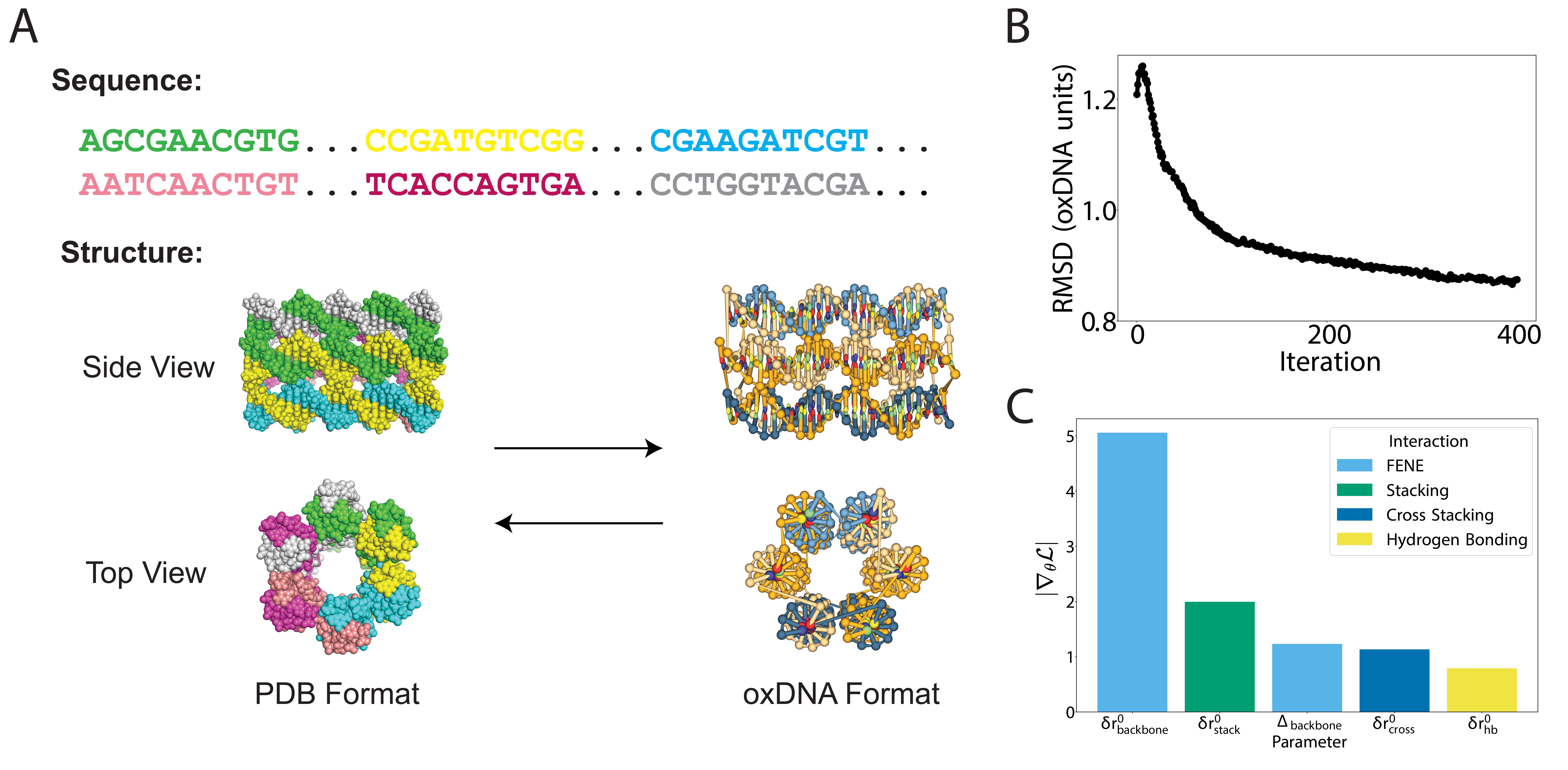}}
\caption{
Fitting oxDNA parameters to a target sequence/structure.
\textbf{A.} The gated DNA nanopore introduced in Ref. \cite{burns2016biomimetic}. 
The structure is depicted both in PDB format and in the coarse-grained representation employed by oxDNA.
\textbf{B.} The average RMSD of the simulated structure compared to the target structure throughout the optimization.
\textbf{C.} Sensitivity analysis for the RMSD optimization, represented by the gradients with the top five largest absolute values from the first iteration.
}
\label{fig:structure_opt}
\end{center}
\end{figure}

\subsection{Propeller Twist}

In a DNA duplex, two paired bases do not typically align in the same plane.
The \emph{propeller twist} of a DNA duplex refers to the rotation of the base pairs around the axis perpendicular to the helical axis of the DNA.

Again, consider a duplex comprised of two single strands, $\alpha$ and $\beta$, each with $n$ nucleotides where the $i^{th}$ base pair involves the pairing of the $i^{th}$ nucleotide of $\alpha$ and the $(n-i)^{th}$ nucleotide of $\beta$.
For the $i^{th}$ base pair, let $\mathbf{a}_{2,i}^{\alpha}$ and $\mathbf{a}_{2,n-i}^{\beta}$ denote the vectors normal to the plane of the base for each of two base paired nucleotides (see above for the definition of $\mathbf{a}_2$).
The propeller twist of the $i^{th}$ base pair is
\begin{align}
    \phi_i = \arccos (\mathbf{a}_{2,i}^{\alpha} \cdot \mathbf{a}_{2,n-i}^{\beta})
\end{align}
The propeller twist of a single state for the is computed as the average twist over all base pairs, 
\begin{align}
    O(\mathbf{R}) = \frac{1}{n}\sum_{i=1}^n \phi_i
\end{align}
As with pitch, the first two base pairs at each end are omitted due to fraying.
The propeller twist $T$ is understood be the expected value computed across many equilibrium states,
\begin{align}
    T = \mathbb{E}[O(\mathbf{R})]_{\mathbf{R} \sim \exp(-\beta U(\mathbf{R}))}
\end{align}

\subsection{Sequence-Structure Pair}

In machine learning, it is typical to train a model on known sequence-structure pairs (e.g. AlphaFold).
We sought to demonstrate the flexibility of our method to accommodate such objective functions.
As a simple example, we consider the gated DNA nanopore system of Burns et al. composed of six circular helices arranged in a ring~\cite{burns2016biomimetic}.
We treated the unrelaxed structure deposited in Nanobase~\cite{poppleton2022nanobase} as the ground truth, target structure (Figure \ref{fig:structure_opt}).
We used the negative simulated RMSD, measured via the oxDNA analysis tools~\cite{poppleton2020design}, as a loss function.
We successfully optimized the current oxDNA parameters to better match the unrelaxed structure as represented by a decrease in the average RMSD (Figure \ref{fig:structure_opt}B).
This example is relatively contrived as the unrelaxed structure contains unphysical bonds between adjacent helices and the optimization relaxes the parameters of the backbone FENE spring to accommodate this (see Figure \ref{fig:structure_opt}C). 
Regardless, this serves as a simple demonstration of our ability to optimize directly for target sequence-structure pairs and to leverage off-the-shelf analysis tools, obviating the need for reimplementation.
Each sampling of reference states involved 28 independent simulations of $10^5$ timesteps with snapshots sampled every $1,000$ steps.

\begin{figure}[t!]
\begin{center}
\centerline{\includegraphics[width=1.0\columnwidth]{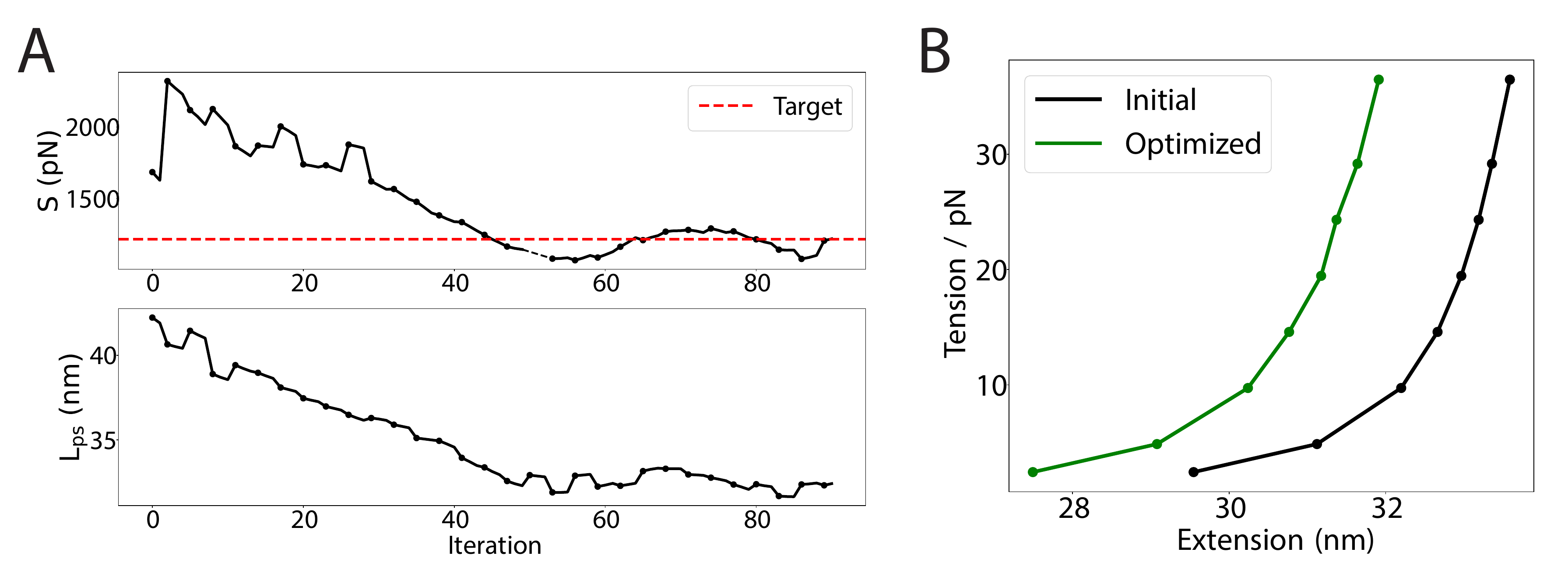}}
\caption{
Optimization of stretch modulus via a worm-like chain fit in which the persistence length is constrained via a passive calculation. 
\textbf{A.} The stretch modulus and corresponding persistence length over time.
\textbf{B.} The force extension curves corresponding to the initial and optimized parameters.
Iterations in which reference state are resampled are represented as scatter points.
}
\label{fig:wlc}
\end{center}
\end{figure}

\section{Optimizing Mechanical Properties}\label{sec:mech-appendix}

\subsection{Persistence Length}
Persistence length is a measure of stiffness, representing the length over which a polymer chain maintains a given direction before bending due to thermal fluctuations. 
We employ two methods for computing DNA persistence length: passively, by considering the decay in correlation length of tangent vectors; and simulating the force response of the DNA duplex and fitting to a worm-like chain (WLC) model. 
These approaches are described respectively below. 

\subsubsection{Passive calculation}

For an infinitely long, semi-flexible polymer for which correlations in alignment decay exponentially with separation, $L_{ps}$ obeys the following equality:
\begin{align}
    \langle \mathbf{l_m} \cdot \mathbf{l_0} \rangle = \exp (-m \langle l_0 \rangle / L_{ps})
\end{align}
where $\mathbf{l_0}$ denotes the vector between the first two monomer and $\langle \mathbf{l_m} \cdot \mathbf{l_0} \rangle$ denotes the correlation between the first and $n^{th}$ monomers~\cite{ouldridgeStructuralMechanicalThermodynamic2011a,cantor1980biophysical}.
In practice, $L_{ps}$ is computed by fitting the slope of the logarithm of this relationship, i.e.
\begin{align}
    \log (\langle \mathbf{l_m} \cdot \mathbf{l_0} \rangle) = -m \langle l_0 \rangle / L_{ps} \label{eq:log-lp}
\end{align}
Thus, $L_{ps}$ can be calculated via simulation by sampling many equilibrium configurations, computing the average correlation of nucleotide tangent vectors as a function of separation, and extracting the persistence length according to Eq.~\ref{eq:log-lp}.

We calculate the tangent vector to the $i^{th}$ base pair, $\mathbf{l_i}$, to be the the (normalized) vector between the midpoints of that base pair and the adjacent base pair in the duplex: 
\begin{align}
    \mathbf{l_i} = \frac{\mathbf{m}_{i+1} - \mathbf{m}_{i}}{||\mathbf{m}_{i+1} - \mathbf{m}_{i}||}
\end{align}
where $\mathbf{m}_{i}$ denotes the midpoint of the base sites interacting in base pair $i$.
For two base pairs $i$ and $j$ separated by $m$ base pairs, the correlation between $i$ and $j$ can be expressed as 
\begin{align}
    c_{ij}  = \mathbf{l_i} \cdot \mathbf{l_j}
\end{align}
For a single snapshot of a duplex of length $n$, there are $n-m$ such pairs of base pairs separated by $m$ base pairs.
Thus, the average correlation between base pairs separated by a distance $m$ is
\begin{align}
    c_m = \frac{1}{n - m}\sum_{ij} c_{ij}\delta(j - i = m)
\end{align}
We compute an expected correlation for nucleotides separated by $m$ base pairs over many snapshots, i.e. $\mathbb{E}[O_m(\mathbf{R})]_{\mathbf{R} \sim \exp(-\beta U(\mathbf{R}))}$.
Given values for $\mathbb{E}[O_m(\mathbf{R})]$ for $1 \leq m \leq n$, we can compute $L_{ps}$ by fitting 
\begin{align}
    \mathbb{E}[O_m(\mathbf{R})] = \exp(-m\langle l_0 \rangle / L_{ps}) \label{eqn:lps-exp}
\end{align}
with $\langle l_0 \rangle$ the average length of the tangent vectors, and extracting the slope, 
\begin{align}
    L_{ps} = -\frac{\ln \mathbb{E}[O_m(\mathbf{R})]}{m\langle \ell_0 \rangle} \label{eqn:lps-slope}
\end{align}

Crucially, the entire calculation of $L_{ps}$ given a set of reference states can be made differentiable by combining DiffTRE and implicit differentiation. 
$\mathbb{E}[O_m(\mathbf{R})]$ is differentiable for all $m$ via reweighting and implicit differentiation can be used to differentiate the line-fitting procedure described in Equations \ref{eqn:lps-exp} and \ref{eqn:lps-slope} in the case where an analytic solution is not available.

Persistence length calculations notoriously require many uncorrelated states to obtain sufficient correlation statistics~\cite{ouldridgeStructuralMechanicalThermodynamic2011a}.
In all optimizations of $L_{ps}$ via this passive calculation, we simulate a 60 bp duplex following Ref. \cite{snodin2015introducing}.
Each calculation of $L_{ps}$ involves 64 parallel simulations each comprising $2.5 \times 10^6$ timesteps, sampling reference states every $10^4$ steps.
We neglect the first 2 base pairs on either end of the duplex due to boundary effects and only fit Equation \ref{eqn:lps-exp} for $m \leq 40$ as large $m$ have poor statistics.
We again use a learning rate of $0.001$ and resample states via the same protocol as described for pitch optimizations.

\begin{figure}[t!]
\begin{center}
\centerline{\includegraphics[width=1.0\columnwidth]{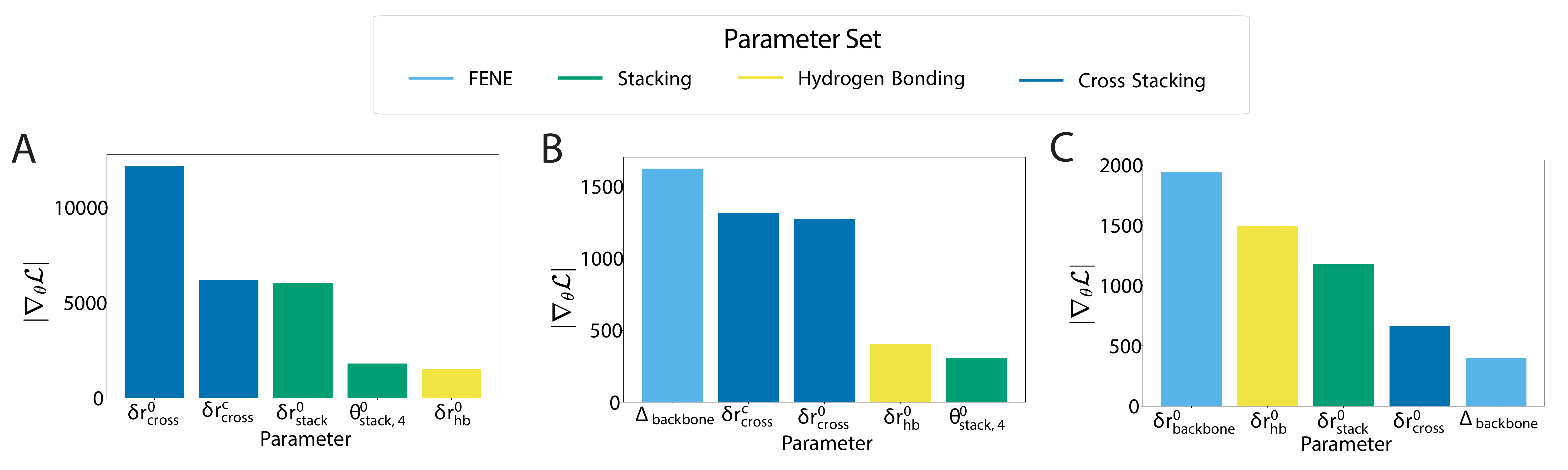}}
\caption{
Sensitivity analysis for the stretch-torsion observables, represented by the five most significant parameters ranked by the absolute value of their gradients, for the \textbf{A.} effective stretch modulus $S_{\text{eff}}$, \textbf{B.} torsional modulus $C$, and \textbf{C.} twist-stretch coupling $g$.
}
\label{fig:moduli_sensitivity}
\end{center}
\end{figure}

\subsubsection{Worm-like-chain (WLC) Fitting}

Under low tension, a long molecule of dsDNA behaves like an extensible wormlike chain (WLC)~\cite{marko1995stretching}. Both persistence length $L_{ps}$ and extensional modulus $S$, which quantifies the elastic resistance of dsDNA to stretching, can be extracted from fits to the WLC expression of Odjik \textit{et al.}: 
\begin{align}
    x = L_0\left(1 + \frac{F}{S} - \frac{\kT}{2F}\left[ 1 + y\coth y\right] \right) \label{eqn:wlc-supp}
\end{align}
with
\begin{align}
    y = \left( \frac{FL_0^2}{L_{ps}\kT}  \right)^{1/2}
\end{align}
where $x$ is the extension of the duplex under force $F$; $L_0$ is the contour length of the duplex; and $\kT$ is the thermal energy unit at temperature T. 

In prior work~\cite{ouldridgeStructuralMechanicalThermodynamic2011a,ouldridge2012coarse}, $S$ was calculated by applying tension to the middle 100 base pairs of a 100 base pair duplex and conducting an unconstrained three-parameter fit to Equation \ref{eqn:wlc-supp} to obtain $S$, $L_{ps}$, and $L_0$.
This process is amenable to gradient-based optimization by again combining DiffTRE and implicit differentiation.
Consider a simulation of such a duplex under pulling force $F_i$ that yields a collection of sampled states $\overrightarrow{X}_i$.
Let $O_i : \mathbb{R}^{3 \times n} \to \mathbb{R}$ be a function that computes the end-to-end distance of the 100 bp region subject to the pulling force $F_i$.
The extension of the duplex is simply the arithmetic mean of extensions over all sampled states and therefore can be expressed as the following reweighted expectation:
\begin{align}
    \mathbb{E}\left[ O_i(X) \right]_{X \sim \exp (-\beta U^{tot}(X))} = \sum_X w_X O_i(X)
\end{align}
with $U^{tot}(X) = U^{ext}(X) + U(X)$ where $U(X)$ denotes the Hamiltonian of the DNA model parameterized by $\theta$ and $U^{ext}(X)$ denotes the energetic contribution corresponding to the constant external force.
This yields a set of weights
\begin{align}
    w_X &= \frac{\exp\left(-\beta \left(U^{tot}_{\theta}(X) - U^{tot}_{\hat{\theta}}(X)\right)\right)}{\sum_{X'} \exp\left(-\beta \left(U^{tot}_{\theta}(X') - U^{tot}_{\hat{\theta}}(X')\right)\right)} \\
    &= \frac{\exp\left(-\beta \left(U^{ext}_{\theta}(X) + U_{\theta}(X) - U^{ext}_{\hat{\theta}}(X) - U_{\hat{\theta}}(X)\right)\right)}{\sum_{X'} \exp\left(-\beta \left(U^{ext}_{\theta}(X') + U_{\theta}(X') - U^{ext}_{\hat{\theta}}(X') - U_{\hat{\theta}}(X')\right)\right)} \\
    &= \frac{\exp\left(-\beta \left(U_{\theta}(X) - U_{\hat{\theta}}(X)\right)\right)}{\sum_{X'} \exp\left(-\beta \left(U_{\theta}(X') - U_{\hat{\theta}}(X')\right)\right)}\label{eq:Uext_cancels}
\end{align}
Importantly, the contribution of $U^{ext}$ to $w_X$ cancels which is crucial to calculating $w_X$ in practice as in general $U^{ext}$ is not known for an arbitrary external force (or torque). 
This is also intuitive as $U^{ext}$ has no \emph{explicit} dependence on the parameters; i.e. $\nabla_{\theta}U^{ext} = 0$.
With our differentiable estimate for the extension $O_i$ under force $F_i$ in hand, we can use nonlinear least squares to solve for $S$ per Equation \ref{eqn:wlc-supp}.
As in the case of the passive persistence length calculation, we use implicit differentiation to differentiate this fitting procedure.

However, in practice, we find this calculation to be highly sensitive to small changes in parameter values which is an undesirable property for gradient-based optimization.
To remedy this instability, we fix the persistence length via the passive calculation and compute $S$ and $L_0$ via the WLC fit.
This yields stable optimization as shown in Figure \ref{fig:wlc}.

We compute the persistence length with the parameters described above.
For the WLC fit, we consider force values of 0.05, 0.1, 0.2, 0.3, 0.4, 0.5, 0.6, and 0.75 (in oxDNA units).
The WLC fit requires a calculation of the extension under each force and therefore we perform a batch of simulations for each force.
For each force, we perform 12 simulations of length $2.5 \times 10^6$ timesteps and sample reference states every $10^5$ steps.
Low forces are particularly hard to resolve as there are larger fluctuations in conformational space, and we therefore simulate $\times 4$ as many independent simulations for low forces (i.e. 0.05 and 0.1 oxDNA units).
For comparison with Ref. \cite{ouldridge2010dna}, we perform this optimization using the oxDNA 1.0 force field.
In total, sampling all reference states requires 232 independent simulations.
We use Ray to distribute this calculation~\cite{moritz2018ray}.
We use a learning rate of $0.001$, and resample reference states with $\lambda = 0.95$ or after three gradient updates.

\begin{figure}[t!]
\begin{center}
\centerline{\includegraphics[width=1.0\columnwidth]{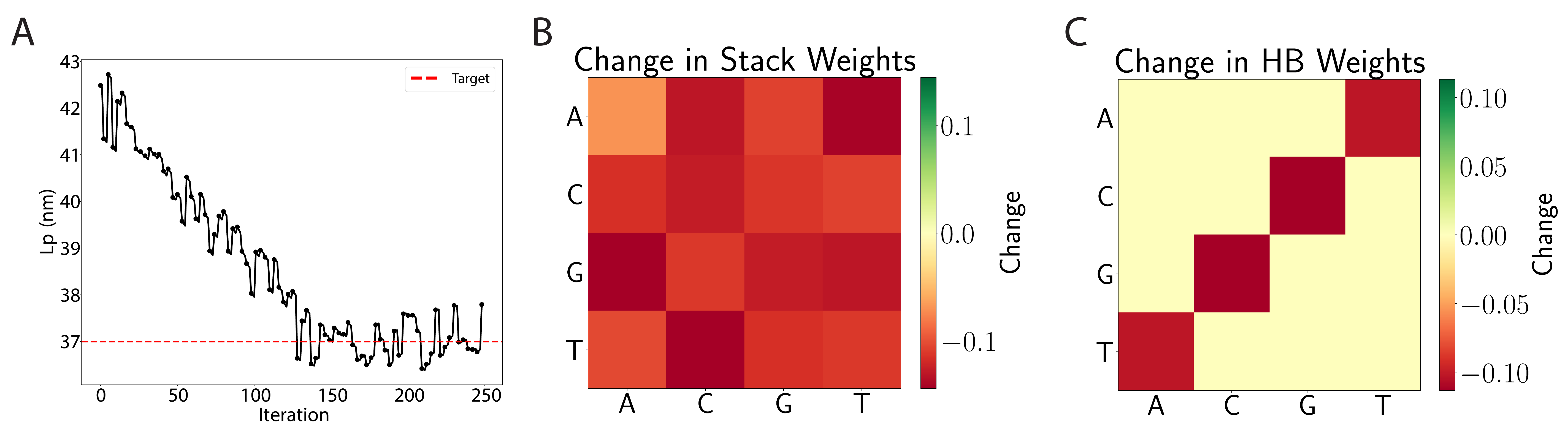}}
\caption{
Fitting the sequence-specific stacking and hydrogen-bonding parameters to a target persistence length.
\textbf{A.} 
The persistence length over time.
Iterations in which reference state are resampled are represented as scatter points.
\textbf{B-C.} The absolute change in the sequence-specific stacking and hydrogen bonding parameters between the initial and optimized values.
We obey the same symmetries for sequence-specific parameters as described in Ref. \cite{vsulc2012sequence}
}
\label{fig:lp-ss}
\end{center}
\end{figure}

To demonstrate the flexibility of our method, we also optimize only the sequence-specific parameters to achieve an arbitrarily chosen persistence length (Figure \ref{fig:lp-ss}).
Sequence specificity is implemented in recent versions of oxDNA via context-dependent scalars of the hydrogen bonding and stacking interactions, and equality between subsets of sequence-specific parameters is enforced to obey symmetries (e.g. 5' to 3' vs. 3' to 5').
We enforce all such symmetries in our optimization.

\subsection{Stretch and Torsional Moduli}

An analogous quantity of interest is the \emph{torsional modulus} $C$, which characterizes a polymer's resistance to twisting under external torque. Under the action of a non-zero external torque $\tau$, a DNA helix will twist some amount $\Delta \theta $ beyond its equilibrium twist value $\theta_0$ of $\sim 35 \si{\degree}$~\cite{bryantStructuralTransitionsElasticity2003}): $\Delta \theta = \theta - \theta_0$, where $\theta$ is the total twist. It can be shown that according to equipartition of energy, at zero torque, one can measure the fluctuations in $\Delta \theta$ and extract the torisional modulus according to~\cite{goreDNAOverwindsWhen2006,assenzaAccurateSequenceDependentCoarseGrained2022}:
\begin{align}
    \langle \Delta \theta^2 \rangle = \frac{\kT L_0}{C}
    \label{eq:equipartition_C}
\end{align}
where $\langle \Delta \theta^2 \rangle$ is the variance in deviation from equilibrium twist over some ensemble of samples and $L_0$ is the equilibrium end-to-end extension of the DNA. 
In the original presentation of oxDNA, $C$ was calculated using Monte Carlo moves chosen to enforce that the base pairs at the end of the central section remained perpendicular to the vector between their midpoints~\cite{ouldridgeStructuralMechanicalThermodynamic2011a}.
This allowed for an unambiguous definition of the axis about which torsion could be applied and twist measured.
However, such MC moves are not implemented in a publicly available software package and such constraints cannot be applied as rigorously in MD, yielding poor estimates of C using Eqn.~\ref{eq:equipartition_C}.

We therefore follow the protocol of Assenza and Pérez~\cite{assenzaAccurateSequenceDependentCoarseGrained2022} for calculating $C$.
This involves subjecting a dsDNA duplex to various external forces and torques; measuring the changes in extension and twist due to external force and torque, respectively; and computing stretch modulus $S$ and torsional modulus $C$ via an analytic formula. This approach also allows for the extraction of the twist-stretch coupling constant $g$.

For a given force $f$ and torque $\tau$, we simulate a 40 bp duplex subjected to the simultaneous action of a pulling force $f$ and torque $\tau$ along the z-axis.
The molecule is initialized with its long axis along the z-axis and a harmonic constraint is applied to one end of the duplex to fix the position of two bottom base pairs.
At the other end of the duplex, a pulling force with constant magnitude $f$ is applied in the z-direction to the center of mass of the second base pair from the end and harmonic constraints are applied to x- and y-coordinates of the same base pair to maintain alignment with the z-axis.
A constant torque $\tau$ is applied to the two base pairs at the same end.
See the section titled ``Stretch–Torsion Simulations'' in Ref.~\cite{assenzaAccurateSequenceDependentCoarseGrained2022} for exact simulation details.

Following Ref.~\cite{assenzaAccurateSequenceDependentCoarseGrained2022}, we first performed simulations at $\tau = 0$ for pulling forces $f$ = 2, 4, 6, 8, 10, 15, 20, 25, 30, 35, and 40 $\si{\pico\newton}$.
For each force $f$, we compute the extension $\Delta L$ as
\begin{align}
    \langle \Delta L \rangle = \langle L \rangle - L_0 \label{eq:dl}
\end{align}
where $\langle L \rangle$ is average end-to-end extension over the set of reference states from the simulation under force $f$ and $L_0$ is the average extension under $f = 0$ and $\tau = 0$.
We then fit the linear dependence of $\Delta L$ on the pulling force $f$ at $\tau = 0$ to extract the slope $A_1$,
\begin{align}
    \langle \Delta L \rangle = A_1 f
\end{align}
Next, we then performed simulations at $f = 2~\si{\pico\newton}$ for values of the torque $\tau = $ 0, 5, 10, 15, 20, 25, and 30 $\si{\pico\newton\cdot\nano\metre}$.
For each simulation, we compute both $\langle \Delta L \rangle$ (as defined above) and $\langle \Delta \theta \rangle$ where
\begin{align}
   \langle \Delta \theta \rangle = \langle \theta \rangle - \theta_0 \label{eq:dtheta}
\end{align}
where $\langle \theta \rangle$ is the expected value of twist over the set of reference states from the simulation under torque $\tau$ and $\theta_0$ is the equilibrium twist under $f = 0$ and $\tau = 0$.
See Ref. \cite{assenzaAccurateSequenceDependentCoarseGrained2022} for a formal definition of $L$ and $\theta$.
We then fit the linear dependence of both the extension and the torsion angle on torque, extracting the slopes $A_3$ and $A_4$, respectively ($A_2$ is omitted for notational consistency with Ref~\cite{assenzaAccurateSequenceDependentCoarseGrained2022}):
\begin{align}
    \langle \Delta L \rangle = \text{constant} + A_3 \tau \label{eq:dl_line} \\
    \langle \Delta \theta \rangle = \text{constant} + A_4 \tau \label{eq:dtheta_line}
\end{align}
Finally, we can compute $C$, $\tilde{S}$, and $g$ from $A_1$, $A_3$, and $A_4$ following the derivation given in Ref~\cite{assenzaAccurateSequenceDependentCoarseGrained2022}:
\begin{align}
    \tilde{S} &= \frac{L_0}{A_1} \\
    C &= \frac{A_1 L_0}{A_4A_1 - A_3^2} \\
    g &= -\frac{A_3 L_0}{A_4A_1 - A_3^2}
\end{align}
where $\tilde{S} \equiv S - g^2/C$ is the effective extensional modulus.
As in the case of the WLC fit and the passive calculation of $L_{ps}$, this entire calculation is made differentiable by reweighting the expectations computed in Equations \ref{eq:dl} and \ref{eq:dtheta} and using implicit derivatives for the line fitting procedures in Equations \ref{eq:dl_line} and \ref{eq:dtheta_line}.
In practice, a linear line fitting procedure has an analytical derivative.

Following Ref. \cite{assenzaAccurateSequenceDependentCoarseGrained2022}, we use LAMMPS to sample reference states for each force-torque pair (see Ref. \cite{assenzaAccurateSequenceDependentCoarseGrained2022} for simulation parameters).
For each force-torque pair, we perform three identical simulations of length $2.5 \times 10^6$ timesteps and sample reference states every $2,500$ steps.
Before sampling reference states, we initiate each simulation with an equilibration period of $50,000$ timesteps.
We resample reference states with $\lambda = 0.95$ or after five gradient updates.

In Figure 3 we also consider the experimental uncertainties of the torsional modulus, effective stretch modulus, and twist-stretch coupling.
Consider an observable $X$ with target value $X^{target}$ and experimental uncertainty $\epsilon_X$.
To account for this in our objective function, we use the following form:
\begin{equation}\label{eq:omega2}
    \mathcal{L}(X^{sim}) = 
    \begin{cases} 
        0 & \text{if } X^{target}_{lo} \leq X^{sim} \leq X^{target}_{hi} \\
        \text{RMSE}(X^{sim}, X^{target}_{lo}) & \text{if } X^{sim} < X^{target}_{lo} \\
        \text{RMSE}(X^{sim}, X^{target}_{hi}) & \text{otherwise} 
    \end{cases}
\end{equation}
where $X^{target}_{lo} = X^{target} - \epsilon_X$, $X^{target}_{hi} = X^{target} + \epsilon_X$, and $X^{sim}$ is the value of $X$ measured via simulation.

\section{Thermodynamic Optimization}\label{sec:thermo-appendix}

In this section, we describe the details of our thermodynamic optimization including enhanced sampling protocols, simulation details, temperature extrapolation, finite size corrections, line fitting procedures, and observable definitions.

\subsection{Duplex Melting Temperature Calculation}

The melting temperature ($T_m$) of a DNA duplex is defined as the temperature at which 50\% of the DNA strands are bound in a duplex and 50\% of the DNA strands remain unbound as single-stranded DNA.
In general, binding and unbinding events occur on long timescales and therefore require enhanced sampling methods to calculate via simulation.

At a high level, melting temperature calculations are performed in oxDNA as follows.
Given a DNA duplex, a discrete order parameter is assigned to each base pair indicating whether or not the two nucleotides are bound.
Thus, a global order parameter indicating the total number of base pairs for a given state can be defined; a state is defined as bound if it has at least one base pair and unbound otherwise.
Next, a biased simulation is run at a single temperature in which an umbrella weight is assigned to each possible number of base pairs.
The simulation results are then unbiased to obtain the number of states visited in the bound and unbound states.
These counts are then extrapolated to a range of temperatures and a finite size correction is applied to the extrapolated counts.
Finally, a melting curve is fit to the ratio of bound to unbound counts at each temperature and the melting temperature is determined as the temperature at which there are 50\% bound and 50\% unbound states.

Formally, consider a system of two complimentary sequences of single-stranded DNA of length $n$ that can base pair to form a duplex.
The $i^{th}$ base pair is comprised of the $i^{th}$ and the $(2n - i + 1)^{th}$ nucleotides with the first and second strands represented by nucleotides indexed by $[1, n]$ and $[n+1, 2n]$, respectively.
Given a configuration $R \in \mathbb{R}^{2n \times 3}$, assume the existence a base-pair level order parameter
\begin{equation}
    B(i) = 
    \begin{cases}
      1 &\text{if $R_i$ and $R_{2n - i + 1}$ are bonded}\\
      0 &\text{otherwise}
    \end{cases}
\end{equation}
We can then define a state-level order parameter that computes the total number of base pairs,
\begin{align}
    N(R) = \sum_{i = 1}^n B(i)
\end{align}
Since there are $n$ possible base pairs, $N(R) \in [0, n]$.
We then define a vector of umbrella weights $\omega \in \mathbb{R}^{n+1}$ where $\omega_i$ denotes the umbrella weight applied to a state with $i$ base pairs.
If $U : \mathbb{R}^{2n \times 3} \to \mathbb{R}$ denotes the oxDNA energy function, umbrella sampling via these weights will sample states from the following probability distribution:
\begin{align}
    p(R) = \frac{\omega_{N(R)} \exp\left(-\beta U\left(R\right)\right)}{Z}
\end{align}
where
\begin{align}
    Z = \int \omega_{N(R')} \exp(-\beta U(R')) dR'
\end{align}
Next, we sample a trajectory of $T$ states $\overrightarrow{R} \in \mathbb{R}^{T \times 2n \times 3}$ from this distribution via virtual-move Monte Carlo (VMMC) at temperature $\tau$.
Given this trajectory, we can compute the biased number of states with $i$ base pairs, $S_i$:
\begin{align}
    S_i^{biased} = \sum_{t = 1}^T \delta(N(\overrightarrow{R}_t) = i)
\end{align}
For an arbitrary state-level observable $A$, the expected value with respect to a set of states sampled from an umbrella sampling simulation can be computed as
\begin{align}
    \langle A \rangle = \frac{\langle A / \omega \rangle_{\overrightarrow{R}}}{ \langle 1 / \omega \rangle_{\overrightarrow{R}}}
\end{align}
where $\omega$ denotes the umbrella weight for a given state.
Thus, we can obtain an unbiased value of $S_i$,
\begin{align}
    S_i^{unbiased} = \sum_{t = 1}^T \frac{1}{\omega_i}\delta(N(\overrightarrow{R}_t) = i)
\end{align}
where we ignore the normalization factor as we only care about ratios of counts when computing $T_m$.
In the same manner, given a simulation at temperature $\tau$, one can treat this simulation as a biased sample of the ensemble at another temperature $\tau '$ and the expected value of an observable $A$ can be corrected in the same way as in umbrella sampling ~\cite{sengar2021primer}:
\begin{align}
    \langle A(x) \rangle_{\tau '} = \langle A(x) \exp\left(U\left(x, \tau\right) / k\tau  - U\left(x, \tau '\right) / k\tau ' \right) \rangle
\end{align}
Note that in oxDNA $U(x, \tau ')$ is different than $U(x, \tau)$ due to the temperature-dependence of the stacking term in the interaction energy.
Thus, given our set of states sampled at temperature $\tau$ with our umbrella weights $\overrightarrow{\omega}$, we can obtain an unbiased value of $S_i$ at temperature $\tau '$:
\begin{align}
    S_{i\text{, }\tau '}^{unbiased} = \sum_{t = 1}^T \frac{1}{\omega_i} \exp(\Delta U(\overrightarrow{R}_t)_{\tau '}) \delta(N(\overrightarrow{R}_t) = i) \label{eqn:unb_count_extrap}
\end{align}
where
\begin{align}
    \Delta U(\overrightarrow{R}_t)_{\tau '} = U(\overrightarrow{R}_t, \tau) / k\tau  - U(\overrightarrow{R}_t, \tau ') / k\tau '
\end{align}
Again, we omit the normalizing factor.
Given $S_{i\text{, }\tau '}^{unbiased}$ for $i \in [0, n]$ for a given temperature $\tau '$, we then compute the total number of bound and unbound states as
\begin{align}
    S_{\text{bound, } \tau '} &= \sum_{i = 1}^n S_{i\text{, }\tau '}^{unbiased} \\
    S_{\text{unbound, } \tau '} &= S_{0\text{, }\tau '}^{unbiased}
\end{align}
and define the relative probability with which bound and unbound states are observed as
\begin{align}
    \Phi_{\tau '} = \frac{S_{\text{bound, } \tau '}}{S_{\text{unbound, } \tau '}}
\end{align}
The fraction of bound pairs can then be expressed as
\begin{align}
    f_{1\text{, }\tau '} = \frac{\Phi}{1 + \Phi} \label{eqn:f1}
\end{align}
Importantly, $f_{1\text{, }\tau '}$ is computed using states sampled from a simulation of two DNA strands in a finite-sized box with periodic boundaries.
In practice, this quantity is affected by fluctuations in the local concentration of reactants and therefore $f_{1\text{, }\tau '} \neq f_{\infty\text{, }\tau '}$ where $f_{\infty\text{, }\tau '}$ is the bulk equilibrium bonding fraction at the same temperature and concentration.
To account for this source of error, we apply the correction introduced by Ouldridge~\cite{ouldridge2012coarse}:
\begin{align}
    f_{\infty\text{, }\tau '} = \left( 1 + \frac{1}{2\Phi} \right) - \sqrt{\left( 1 + \frac{1}{2\Phi} \right)^2 - 1} \label{eqn:finf}
\end{align}
For a set of evenly-spaced temperatures $\{\tau_{lo}, \cdots, \tau_{hi}\}$, we apply Equations \ref{eqn:unb_count_extrap}-\ref{eqn:finf} to compute $\{ f_{\infty\text{, }\tau_{lo}}, \cdots, f_{\infty\text{, }\tau_{hi}} \}$. 
Finally, we interpolate to find the temperature $\tau^*$ at which $f_{\infty\text{, }\tau^*} = 0.5$.
We can similarly interpolate to compute the width of the melting curve where
\begin{align}
    T_{\text{width}} = \tau_{0.2} - \tau_{0.8}
\end{align}
where $\tau_{0.2}$ and $\tau_{0.8}$ are the temperatures at which the bulk equilibrium bonding fraction is $0.2$ and $0.8$, respectively.

\begin{figure}[t!]
\begin{center}
\centerline{\includegraphics[width=1.0\columnwidth]{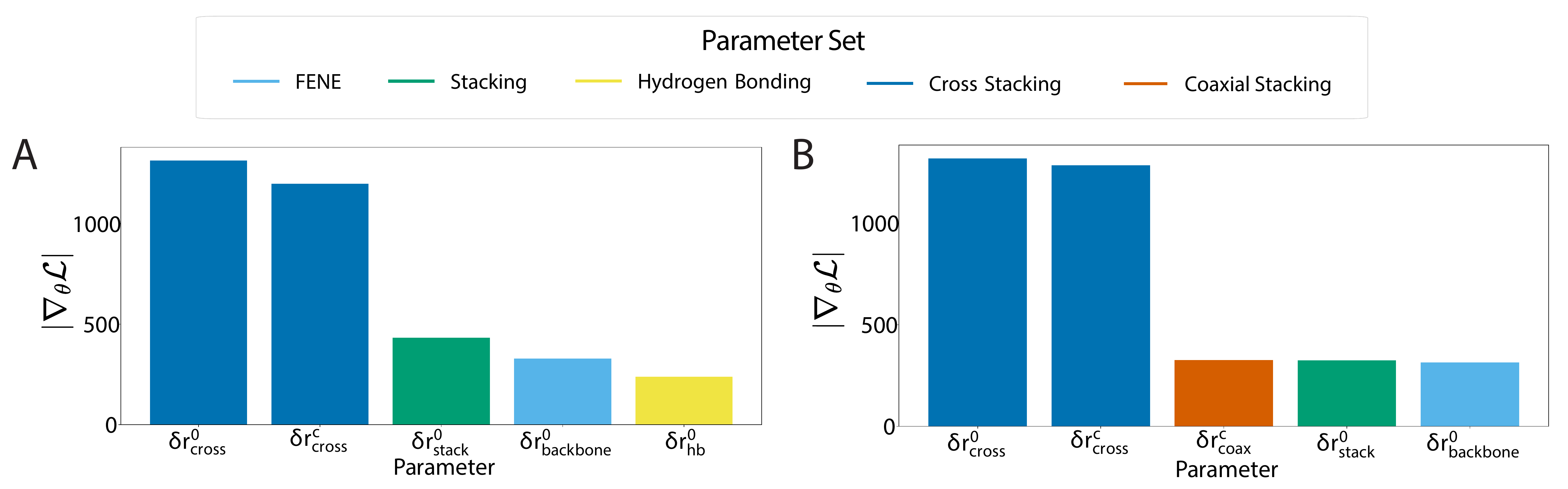}}
\caption{
Sensitivity analysis for the \textbf{A.} duplex and \textbf{B.} hairpin $T_m$ optimizations, represented by the five most significant parameters ranked by the absolute value of their gradients.
}
\label{fig:thermo_grads}
\end{center}
\end{figure}

Given a set of reference states sampled at a single temperature $\tau$ subject to umbrella weights $\omega$, the calculation of the melting temperature (and melting curve width) are differentiable via trajectory reweighting and implicit differentiation.
Consider the calculation of $S_{i\text{, }\tau '}^{unbiased}$ via Equation \ref{eqn:unb_count_extrap}.
As this quantity is a sum rather than an expected value and therefore the contribution of each state is scaled by $1$ rather than $1/N$ in the case where $\theta = \hat{\theta}$, we rewrite Equation \ref{eqn:unb_count_extrap} as
\begin{align}
    S_{i\text{, }\tau '}^{unbiased} = \sum_{t = 1}^T \frac{T}{w_i}\frac{1}{\omega_i} \exp(\Delta U(\overrightarrow{R}_t)_{\tau '}) \delta(N(\overrightarrow{R}_t) = i) \label{eqn:diff-sum}
\end{align}
where $w_i$ is the DiffTRE weight for the $i^{th}$ state.
Importantly, $\frac{N}{w_i} = 1$ when $\theta = \hat{\theta}$.
Additionally, as $\nabla_{\theta} \omega = 0$, such prefactors can be computed.
However, given the temperature dependent stacking term in the oxDNA potential, $\nabla_{\theta} \exp(\Delta U(\overrightarrow{R}_t)_{\tau '}) \neq 0$ in the general case where $\theta$ includes stacking parameters.
The oxDNA standalone code uses an energy threshold for computing $B(i)$ and therefore $\nabla_{\theta} B(i) \neq 0$ but as this quantity is thresholded (and not continuous) we neglect this term; we find that this is a reasonable approximation in practice.
Given our differentiable formulation of $S_{i\text{, }\tau '}^{unbiased}$ in Equation \ref{eqn:diff-sum}, the remaining calculation of the melting temperature and width require elementary differentiable operators, including the process of one-dimensional linear interpolation.

Above, we describe our method of differentiable melting curve calculation via a single order parameter (i.e. number of base pairs).
In practice, we use a second order parameter, the inter-strand center of mass distance, to accelerate sampling.
Equation \ref{eqn:diff-sum} can flexibly accomodate an arbitrary number of order parameters as the delta function can be evaluated for an arbitrary subset of order parameters while the weight for a given state $\omega_i$ depends on all biased order parameters used in sampling.

Computing $T_m$ values requires sampling many uncorrelated states. 
To sample reference states, we perform 34 individual simulations of $5 \times 10^7$ virtual-move Monte Carlo (VMMC) steps, sampling snapshots every $2.5 \times 10^5$ steps.
We perform biased simulations at $307.15 \text{ K}$.
We resample reference states with $\lambda = 0.95$ or after three gradient updates.
All duplex $T_m$ calculations are done with the oxDNA 1.0 force field rather than oxDNA 2.0.

\subsection{Hairpin Melting Temperature Calculation}

The (differentiable) calculation of hairpin melting temperatures and melting curve widths is nearly identical to that of duplexes.
However, since hairpin melting is the behavior of a single strand rather than of two independent strands, the melting behavior is not affected by fluctuations in the local concentration of single strands.
Therefore, $f_{1\text{, }\tau '} \neq f_{\infty\text{, }\tau '}$ and Equation \ref{eqn:finf} is not be applied.
As in the case of duplex melting, we use two order parameters for umbrella sampling -- (1) the number of base pairs and (2) the binned minimum distance between all complementary nucleotides.
We referenced \url{https://github.com/WillTKaufhold1/oxdna-tutorial} for an example implementation of such a simulation.

As with duplex $T_m$'s, accurately computing hairpin $T_m$'s requires sampling many uncorrelated states.
To sample reference states, we perform 32 independent simulations of $2.5 \times 10^7$ VMMC steps, sampling snapshots every $50,000$ steps.
We perform biased simulations at $330.15 \text{ K}$.
Reference states are resampled via the same criteria as with duplex $T_m$'s.

\begin{figure*}[t!]
\begin{center}
\centerline{\includegraphics[width=0.95\textwidth]{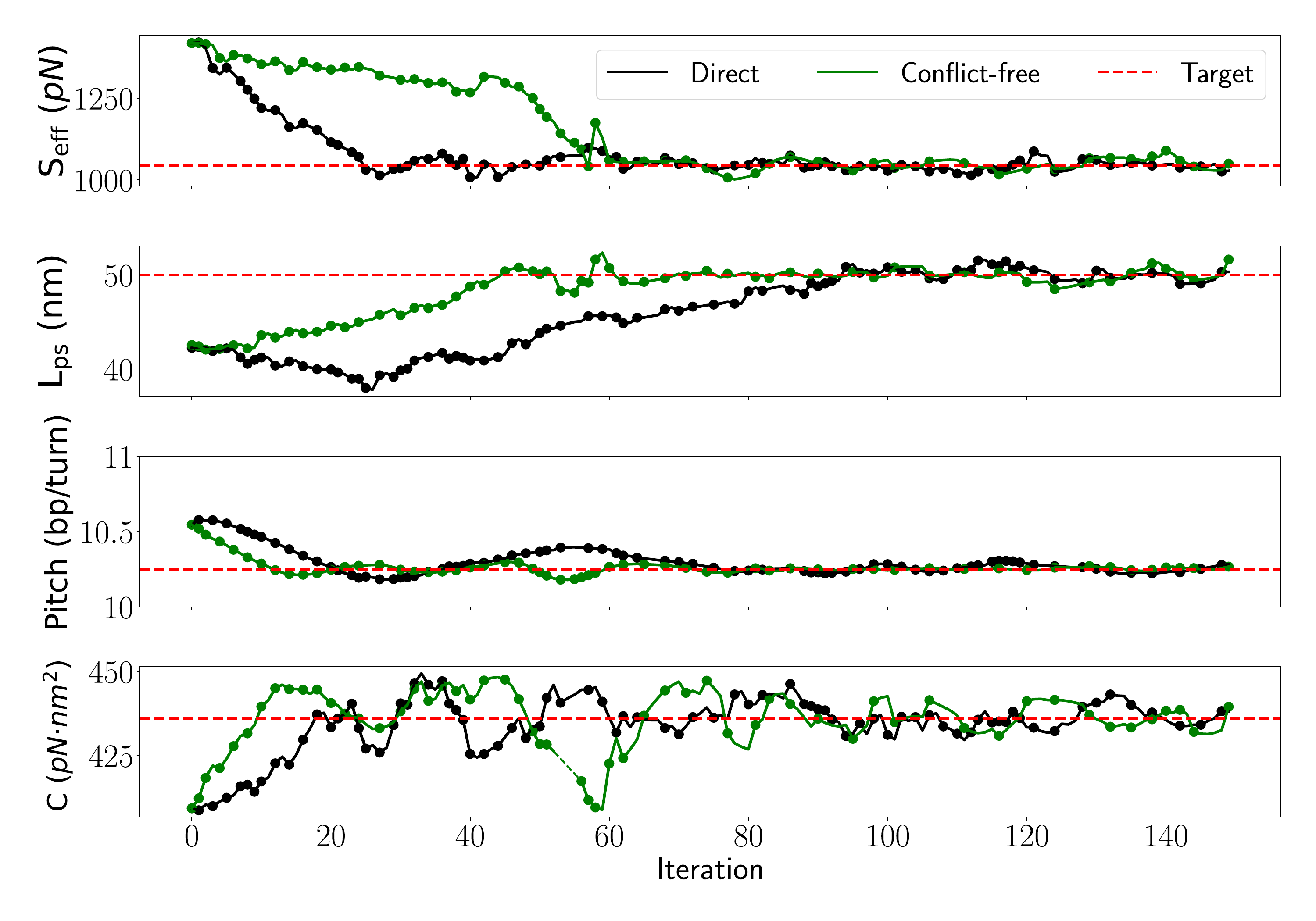}}
\caption{
Fitting oxDNA parameters to $S_{\text{eff}}$, $L_{ps}$, pitch, and $C$, with and without conflict-free updates.
The target pitch is $10.25$ bp/turn, the target $S_{\text{eff}}$ is $1045$ pN, and the target $L_{ps}$ is $50$ nm, and the target $C$ is $436 \text{ pN}\cdot\text{nm}^2$.
Outliers in the torsional modulus $C$ at iterations 53-55 with conflict-free updates are omitted and replaced with a dashed line.
Iterations in which reference states are resampled are represented as scatter points.
}
\label{fig:seff-lp-c-pitch}
\end{center}
\end{figure*}

\end{document}